\documentclass[12pt]{article}

\usepackage[utf8]{inputenc}
\usepackage[T1]{fontenc}
\usepackage{amsmath,amssymb,mathtools}
\usepackage{slashed}
\usepackage{graphicx}
\usepackage{booktabs,multicol,multirow}
\usepackage[shortlabels]{enumitem}
\usepackage{hyperref}
\usepackage[capitalize]{cleveref}
\usepackage{xspace}
\usepackage[noblocks]{authblk}
\usepackage[dvipsnames]{xcolor}
\usepackage{soul}
\usepackage[textsize=tiny]{todonotes}
\usepackage{subcaption}
\usepackage[normalem]{ulem}
\usepackage{nicefrac}
\usepackage{bm}
\usepackage{units}
\usepackage{makecell}
\usepackage[frozencache=true]{minted}
\definecolor{bg}{rgb}{0.95,0.95,0.95}
\usepackage[sorting=none,%
  citestyle=numeric-comp,%
  bibstyle=mynumeric,%
  giveninits=true]{biblatex}

\usepackage[compat=1.1.0]{figs/feynman/tikzfeynman/tikzfeynman}

\newcommand{\MS}{{\ensuremath{\overline{\text{MS}}}}\xspace}
\newcommand{\DR}{{\ensuremath{\overline{\text{DR}}}}\xspace}
\newcommand{\OS}{{\ensuremath{\text{OS}}}\xspace}
\newcommand{\cp}{\ensuremath{\mathcal{CP}}\xspace}
\newcommand{\sw}{\sin\theta_w}

\newcommand{\sws}{s_{\theta_w}}
\newcommand{\cws}{c_{\theta_w}}
\newcommand{\sing}{\xspace\text{sng}\xspace}
\newcommand{\Ztwo}{\mathbb{Z}_2}
\newcommand{\lag}{\mathcal{L}}

\newcommand{\nn}{\nonumber}

\newcommand{\lamhhh}{\ensuremath{\lambda_{hhh}}\xspace}
\newcommand{\kaplam}{\ensuremath{\kappa_\lambda}\xspace}
\newcommand{\vev}{\ensuremath{\text{VEV}}\xspace}

\newcommand{\tev}{\,\, \mathrm{TeV}}
\newcommand{\gev}{\,\, \mathrm{GeV}}

\newcommand{\ie}{\textit{i.e.}\xspace}
\newcommand{\eg}{\textit{e.g.}\xspace}
\newcommand{\cf}{\textit{cf.}\xspace}

\newcommand{\anyH}{\texttt{anyH3}\xspace}
\newcommand{\anyBSM}{\texttt{anyBSM}\xspace}

\newcommand{\anyPU}{\texttt{anyPerturbativeUnitarity}}
\newcommand{\SARAH}{\texttt{SARAH}\xspace}
\newcommand{\SPheno}{\texttt{SPheno}\xspace}
\newcommand{\FeynRules}{\texttt{FeynRules}\xspace}
\newcommand{\FeynArts}{\texttt{FeynArts}\xspace}
\newcommand{\FormCalc}{\texttt{FormCalc}\xspace}
\newcommand{\UFO}{\texttt{UFO}\xspace}
\newcommand{\py}{\texttt{Python}\xspace}
\newcommand{\pyCollier}{\texttt{pyCollier}\xspace}
\newcommand{\COLLIER}{\texttt{COLLIER}\xspace}
\newcommand{\mat}{\texttt{Mathematica}\xspace}

\newcommand{\docs}{\href{https://anybsm.gitlab.io/}{online documentation}\xspace}
\newcommand{\examples}{\href{https://anybsm.gitlab.io/examples.html}{examples}\xspace}
\newcommand{\modelsrepo}{\href{https://anybsm.gitlab.io/models.html}{model repository}\xspace}
\newcommand{\examplesrepo}{\href{https://anybsm.gitlab.io/examples.html}{examples repository}\xspace}

\newcommand{\TSM}[1]{\text{TSM}$_{Y= #1}$\xspace}

\NewBibliographyString{refname}
\NewBibliographyString{refsname}
\DefineBibliographyStrings{english}{%
  refname = {Ref\adddot},
  refsname = {Refs\adddot}
}
\DeclareCiteCommand{\ccite}
  {%
  \ifnum\thecitetotal=1
    \bibstring{refname}%
  \else%
    \bibstring{refsname}%
  \fi%
  \addnbspace\bibopenbracket%
  \usebibmacro{cite:init}%
   \usebibmacro{prenote}}
  {\usebibmacro{citeindex}%
   \usebibmacro{cite:comp}}
  {}
  {\usebibmacro{cite:dump}%
   \usebibmacro{postnote}%
   \bibclosebracket}
\newrobustcmd*{\Ccite}{\bibsentence\ccite}

\newmintinline[code]{python}{}

\evensidemargin \oddsidemargin
\begin{document}

\thispagestyle{empty}
\def\thefootnote{\fnsymbol{footnote}}

\begin{flushright}
\texttt{DESY-23-042}\\
\texttt{EFI-23-1}
\end{flushright}
\vspace{-1.5cm}\includegraphics[height=3cm]{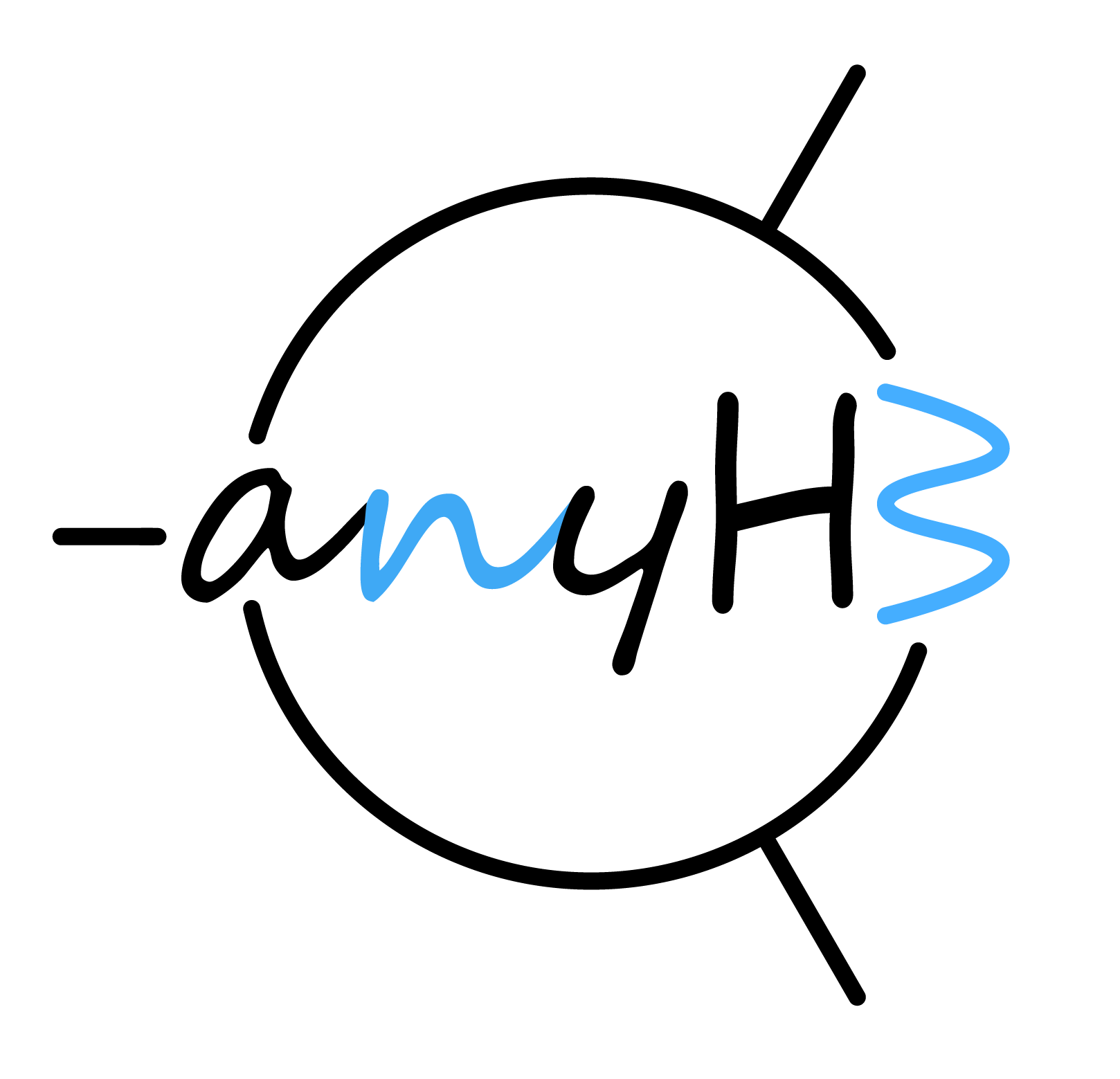}
\vspace{.5cm}
\begin{center}
{\large{\bf
\anyH: precise predictions for the trilinear Higgs coupling\\[.5em]
in the Standard Model and beyond
}}
\\
\vspace{3em}
 {
Henning Bahl$^{1}$\footnotetext[0]{hbahl@uchicago.edu, 
johannes.braathen@desy.de, martin.gabelmann@desy.de,
georg.weiglein@desy.de},
Johannes Braathen$^{2}$,
Martin Gabelmann$^{2}$,
Georg Weiglein$^{2,3}$
 }\\[2em]
 {\sl $^1$ University of Chicago, Department of Physics and Enrico Fermi Institute,\\[0.2em] 5720 South Ellis Avenue, Chicago, IL 60637 USA}\\[0.2em]
 {\sl $^2$    Deutsches Elektronen-Synchrotron DESY, Notkestr.~85, 22607 Hamburg, Germany}\\[0.2em]
 {\sl $^{3}$ II.\  Institut f\"ur  Theoretische  Physik, Universit\"at  Hamburg, Luruper Chaussee 149,\\ 22761 Hamburg, Germany}
\def\thefootnote{\arabic{footnote}}
\setcounter{page}{0}
\setcounter{footnote}{0}
\end{center}
\vspace{2ex}

\begin{abstract}

The trilinear Higgs coupling $\lambda_{hhh}$ of the detected Higgs boson is an important probe for physics beyond the Standard Model. Correspondingly,
improving the precision of the theoretical predictions for this coupling as well as the experimental constraints on it are among the main goals of particle physics in the near future. In this article, we present the public \py code \anyH, which provides precise theoretical predictions for $\lambda_{hhh}$. The program can easily be used for any renormalisable model, where for the input the \texttt{UFO} format is adopted. It allows including corrections up to the full one-loop level with arbitrary values of the external squared momenta and features a semi-automatic and highly flexible renormalisation procedure. The code is validated against known results in the literature. Moreover, we present new results for $\lambda_{hhh}$ in models which so far have not been investigated in the literature. 

\end{abstract}
\setcounter{footnote}{0}
\renewcommand{\thefootnote}{\arabic{footnote}}

\newpage
\tableofcontents
\newpage

\section{Introduction}
The discovery of a Higgs boson at the CERN LHC~\cite{Chatrchyan:2012xdj,Aad:2012tfa} has confirmed that the Higgs potential plays a crucial role in the electroweak symmetry breaking (EWSB). While the measured properties of this Higgs boson are so far compatible with the predictions of the Standard Model (SM) within the experimental and theoretical uncertainties, the structure of the Higgs sector and of its potential remain to be determined. Furthermore, in spite of the successes of the SM, it is clear that new, Beyond-the-Standard-Model (BSM) physics is needed to address deficiencies of the SM --- such as for instance the lack of an explanation for the baryon asymmetry of the Universe. 

In this context, a key quantity to investigate is the trilinear Higgs coupling \lamhhh. This coupling determines the shape of the Higgs potential away from the electroweak (EW) minimum and in turn controls the nature and strength of the EW phase transition (EWPT). For instance, a strong first-order EWPT, which is a requirement~\cite{Sakharov:1967dj} for the scenario of EW baryogenesis~\cite{Kuzmin:1985mm,Cohen:1993nk},
is typically associated with a sizeable deviation of \lamhhh from its SM prediction, as was discussed first in Refs.~\cite{Grojean:2004xa,Kanemura:2004ch} (see also Ref.~\cite{Basler:2017uxn} for a more recent example). Even beyond its crucial role in the context of the EWPT, \lamhhh provides a unique opportunity to find signs of BSM physics arising from extended Higgs sectors. In particular, the loop contributions in models with additional scalars can cause the trilinear Higgs coupling to deviate by up to several hundred percent from its prediction in the SM if there is a substantial mass splitting between the BSM mass scales. This was found first at the one-loop level for the case of a Two-Higgs-Doublet Model (THDM) in \ccite{Kanemura:2002vm,Kanemura:2004mg} but is now known to occur for a wide range of BSM models. The genuine physical nature of these effects was confirmed in \ccite{Braathen:2019pxr,Braathen:2019zoh}, where next-to-leading order (NLO), \ie\ two-loop, corrections to those large one-loop effects were investigated and found to obey the expected perturbative 
behaviour. Unlike what is the case for most couplings of the SM-like Higgs boson at 125~GeV, large deviations in \lamhhh are possible even in scenarios where all its couplings are very close to the SM values at tree level, such as in aligned scenarios~\cite{Gunion:2002zf}. Meanwhile, it has recently been shown in \ccite{Bahl:2022jnx} that
the experimental limits (discussed in further detail below) have become sufficiently strong to probe these potentially large loop effects, and thus the comparison of the predictions for \lamhhh with the latest experimental bounds constitutes a powerful new method for constraining the parameter space of BSM theories (probes of BSM parameter space using the di-Higgs production cross-section directly have also been discussed in $e.g.$ Ref.~\cite{Abouabid:2021yvw}). In this context it is important to keep in mind that \lamhhh cannot be directly measured experimentally. The crucial experimental quantity where \lamhhh enters at leading order is the process of Higgs pair production. The computation of the trilinear Higgs coupling \lamhhh constitutes a necessary intermediate result for the prediction of di-Higgs boson production. In fact, for the case where \lamhhh receives large loop corrections, the additional contributions to the di-Higgs production process may be of sub-leading order~\cite{Bahl:2022jnx}. More generally, computations of trilinear Higgs couplings are also important for investigating decays of BSM Higgs bosons and BSM decays of the SM-like Higgs boson. 

The present experimental information on the trilinear Higgs coupling \lamhhh is by far not as precise as what has been achieved for other couplings of the Higgs boson~\cite{ATLAS:2022vkf,CMS:2022dwd}. Indeed, the current best limits on \lamhhh\ were obtained by the ATLAS collaboration using a combination of data from searches for (non-resonant) di-Higgs production and from the experimental results for single-Higgs production processes; they bound the ratio $\kappa_\lambda$, defined as
\begin{align}
\label{eq:def_kaplam}
    \kappa_\lambda\equiv \frac{\lambda_{hhh}}{(\lambda_{hhh}^\text{SM})^{(0)}}\,,
\end{align}
to be within the range $-0.4<\kappa_\lambda<6.3$ at the 95\% confidence level (C.L.)~\cite{ATLAS:2022kbf,ATLAS:2022jtk}. In \cref{eq:def_kaplam}, $(\lambda_{hhh}^\text{SM})^{(0)}$ denotes the tree-level prediction for the trilinear coupling in the SM. The CMS collaboration has obtained similar results~\cite{CMS:2022dwd}, namely $-1.24<\kappa_\lambda<6.49$. The quoted limits were obtained under the assumption that besides a variation of $\kappa_\lambda$ all other couplings entering the analyses are fixed to their SM values. The current experimental limits leave ample room for BSM deviations, which would so far remain unobserved, but could be accessed in the foreseeable future, given the expected prospects for probing \lamhhh at the LHC and future colliders --- see \ccite{deBlas:2019rxi} for a review. Specifically, at the high-luminosity upgrade of the LHC (the HL-LHC), the projection for $\kappa_\lambda$ at 95\% C.L.\ is $0.1<\kappa_\lambda<2.3$~\cite{Cepeda:2019klc}. At a future $e^+e^-$ linear collider and a 100-TeV hadron collider it is expected that $\kappa_\lambda$ can be determined at the level of ${\cal O}(10\%)$~\cite{deBlas:2019rxi, Fujii:2015jha, Fujii:2017vwa, Roloff:2019crr,Goncalves:2018yva, Chang:2018uwu}. It should be noted that these projections were obtained under the assumption that $\kappa_\lambda = 1$ is realised in nature and may significantly change if the actual value of $\kappa_\lambda$ is different. In particular, for an enhanced value of $\kappa_\lambda$ the prospects for extracting \lamhhh from the process $e^+e^- \to Zhh$ at a linear collider with about 500~GeV would improve, while as a consequence of destructive interference contributions the prospects at the HL-LHC would deteriorate, see \eg\ \ccite{Durig:2016jrs,LCCPhysicsWorkingGroup:2019fvj,Biekotter:2022kgf}.

As stressed above, the most direct probe of the trilinear Higgs coupling are searches for di-Higgs production, because this process involves \lamhhh already at the leading order (LO). Single-Higgs production involves contributions of the trilinear Higgs coupling starting at the next-to-leading order (NLO) and EW precision observables at the next-to-next-to-leading order (NNLO) --- see for instance \ccite{Degrassi:2016wml,Degrassi:2017ucl}.  Of course, a general analysis should not be restricted to the case where BSM contributions enter exclusively via the trilinear Higgs coupling. On the other hand, in scenarios where large loop corrections to \lamhhh constitute the leading contributions to di-Higgs production, an effective coupling approach where the dominant corrections are incorporated into \kaplam provides a convenient framework to efficiently constrain BSM models with available experimental results, as discussed \eg\ in \ccite{Kanemura:2016lkz,Bahl:2022jnx}. We will discuss in this paper in more detail the applicability of experimental constraints set on \lamhhh.

A number of computations of the trilinear Higgs couplings in BSM theories have been carried out in the literature. At one-loop order, corrections were first computed in the SM and the Minimal Supersymmetric Standard Model (MSSM) in \ccite{Barger:1991ed,Hollik:2001px,Dobado:2002jz} (see also \ccite{Williams:2007dc,Williams:2011bu} for the case of the MSSM with complex parameters). One-loop calculations of \lamhhh have since also been performed in the Next-to-MSSM (NMSSM) in \ccite{Nhung:2013lpa} as well as for various non-supersymmetric extensions of the SM: with singlets~\cite{Kanemura:2015fra,Kanemura:2016lkz,He:2016sqr,Kanemura:2017wtm}, additional doublets~\cite{Kanemura:2002vm,Kanemura:2004mg,Kanemura:2015mxa,Arhrib:2015hoa,Kanemura:2016sos,Kanemura:2017wtm,Falaki:2023tyd}, and triplets~\cite{Aoki:2012jj,Chiang:2018xpl,Bahl:2022gqg}. For some of these models, specific results for \lamhhh are available in the public programs \texttt{H-COUP}~\cite{Kanemura:2017gbi,Kanemura:2019slf} and \texttt{BSMPT}~\cite{Basler:2018cwe,Basler:2020nrq}. At two-loop order, \ccite{Senaha:2018xek,Braathen:2019pxr} obtained the two-loop $\mathcal{O}(\alpha_t\alpha_s)$ and $\mathcal{O}(\alpha_t^2)$ corrections to \lamhhh in the SM. In supersymmetric theories,  \ccite{Brucherseifer:2013qva,Muhlleitner:2015dua} investigated the $\mathcal{O}(\alpha_t\alpha_s)$ corrections to \lamhhh in the MSSM and the NMSSM, respectively, and recently \ccite{Borschensky:2022pfc} extended the NMSSM calculation to include also $\mathcal{O}(\alpha_t^2)$ effects. Regarding non-supersymmetric models, the leading two-loop BSM contributions to \lamhhh (arising from BSM scalars and, potentially, top quarks) are known for the Inert Doublet Model (IDM)~\cite{Senaha:2018xek,Braathen:2019pxr,Braathen:2019zoh}, THDMs~\cite{Braathen:2019pxr,Braathen:2019zoh}, $O(N)$-symmetric real-singlet extensions of the SM~\cite{Braathen:2019zoh,Braathen:2020vwo}, and for various models with classical scale invariance~\cite{Braathen:2020vwo}. 

In this work, we present the \py package \anyH, which takes a big step forward in facilitating the prediction of \lamhhh.\footnote{We note that for the remainder of this paper \lamhhh is defined to refer specifically to the renormalised one-loop corrected trilinear coupling of the SM-like Higgs boson at 125~GeV.}
\anyH allows the analytic and/or numerical computation of the trilinear Higgs coupling for general renormalisable theories  to full one-loop order. For user convenience, the model definitions needed in \anyH in order to enable the application of generic results to specific theories can be provided in the form of the widely-employed \UFO format~\cite{Degrande:2011ua,Darme:2023jdn}. \anyH also offers a high level of flexibility in the renormalisation schemes used in calculations --- with pre-defined commands for standard scheme choices and the additional possibility for the user to define other choices of counterterms. Furthermore, the tool allows the user to modify the treatment of tadpole contributions (for recent discussions see \eg\ \ccite{Dudenas:2020ggt,Braathen:2021fyq,Dittmaier:2022maf,Dittmaier:2022ivi}, Appendix A of \ccite{Krause:2016oke}, and section~4 of \ccite{Krause:2017mal}). \anyH is part of the wider \anyBSM framework, where developments for further (pseudo-)observables are foreseen in the future. An additional \anyBSM feature that is already available is the module \anyPU, which allows efficient and reliable verifications of perturbative unitarity constraints (at leading order, and in the high-energy limit). 

This paper is organised as follows: we start by discussing in \cref{sec:genericcalc} the main elements of our automated computation of \lamhhh, as well as the interpretation of the obtained results. Next, we present in \cref{sec:flow} the workflow of \anyH before presenting a brief tutorial of the program in \cref{sec:tutorial}. In \cref{sec:crosschecks}, we discuss the cross-checks that were performed for the various models that are installed along with \anyH. Finally, we present in \cref{sec:applications} examples of applications of \anyH with an emphasis on new results. 
We summarise our results in \cref{sec:conclusions}. A number of Appendices provide additional details on the program and the considered models: \cref{app:generic} contains our conventions for general renormalisable models and for the generic expressions included in \anyH as well as the conventions or restrictions on \texttt{UFO} model files; \cref{app:newmodels} discusses the various ways of generating compatible \UFO models for new BSM theories; \cref{app:models} presents the different models discussed in this paper, including details on renormalisation prescriptions and the treatment of tadpole contributions; \cref{app:additionalufo} lists modifications of the \UFO standard that are applied by \anyBSM internally; \cref{app:cache} explains the caching that is available in \anyH, while \cref{app:pycollier} describes the \py interface \pyCollier to the \texttt{Fortran} library \texttt{COLLIER} employed for computing loop functions numerically.


\section{A generic approach to the trilinear Higgs coupling}
\label{sec:genericcalc}

In the following we outline the steps for the calculation of the trilinear Higgs coupling where the SM-like Higgs boson $h$ appears at each of the three external legs. A generalisation to trilinear Higgs couplings involving one or more BSM Higgs boson(s) is left for future work.

All calculations performed in the program are based on results for general renormalisable theories. These generic results were obtained by the following steps:
\begin{itemize}
    \item For the different types of contributions entering the calculation, as specified in \cref{eq:lambdahhh} below,     all possible Feynman diagram topologies were identified. 
    \item For each of these topologies, all possible insertions of generic fields (scalars, fermions, vector bosons, ghosts) were processed.
    \item Each of these generic diagrams was calculated using a generic Lagrangian with the help of \FeynArts~\cite{Kublbeck:1990xc,Eck:1992ms,Hahn:2000kx} and \FormCalc~\cite{Hahn:1998yk}.
\end{itemize}
The resulting \textit{generic} expressions were hard-coded into the \py program code to be applied to a specific model upon run-time.

The steps described above include the introduction of several conventions in how the generic Lagrangian and its resulting Feynman rules are written. For instance, all fermion-fermion-scalar operators, $F_1 F_2 S_3$, are written in terms of left- and right-handed projectors, $(c_L^{123}P_L +c_R^{123}P_R)F_1\bar{F}_2 S_3$, where the explicit form of $c_{L,R}^{123}$ (depending on the corresponding operator within the considered model) is yet unspecified. A detailed description of all used conventions is given in \cref{app:generic}. The generic results, which are expressed in terms of generic couplings, have been implemented into the \py code. Upon run-time of the program, the couplings of the specified model are mapped onto the generic Lagrangian allowing one to directly obtain results for all contributing Feynman diagrams. A similar approach is followed for example in \SARAH~\cite{Staub:2008uz,Staub:2009bi,Staub:2010jh,Staub:2012pb,Staub:2013tta} and \texttt{TLDR}~\cite{Goodsell:2019zfs}.

Currently \anyBSM contains generic results for the scalar three-point function (where in the present implementation the three external scalars are assumed to be identical), the scalar two-point function, the scalar one-point function, the vector-boson two-point function, and the vector-boson--scalar two-point function. An overview of all topologies can be found in \cref{app:diagrams}. 
Making use of these building blocks, the trilinear Higgs coupling $\lambda_{hhh}$ is calculated by the sub-module \anyH at the one-loop level,
\begin{align}\label{eq:lambdahhh}
    \lambda_{hhh} ={}& -\hat\Gamma_{hhh}(p_1^2,p_2^2,p_3^2) = \lambda^{(0)}_{hhh} + \delta^{(1)}_\text{genuine}\lambda_{hhh} + \delta^{(1)}_\text{tadpoles}\lambda_{hhh}+ \delta^{(1)}_\text{WFR}\lambda_{hhh} + \delta^{(1)}_\text{CT}\lambda_{hhh} ,
\end{align}
where $\hat\Gamma_{hhh}$ is the renormalised Higgs-boson three-point function. The superscripts indicate the loop order, namely $\lambda_{hhh}^{(0)}$ denotes the tree-level result for the trilinear Higgs coupling, while $\delta^{(1)}_\text{genuine}\lambda_{hhh}$, $\delta^{(1)}_\text{tadpoles}\lambda_{hhh}$, $\delta^{(1)}_\text{WFR}\lambda_{hhh}$, and  $\delta^{(1)}_\text{CT}\lambda_{hhh}$ are one-loop contributions. Specifically $\delta^{(1)}_\text{genuine}\lambda_{hhh}$ denotes the genuine vertex corrections, while $\delta^{(1)}_\text{tadpoles}\lambda_{hhh}$ and $\delta^{(1)}_\text{WFR}\lambda_{hhh}$ correspond to contributions involving tadpole insertions and external-leg corrections, respectively. The last term of \cref{eq:lambdahhh} denotes the counterterm contribution (see below), while the second-last term encodes the contribution from external leg corrections. As indicated in \cref{eq:lambdahhh}, \anyH is able to handle arbitrary external momenta. For all the one-loop pieces appearing in \cref{eq:lambdahhh}, we work in \anyH only with UV-finite parts, unless otherwise specified. Explicit checks of UV-finiteness can be performed in \anyH, but have been done separately. 

We also note that \anyH allows independent calculations of self-energies and tadpoles. The necessary generic results were derived following the same steps as described above.


\subsection{Renormalisation}
\label{sec:genericcalc:ren}

Renormalisation is a fundamental ingredient of loop calculations. Minimal subtraction schemes like \MS are arguably the easiest schemes to automate since their implementation basically boils down to setting the divergent parts of the appearing loop integrals to zero. 

However, for many processes it is known that \MS schemes may result in undesirable features like artificially large loop corrections or gauge dependencies. Also in the specific context of the trilinear Higgs coupling it has been observed that a potentially large part of the one-loop corrections can be absorbed into the Higgs-boson mass, which appears at the tree level, for instance by renormalising it in the on-shell (OS) scheme~\cite{Hollik:2001px, Dobado:2002jz, Braathen:2019zoh}.

For this reason, \anyH allows the specification of different renormalisation schemes. All bosonic (\ie, of particles with spin zero or spin one) masses appearing in the tree-level expression for the trilinear Higgs coupling can optionally be renormalised in the OS scheme. Moreover, also the vacuum expectation value (\vev) entering at lowest order can be renormalised in the OS scheme (see \cref{app:sm:vevren} for more details). In addition, \anyH allows the user to define custom counterterms, which are then included in the calculation of the trilinear Higgs coupling. In summary, the counterterm contribution to $\lambda_{hhh}$ reads
\begin{align}\label{eq:lambdahhhCT}
    \delta^{(1)}_\text{CT}\lambda_{hhh} ={}& \sum_i \frac{\partial}{\partial m_i^2}\lambda^{(0)}_{hhh} \cdot \delta^{(1)}_\text{CT}m_i^2 + \frac{\partial}{\partial v}\lambda^{(0)}_{hhh} \cdot \delta^{(1)}_\text{CT}v + \delta^{(1)}_{\text{custom-CT}}\lambda_{hhh},
\end{align}
where $h$ is used to denote the SM-like Higgs boson, $m_i$ denotes all scalar or vector boson masses appearing in $\lambda_{hhh}^{(0)}$, and $v$ is used to denote the electroweak \vev. The notations $\delta^{(1)}_\text{CT}x$ denote the one-loop counterterms for the parameters $x$. If one of the masses $m_i$ is chosen to be renormalised in the OS scheme, its counterterm is determined via
\begin{align}
    \delta^{(1)}_\text{CT}m_i^2 = -\mathrm{Re}\Sigma^{(1)}_{ii}(p^2=m_i^2),
\end{align}
where $\Sigma_{ii}$ is the self-energy of the scalar/vector particle $i$ (with mass $m_i$),
which is defined according to the conventions of e.g.\ \ccite{Martin:2003it,Goodsell:2019zfs}.\footnote{We note that this is also the sign convention employed by \anyH internally. 
For separate calculations of self-energies a flag can be used to switch the overall sign convention for the results, see the \docs for more details.} For the determination of the electroweak \vev~counterterm, we refer to \cref{app:sm:vevren}. We note, finally, that if no default or user-defined schemes are provided for a certain parameter, then an \MS renormalisation is employed. 


\subsection{Tadpole contributions}
\label{sec:genericcalc:tadpoles}

Besides the renormalisation of the masses and the electroweak \vev~appearing at the tree level, also tadpole contributions need to be taken into account. Since the \UFO\ standard does not provide a unified notation to store information about the minimisation of the Higgs potential\footnote{The dependence of masses and vertices on tadpole terms is at present not stored in the \UFO model files.}, we use as default setting the Fleischer--Jegerlehner tadpole scheme of \ccite{Fleischer:1980ub} (using $\MS$ tadpole counterterms). As a consequence, all tadpole diagrams have to be calculated explicitly --- see \cref{app:sm:tadren} for a detailed discussion. This does not only include explicit tadpole contributions to $\lambda_{hhh}$, denoted as $\delta^{(1)}_{\text{tadpoles}} \lambda_{hhh}$, but also to $\delta^{(1)}_{\text{CT}} \lambda_{hhh}$ and $\delta^{(1)}_{\text{WFR}} \lambda_{hhh}$. It should be stressed that the treatment of the tadpole contributions can be adapted by the user. In particular, it is possible to avoid the explicit appearance of tadpole diagrams by an appropriate choice of $\delta^{(1)}_{\text{custom-CT}}\lambda_{hhh}$ using \eg\ \OS tadpole counterterms.\footnote{Note that $\delta^{(1)}_{\text{custom-CT}}\lambda_{hhh}$ can be defined in terms of one-, two-, or three-point functions (and derivatives thereof), which are computed automatically by the code.}


\subsection{External leg corrections}
\label{sec:extleg}

\anyH also includes external leg corrections to ensure the proper normalisation of the external scalars. These corrections are given by
\begin{align}\label{eq:lambdahhhWFR}
    \delta^{(1)}\lambda_{hhh}^\text{WFR} ={}& \sum_i\left(\frac{1}{2}\Sigma_{hh}^\prime(p_i^2)\lambda^{(0)}_{hhh} + \sum_{j,h_j\neq h}\frac{\Sigma_{h h_j}(p_i^2)}{p_i^2-m_{h_j}^2}\lambda^{(0)}_{h_jhh} \right)\nn\\
    \equiv{}& \sum_i\left(\frac{1}{2}\delta^{(1)}Z_h(p_i^2)\lambda^{(0)}_{hhh} + \sum_{j,h_j\neq h}\delta^{(1)}Z_{hh_j}(p_i^2)\lambda^{(0)}_{h_jhh} \right)\,,
\end{align}
where the prime indicates a derivative with respect to the external momentum squared. The second line of this equation serves to define the notations $\delta^{(1)}Z(p^2)$, which we will employ later in this paper. For the on-shell case, $p_i^2 = m_h^2$, the first term on the right-hand side yields the usual LSZ factor as it occurs for the case without mixing between different Higgs bosons, while the second term accounts for the contributions from possible scalar mixing effects on the external legs. As explained above, the self-energies appearing in \cref{eq:lambdahhhWFR} are meant to contain only the UV-finite contributions. It should be noted that \anyH by default evaluates the external leg corrections at the same momenta as the vertex corrections. In order to implement different choices of the field renormalisations together with their appropriate wave function normalisation contributions, one can alternatively choose to turn off the automatic calculation of external-leg corrections and re-introduce the corresponding contributions in $\delta^{(1)}_{\text{custom-CT}}\lambda_{hhh}$. This is in particular needed for the case where the result for the trilinear Higgs coupling obtained with \anyH is meant to be incorporated into the prediction for the cross section for di-Higgs production. In the di-Higgs production process the trilinear Higgs coupling enters with two on-shell external legs, while the third leg is an off-shell internal line of the amplitude for di-Higgs production, see \cref{sec:momdep} and \cref{app:ssm} below for more details.


\subsection{Interpretation of the result for \texorpdfstring{\lamhhh}{lambdahhh}}

The trilinear Higgs coupling \lamhhh itself is not a physical observable. Its experimental determination from a physical process (typically di-Higgs production, but via higher-order contributions also single-Higgs production provides some sensitivity) relies on assumptions on the other couplings and particles involved in the processes. The current limits on \lamhhh from ATLAS~\cite{ATLAS:2022kbf} and CMS~\cite{CMS:2022dwd} were obtained under the assumption that all other Higgs couplings that are relevant for the respective processes have the SM values and that no other BSM particles contribute to these processes. Moreover, \lamhhh has been treated as a constant which does not depend on the inflowing momenta. When comparing the predictions obtained with \anyH in the considered model to the experimental limits, the user of \anyH should ensure that these assumptions are fulfilled sufficiently well.  

Alternatively, the output of \anyH can be used as input for other codes calculating di-Higgs production cross sections. In this case, the user can choose to include the momentum dependence and switch-off the external-leg corrections 
for the internal Higgs propagator (an explicit example for this case is discussed in \cref{fig:Yeq1_triple_p2_dep} and also in \cref{app:ssm}).


\section{User- and Program-flow}
\label{sec:flow}

\begin{figure}
    \centering
    \includegraphics[width=
    \textwidth]{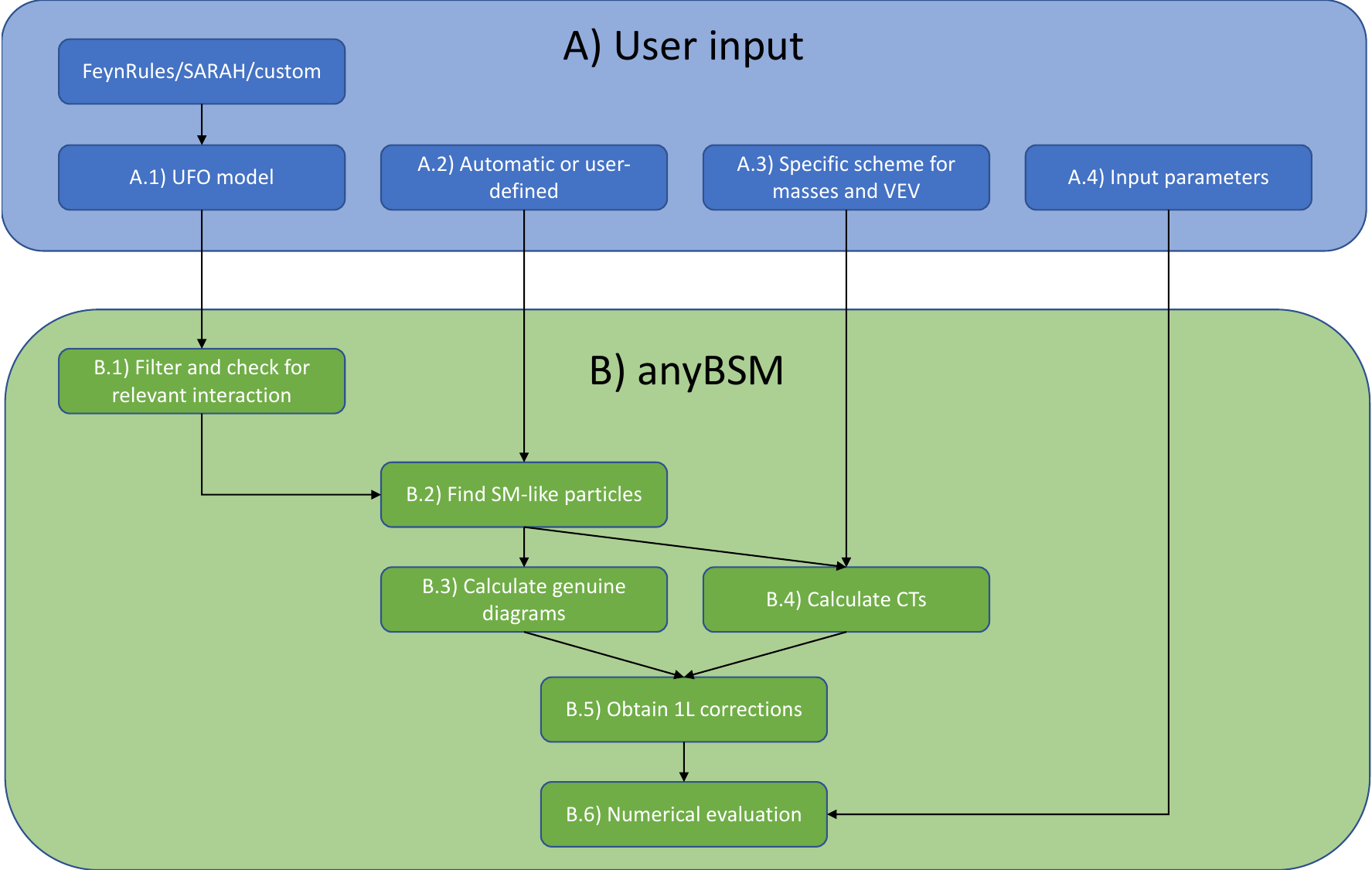}
    \caption{Schematical workflow for the calculation of $\lamhhh$.}
    \label{fig:workflow}
\end{figure}

The main objective of this work is to provide one-loop corrections to the trilinear Higgs coupling in wide classes of BSM models. A typical work-flow of how this is organised is shown in \cref{fig:workflow}. We distinguish two major sections for demonstrative purposes: A) User input and B) the actual program flow. User input is required in section A). In addition, it is also possible to control each of the steps discussed in B) by using the $\anyBSM$ library. The latter will be discussed in more detail below. 

At the user level (A) several inputs are required:
    \begin{itemize}
    \item A.1) The model-specific information such as particle content and Feynman rules relies on the \UFO standard. In order to obtain a \UFO description of the model of interest one can use \SARAH, \FeynRules, or \UFO files obtained with any other tool. For the latter two cases, we provide a converter, which is discussed in \cref{app:generic:converter}, while in the case of \SARAH the conventions for all relevant Lorentz structures match the \anyBSM conventions. A detailed description of the \UFO format can be found in \ccite{Degrande:2011ua,Darme:2023jdn}.
    \item A.2) In order to renormalise the relevant input parameters consistently, the tool needs to know which of the particles defined in the \UFO model correspond to the SM particles. This information is specified in an auxiliary file called \texttt{schemes.yml}.
    \item A.3) Once the SM parameters and particles (along with their masses) are identified, one needs to specify in which renormalisation scheme they are given. This is also done in the file  \texttt{schemes.yml}. An example specification of this file is shown in \cref{sec:documentation:ren}.
    \item A.4) Numerical values for all input parameters. Additionally, the \UFO model may provide analytic relations between input parameters (so called ``external'' parameters) and \eg\ Lagrangian parameters or mixing angles (so called ``internal'' parameters). The program automatically resolves these dependencies and writes all internal parameters in terms of external parameters. The numerical values are required for the numerical evaluation of the analytically obtained results for \lamhhh and are by default read from the \UFO model. Moreover, one can change the default parameter values individually or all at once by specifying \eg\ a {\tt SLHA}~\cite{Skands:2003cj,Allanach:2008qq} input file (see \cref{sec:tutorial:setparams} for examples). 
    \end{itemize}
It should be stressed that the relations between internal and external parameters given via the \UFO model in A.4) are essential for the renormalisation procedure. In particular \cref{eq:lambdahhhCT} will be evaluated once all parameter dependencies have been applied. Thus, if any of the Higgs masses is not defined as an input parameter in the \UFO~model, it cannot be renormalised in the \OS scheme automatically. Instead, the corresponding counterterm contribution would need to be provided manually via the custom counterterm $\delta^{(1)}_{\text{custom-CT}}\lambda_{hhh}$. It is, therefore, recommended to align the chosen parametrisation for input parameters in the \UFO model along with the chosen renormalisation schemes.

The following steps are performed automatically using the information gathered before:
\begin{itemize}
    \item B.1) The \UFO model is loaded and several checks are performed:
      \begin{itemize}
          \item Whether all relevant couplings are present (especially quartic scalar couplings which are sometimes excluded in \UFO outputs).
          \item Whether all relevant couplings are defined through the same Lorentz structures that are also used by \anyBSM. Otherwise, one can use the model converter discussed in \cref{app:generic:converter}.
      \end{itemize}
    \item B.2) Definition of the SM-like particles and parameters based on the inputs made in A.2) in the file \texttt{schemes.yml}. The program is also capable of finding the SM particles and parameters automatically based on their PDG codes and numerical (mass) values. This functionality is also used to cross-check the user-input in order to avoid erroneous configurations.
    \item B.3) and B.4) All possible field-insertions into the generic diagrams are determined. The corresponding couplings and masses are inserted into the generic results. The calculation of any $n$-point function involved in the counterterm contributions follows the same procedure. Finally, the result for every $n$-point function is stored on-disk for caching/later use (see \cref{app:cache}).
    \item B.5) Collection of the individual results and construction of the expression for the renormalised \lamhhh.
    \item B.6) Numerical or analytical evaluation. For diagrams with non-zero external momentum, the loop functions are evaluated using \pyCollier (see \cref{app:pycollier}), which is a \py interface for \COLLIER~\cite{Denner:2016kdg}. The analytical evaluation can be simplified/modified using \texttt{SymPy}~\cite{10.7717/peerj-cs.103}.
\end{itemize}
We want to stress that this particular strategy for obtaining a prediction for \lamhhh in a given model has a number of ingredients in common with the calculation of many other observables. For this reason, the code \anyH for calculating \lamhhh is embedded into a larger program called \anyBSM. The program \anyBSM provides many utilities capable of performing the steps described above to set up the calculation of a particular observable. Utilities of this kind are for instance the interface to \UFO or the insertion of \UFO particles and Feynman rules into generic Feynman diagrams. These ingredients are used in submodules --- of which \anyH is the first 
one that has been implemented --- to define actual quantities to compute. In fact, \anyH only takes care of B.5) of the program points mentioned above while all other steps are taken care of in decoupled classes/modules of the program \anyBSM.

\begin{figure}
    \centering
    \includegraphics[width=0.6
    \textwidth]{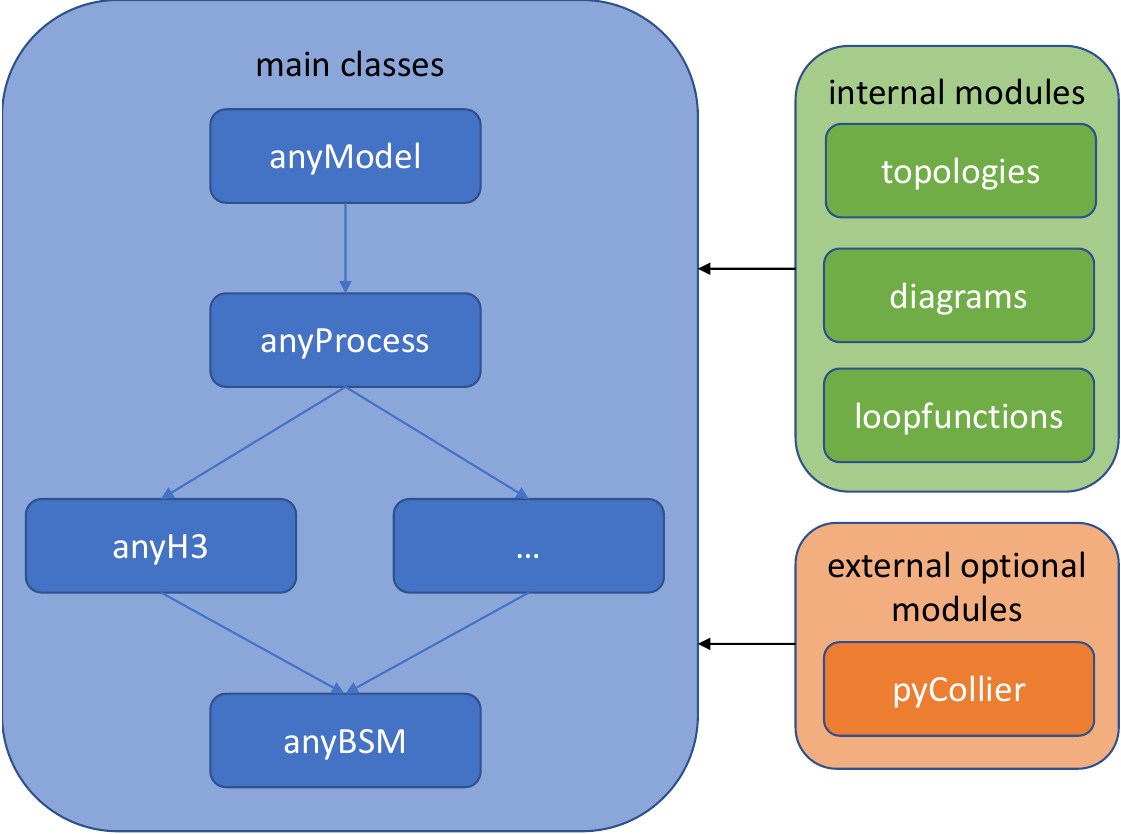}
    \caption{Class and module structure of \anyBSM. The ellipsis denotes additional observables which can be implemented in the future.}
    \label{fig:class_structure}
\end{figure}

This class and module structure is depicted in \cref{fig:class_structure}. The class \texttt{anyModel} encodes all information about the used model (Feynman rules, particles, interactions, etc.). This information is then inherited by the \texttt{anyProcess} class, which is used to calculate generic quantities like two- or three-point functions. This class makes use of various internal and external modules to derive the necessary diagrams and loop functions. The generic results of the \texttt{anyProcess} class are then used by the \anyH class to specifically calculate the trilinear Higgs coupling. Future extensions of \anyBSM featuring the calculation of new observables can be built on the basis of the \texttt{anyProcess} class (indicated by the ellipsis). In the end, the \anyBSM class collects the classes for the different observables into a single object.


\section{Program Tutorial}
\label{sec:tutorial}

In this Section, we describe the basic features of \anyBSM and \anyH. As explained above \anyH is part of \anyBSM, which provides a flexible \py framework for precision calculations (not only of \lamhhh). This Section is not meant to be a detailed manual but instead is intended to give an overview of the overall functionality. A detailed and up-to-date description of all available methods and options is available in the online manual, which can be found at
\begin{center}
    \url{https://anybsm.gitlab.io/anybsm}.
\end{center}


\subsection{Installation}
\label{sec:tutorial:installation}
The \anyBSM source code is hosted at
\begin{center}
  \url{https://gitlab.com/anybsm/anybsm}.
\end{center}
Running the code requires at least \py version 3.5. The code is most easily used by installing the corresponding \py package by running
\begin{minted}[bgcolor=bg]{python}
pip install anyBSM
\end{minted}
which will automatically download and install \anyBSM as well as all necessary dependencies. One necessary requirement not handled automatically by \texttt{pip} is the presence of a \texttt{Fortran} compiler (required for the compilation of \COLLIER for the \pyCollier dependency) such as \texttt{gfortran} or \texttt{CLANG} which can be installed from the systems package repository.
Upon the first run of \anyBSM, the model repository will be downloaded and saved into a user-specified location.\footnote{The path to this location is defined in the config file which can be found at \texttt{\string~/.config/anyBSM/anyBSM\_config.yaml} (Linux) or \texttt{\string~/Library/Preferences/anyBSM/anyBSM\_config.yaml} (Mac OS). By default, it is set to the folder \texttt{models} at the same location as the config file (\eg, \texttt{\string~/.config/anyBSM/models} for Linux.}. The model repository is available online at
\begin{center}
    \url{https://gitlab.com/anybsm/anybsm_models}
\end{center}
and can be updated using the version control system \texttt{git}. The model repository will be expanded over time, and community contributions, in particular \texttt{git} merge requests, for tested models are welcome.  


\subsection{Basic syntax}

\anyBSM can either be integrated into \py scripts as a \py package or run directly from the command line. In addition, a \mat interface exists as well.


\subsubsection{\py package mode}
\label{sec:pypackage}
After starting a new \py session and importing \anyBSM via
\begin{minted}[bgcolor=bg]{python}
from anyBSM import anyBSM
\end{minted}
a model --- here for instance the SM --- can be initialised via
\begin{minted}[bgcolor=bg]{python}
SM = anyBSM('SM')
\end{minted}
Alternatively, a path to a \UFO model directory can also be given. As an overview, the dictionary \code{anyBSM.built_in_models} contains a list of all pre-installed \UFO models and their installation directories. During the initialisation step, \anyBSM will try to automatically identify the SM-like Higgs boson, for which \lamhhh is calculated, as well as all other SM particles.

After the model initialisation, \lamhhh can be calculated by running
\begin{minted}[bgcolor=bg]{python}
SM.lambdahhh()
\end{minted}
which returns
\begin{minted}[bgcolor=bg]{python}
{'total': (176.22855628707978+0j),
 'treelevel': (187.28177740658242-0j),
 'genuine': (-16.63170122055135-0j),
 'wfr': (3.8805860322813865-0j),
 'tads': (-10.63364669685712-0j),
 'massren': (20.994575921900807+0j),
 'vevren': (-8.663035156276381+0j),
 'customren': 0}
\end{minted}
Here, ``total'' denotes the total value for \lamhhh in GeV; ``treelevel'', the tree-level value; ``genuine'', the genuine one-loop contribution; ``wfr'', the contribution from external-leg corrections (diagonal and off-diagonal); ``tads'', the contribution from tadpole diagrams; ``massren'', the contribution from mass renormalisation; ``vevren'', the contribution from the renormalisation of the electroweak vev; and, ``customren'', the contribution from a custom counterterm. It should be noted that by default \lamhhh is evaluated for vanishing external momenta. The option for using non-zero external momenta is described in \cref{sec:tutorial:setparams}.


\subsubsection{Command line mode}

As an alternative to using \anyBSM as a \py package, it can also be called directly from the command line. A simple example is
\begin{minted}[bgcolor=bg]{bash}
anyBSM SM
\end{minted}
which returns
\begin{minted}[bgcolor=bg]{text}
\lambda_hhh           = 176.22855628707978
(   tree-level           =  187.2818;
    one-loop-genuine     =  -16.6317;
    one-loop-WFRs        =    3.8806;
    tadpoles             =  -10.6336;
    counterterm (masses) =   20.9946;
    counterterm (VEV)    =   -8.6630;
    counterterm (custom) =    0.0000)
\end{minted}
An overview of the available options for the command line can be displayed by running
\begin{minted}[bgcolor=bg]{bash}
anyBSM -h
\end{minted}
To view more options and details about a specific model one can also add the \texttt{-h} flag to the model name. For example
\begin{minted}[bgcolor=bg]{bash}
anyBSM SM -h
\end{minted}
lists the particle content of the SM and the options for setting numerical values of all parameters (such as the top-quark mass) and their default values from the \UFO model. 

It should be noted that the command line tool provides access only to the basic functionalities of the \anyBSM library, unlike the \py package mode and the \mat mode discussed below.

\subsubsection{\mat mode}
The \mat interface can be conveniently installed as follows:
\begin{minted}[bgcolor=bg]{python}
Import["https://gitlab.com/anybsm/anybsm/-/raw/main/install.m"]
InstallAnyBSM[]
\end{minted}
which checks for all requirements and adds the \anyBSM interface to {\mat}'s \texttt{\$Path} variable. Afterwards, the interface can be used as follows:
\begin{minted}[bgcolor=bg]{Mathematica}
<<anyBSM`
LoadModel["SM"]
lambda = lambdahhh[]
\end{minted}
The result \texttt{lambda} is a \mat \texttt{Association} object similar to the \py \texttt{dictionary} object obtained in \cref{sec:pypackage} using the \py library. However, by default the analytical rather than numerical results are returned, after conversion to valid \mat expressions. A list of all available functions (such as for \eg\ the calculation of self-energies and tadpoles) within \mat is stored in the variable \texttt{\$AnyFunctions}. More information is given in the \docs. Furthermore, comprehensive \mat notebooks that demonstrate the use of \anyBSM's \mat mode are provided in the \examplesrepo. 

The \mat mode has access to the full functionalities of the \py backend (\ie the \anyBSM library). For the sake of clarity, we will restrict ourselves to a description of the \py package in the next sections.


\subsection{Setting parameters}
\label{sec:tutorial:setparams}

\anyBSM uses the default parameters defined in the respective \UFO model. To change \eg\ the value of the top-quark mass in the example SM calculation discussed above, we can run
\begin{minted}[bgcolor=bg]{python}
SM.setparameters({'Mu3': 165})
\end{minted}
where in this example an effective (running) top-quark mass of 165~GeV is used. Alternatively, a \texttt{LHA} file can be used as input,
\begin{minted}[bgcolor=bg]{python}
SM.setparameters('/path/to/LHA_file')
\end{minted}
The program also defines a few additional \UFO parameters in case they are not found in the \UFO model. For instance, if the model does not define an external parameter named \code{Qren} (used for the renormalisation scale $Q_{\text{ren.}}$), the code introduces it internally with a default value of $\texttt{Qren}=\unit[172.5]{GeV}$. The full list of additionally introduced parameters is discussed in \cref{app:additionalufo} as well as in the \docs.

The external momenta entering the computation of \lamhhh can be specified by passing the \code{momenta} attribute to the \code{lambdahhh} function, e.g.
\begin{minted}[bgcolor=bg]{python}
SM.lambdahhh(momenta = [500**2, 'Mh**2', 'Mh**2'])
\end{minted}
where \code{Mh} is the Higgs mass parameter defined in the SM \UFO model file and is automatically replaced by its numerical value. By default, \code{momenta = [0, 0, 0]} is chosen.


\subsection{Renormalisation}
\label{sec:documentation:ren}

Information about the renormalisation is saved in the file \texttt{schemes.yml} in the model directory. A simple example file for the SM is
\begin{minted}[bgcolor=bg]{yaml}
# default names for SM fields and parameters
SM_names:
  Top-Quark: u3
  W-Boson: Wp
  Z-Boson: Z
  Higgs-Boson: h
  VEV: vvSM

default_scheme: OS

renormalization_schemes:
  OS:
    mass_counterterms:
      h: OS
    VEV_counterterm: OS
  MS:
    mass_counterterms:
      h: MS
    VEV_counterterm: MS
\end{minted}
Here, the first \code{SM_names} block defines the names of various SM fields and parameters. The \code{renormalization_schemes} block can be used to define different renormalisation schemes. In the present example, the scheme \code{OS} is defined such that the mass of the field \code{h} as well as the \vev~counterterm are renormalised in the OS scheme. This scheme is set as default scheme via the \code{default_scheme} directive. In addition, the scheme \code{MS} is defined so that the mass of the field \code{h} as well as the \vev~counterterm are renormalised in the \MS scheme. It should be stressed that this does not mean that all inputs are converted from \OS to \MS parameters but rather that the physical interpretation of these parameters is changed from \OS to \MS. However, for a consistent conversion of the parameters, all ingredients (\ie\ two-point functions) are provided by the program. A proper conversion between the schemes will be demonstrated in \cref{sec:applications:uncert}. 

If a non-default scheme should be used, this can \eg\ be specified during the model initialisation:
\begin{minted}[bgcolor=bg]{python}
SM = anyH3('SM', scheme = 'MS')
\end{minted}
As an alternative to using schemes predefined in \texttt{schemes.yml}, renormalisation schemes can also be generated interactively during the run time by using a new name that is not yet used in the \texttt{schemes.yml} file for the \texttt{scheme} directive during the model initialisation or by calling \eg\ \code{SM.add_renormalization_scheme('MS')} afterwards. The new scheme will then be saved into the \texttt{schemes.yml} file. It is also possible to change the renormalisation scheme, \eg\ between two calls of \code{SM.lambdahhh()}, using the appropriate method:
\begin{minted}[bgcolor=bg]{python}
SM.load_renormalization_scheme('OS')
\end{minted}
If no \texttt{schemes.yml} file is present in the \UFO model directory, it will be generated automatically upon the first creation of a renormalisation scheme which automatically searches for all SM-like parameters (particles) based on their numerical (mass) values (and PDG identifiers), \cf\ \cref{sec:flow}.


\subsection{Evaluation modes and output formats}

\anyBSM supports three different evaluation modes:
\begin{itemize}
\item \code{abbreviations}: all results are given in analytical form using the \UFO coupling abbreviations (\code{GC_1}, \code{GC_2}, etc.);
\item \code{analytical}: all results are given in analytical form using the full analytical form for all couplings;
\item \code{numerical}: the numerical values for all parameters are inserted, and a numerical result is returned.
\end{itemize}
The evaluation mode can be set \eg\ via
\begin{minted}[bgcolor=bg]{python}
SM.set_evaluation_mode('analytical')
\end{minted}
if using \anyBSM as a \py package. The default evaluation mode is \code{numerical}.

A detailed breakdown of the results (including results for individual diagrams) in the form of a PDF document can be produced by using \code{draw = True}
as an additional argument for the \code{lambdahhh} function or via the \texttt{-t} option when using the command line interface
\begin{minted}[bgcolor=bg]{python}
SM.lambdahhh(draw = True)
\end{minted}
The individual results listed along with the diagrams are represented in a way which depends on the chosen evaluation mode (e.g.\ numerical or analytical/using abbreviations). The resulting PDF file is saved to the current working directory as well as the model directory. In order to make use of this feature, \LaTeX\ needs to be installed.

In addition to the \mat mode, analytical expressions can be exported from within a \py session to \mat with the help of \texttt{SymPy}
\begin{minted}[bgcolor=bg]{python}
from sympy import mathematica_code
mathematica_code(<sympy_expression>)
\end{minted}
Note that \anyBSM includes a caching system which automatically saves the analytic results into \texttt{json} files (into the \texttt{cache} directory located in the model directory). This leads to a significant speed-up of consecutive runs, see \cref{app:cache}.


\subsection{Getting Help}

All \py classes and methods defined in \anyBSM and \anyH have meaningful doc-strings which can be issued by e.g.\ 
\begin{minted}[bgcolor=bg]{python}
from anyBSM import anyH3
help(anyH3)
help(anyH3.lambdahhh)
\end{minted}
or directly using existing class instances (such as \code{help(SM.lambdahhh)} in the examples above). In addition, the \docs makes use of these doc-strings and provides a search functionality. 

The usage of the command line tool \anyBSM is returned by the command \texttt{\anyBSM\xspace-h}. For a given model, one can obtain further help by issuing the command \texttt{\anyBSM <model name> -h} from the command line. The \mat interface of \anyBSM also provides documentation for all its functions by issuing \code{?<function name>} such as \eg \code{?lambdahhh}. Furthermore, it provides a list of available functions stored in the variable \texttt{\$AnyFunctions}.

In addition, the \anyBSM \examplesrepo provides basic and concrete examples for all three interfaces and for the generation of new model files.


\section{Built-in models and cross-checks}
\label{sec:crosschecks}

The models currently distributed alongside \anyH are
\begin{itemize}
    \item the Standard Model (SM);
    \item the real-singlet extension of the SM (SSM);
    \item the Two-Higgs-Doublet Model (THDM) --- all four Yukawa types;
    \item the Inert-Doublet Model (IDM);
    \item the Next-to-Two-Higgs-Doublet Model (NTHDM) --- \ie the real-singlet extension of the THDM;
    \item triplet extensions of the SM with either a real triplet with hypercharge $Y=0$ or a complex triplet with $Y=1$. The two theories are denoted respectively TSM$_{Y=0}$ and TSM$_{Y=1}$;
    \item the Georgi-Machacek model (a general version, as well as an aligned version);
    \item a $U(1)_{B-L}$ extension of the SM (BmLSM);
    \item the Minimal Supersymmetric Standard Model (MSSM). 
\end{itemize}
These models, and associated conventions, are described in more detail in \cref{app:models}. We emphasise again that additional models can also be included by the user in a convenient and fast way, as described in \cref{app:newmodels}. In this section, we present details about a variety of analytical and numerical cross-checks we performed to validate \anyH. 


\subsection{Cross-checks using analytical computations}

To cross-check the routines implemented in \anyH, we compared the analytical results to calculations performed using \FeynArts and \FormCalc.\ We found full agreement for the following pieces of the calculation performed in \anyH:
\begin{itemize}
    \item Higgs and Goldstone boson self-energies (and momentum derivatives thereof) including self-energies with two distinct external scalars as well as charged scalar self-energies;
    \item Higgs tadpoles;
    \item vector boson self-energies including mixing self-energies (\eg\ $\gamma-Z$ mixing);
    \item genuine one-loop corrections to scalar three-point functions; 
    \item one-particle-reducible contributions to scalar three-point functions.
\end{itemize}
These checks have been performed in the SM, meaning that all contributions to the renormalised trilinear Higgs coupling arising in the SM have been cross-checked. Contributions that do not exist in the SM (\eg\ scalar-mixing self-energies) have been cross-checked in the THDM.

As an additional cross-check, we have verified the cancellation of ultraviolet divergences in the SM, the THDM, the SSM, the \TSM0, and the \TSM1. Moreover, we have found full agreement for the overall one-loop result for \lamhhh with independent calculations in the SM and the THDM, performed with \texttt{FeynArts} and \texttt{FormCalc}, as well as with the results for the \TSM1 from \ccite{Bahl:2022gqg} (see \cref{app:models} for a more detailed description of the models).

\subsection{Numerical cross-checks}

In addition to analytical cross-checks, we have performed a series of numerical cross-checks by reproducing results from the literature.


\subsubsection{SSM}

\begin{figure}
    \centering
    \includegraphics[width=\textwidth]{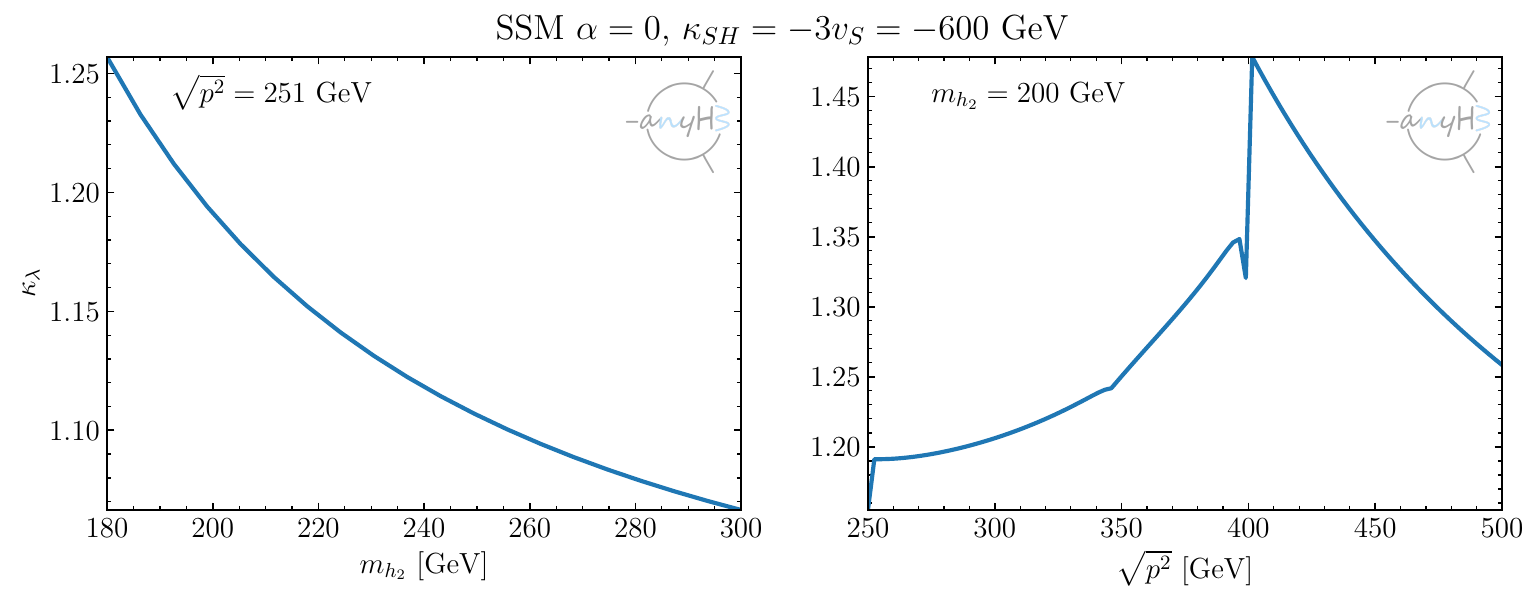}
    \caption{\kaplam in the SSM as a function of the singlet mass (left) and one off-shell external momentum (right). The results match those of \ccite{He:2016sqr} Fig.~6 and~7 (upper-right), respectively.}
    \label{fig:SSMcheck}
\end{figure}

As a first check, we reproduced the SSM results for \kaplam derived in \ccite{He:2016sqr} (following the choice made in this reference and for the sake of comparison, we set the mixing between the \cp-even states to zero). This reproduction is shown in \cref{fig:SSMcheck} which is to be compared with Fig.~6 and~7 (upper-right) of \ccite{He:2016sqr}. In this figure, the momentum of two of the three external Higgs boson legs is always set on-shell $\sqrt{p_1^2}=\sqrt{p_2^2}=m_{h_1}=\unit[125]{GeV}$. In the left plot of \cref{fig:SSMcheck}, the momentum of the third external Higgs leg is fixed at $\sqrt{p_3^2}\equiv\sqrt{p^2}=\unit[251]{GeV}$ and \kaplam is shown as a function of the singlet mass. In the right plot, the singlet mass is fixed to \unit[200]{GeV} and the external momentum of the third Higgs boson leg is varied. 

The behaviour of both plots reproduces the behaviour found in \ccite{He:2016sqr}. However, the exact numerical values in the left panel are shifted due to different treatments of the external-leg corrections. The different treatments of external momenta also lead to a slightly different peak structure in the right plot. However, at $p^2=m_{h_1}^2$ the different treatments coincide. To show this equality we use the shift to \kaplam caused by the BSM sector, $\delta_{hhh}^{(1)} = \kaplam - \nicefrac{\lamhhh^{(1),\,\text{SM}}}{\lamhhh^{(0),\,\text{SM}}}$, which was introduced in Eq.\ (25) of \ccite{He:2016sqr}.
The external leg contribution to $\delta_{hhh}^{(1)}$ in the two different treatments reads 
\begin{align}
\left. \delta_{hhh}^{(1)}(p^2) \right|^{\anyH}_\text{ext.\ leg} & 
        = \frac{1}{2}\big(\underbrace{\delta^{(1)} Z_{h_1}(p^2)}_{\text{off-shell leg}} + \underbrace{2\delta^{(1)} Z_{h_1}(m_{h_1}^2)}_{\text{on-shell legs}}\big) = -\frac{\lambda_{SH}^2 v^2}{16 \pi^2} \left( \frac{\partial \text{B}_0(p^2)}{\partial p^2}  + 2\left.\frac{\partial  \text{B}_0(p^2)}{\partial p^2}\right|_{p^2=m_{h_1}^2} \right) \nn \\
\left. \delta_{hhh}^{(1)}(p^2) \right|^{\text{Ref.~\cite{He:2016sqr}}}_\text{ext.\ leg} & 
        = -\frac{\lambda_{SH}^2 v^2}{16 \pi^2}\left(2\frac{\text{B}_0(p^2)-\text{B}_0(m_{h_1}^2)}{p^2-m_{h_1}^2} + \left.\frac{\partial \text{B}_0(p^2)}{\partial p^2} \right|_{p^2=m_{h_1}^2} \right)\,,
\end{align}
where $\text{B}_0(p^2)\equiv B_0(p^2,m_{h_2}^2, m_{h_2}^2)$ is the one-loop Passarino-Veltman two-point function \cite{Passarino:1978jh,Denner:1991kt}. 
Thus, the two approaches yield the same external leg correction factors in the limit of $p^2\to m_{h_1}^2$. We made use of this relation to cross-check the full analytical result obtained with \anyH with the result derived in \ccite{He:2016sqr} and found full agreement at $p^2=m_{h_1}^2$. For demonstrative purposes, we provide this cross-check using the \mat interface of \anyBSM (\cf\ \cref{sec:tutorial}) as an example usage in the \anyBSM \examplesrepo.


\subsubsection{\texorpdfstring{\TSM1}{TSM (Y=1)}}

\begin{figure}
    \centering
    \includegraphics[width=.49\textwidth]{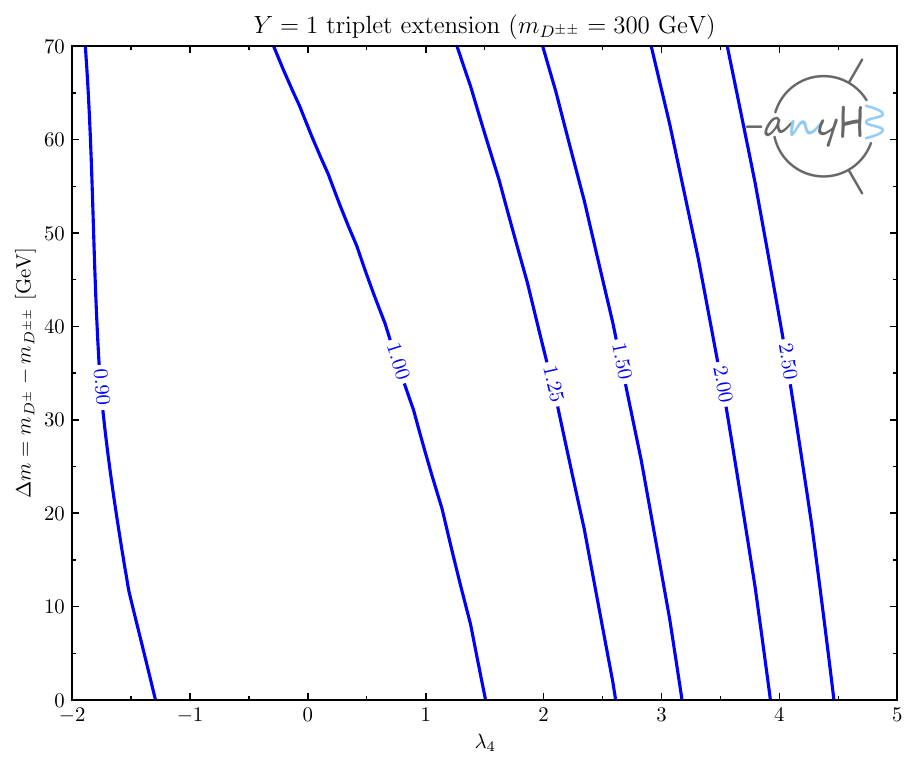}
    \includegraphics[width=.49\textwidth]{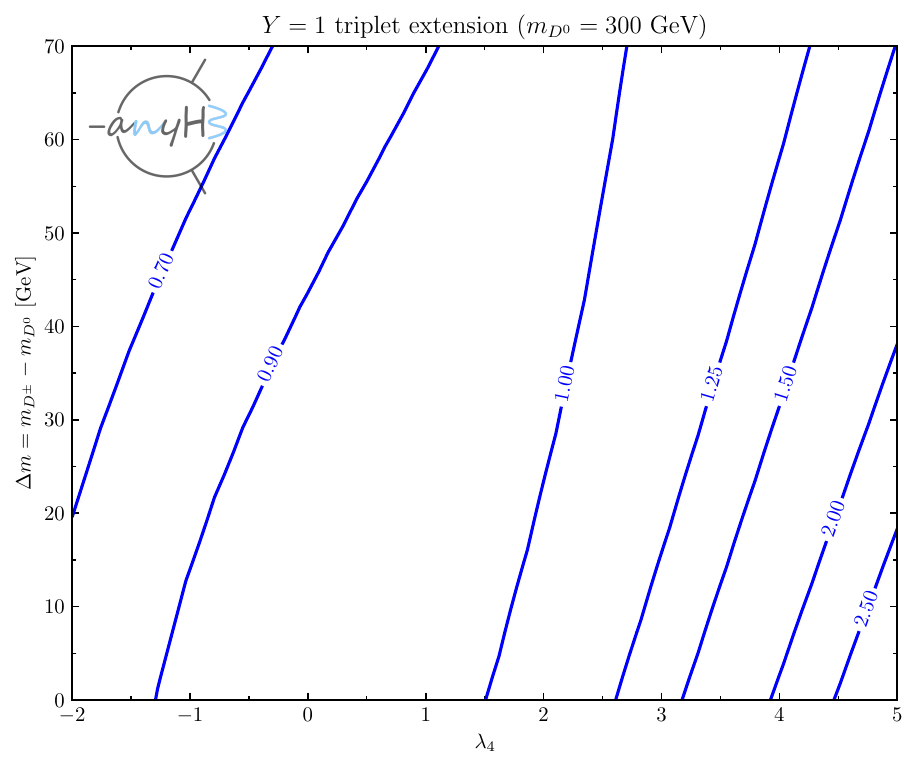}
    \caption{Reproduction of Fig.~11 of \ccite{Aoki:2012jj} showing results for $\kappa_\lambda$ for the $Y=1$ triplet model. \textit{Left}: $\kappa_\lambda$ contours are shown in the $(\lambda_4,\Delta m)$ parameter plane with $\Delta m = m_{D^\pm} - m_{D^{\pm\pm}}$ and $m_{D^{\pm\pm}} = 300\gev$. \textit{Right}: $\kappa_\lambda$ contours  are shown in the $(\lambda_4,\Delta m)$ parameter plane with $\Delta m = m_{D^\pm} - m_{D^0}$ and $m_{D^0} = 300\gev$.}
    \label{fig:Yeq1_triplet}
\end{figure}

As a further verification of \anyH, we reproduced results in the literature for the \TSM1 model~\cite{Aoki:2012jj}. Fig.~11 of \ccite{Aoki:2012jj} shows deviations of \lamhhh from the SM prediction in the plane of the coupling $\lambda_4$ and the mass difference between the lightest and second-lightest BSM states (see \cref{app:tsm1} for further details about the model). Our reproduction of this Figure is shown in \cref{fig:Yeq1_triplet}. In the left panel, the lightest BSM states are the doubly-charged Higgs bosons; in the right panel, the lightest BSM states are the two neutral BSM Higgs bosons. Overall, we observe a very good agreement between our results and the results presented in \ccite{Aoki:2012jj}. The remaining small differences can be traced back to different SM input parameters used in \ccite{Aoki:2012jj}.


\subsubsection{MSSM}

\begin{figure}
    \centering
    \includegraphics[width=\textwidth]{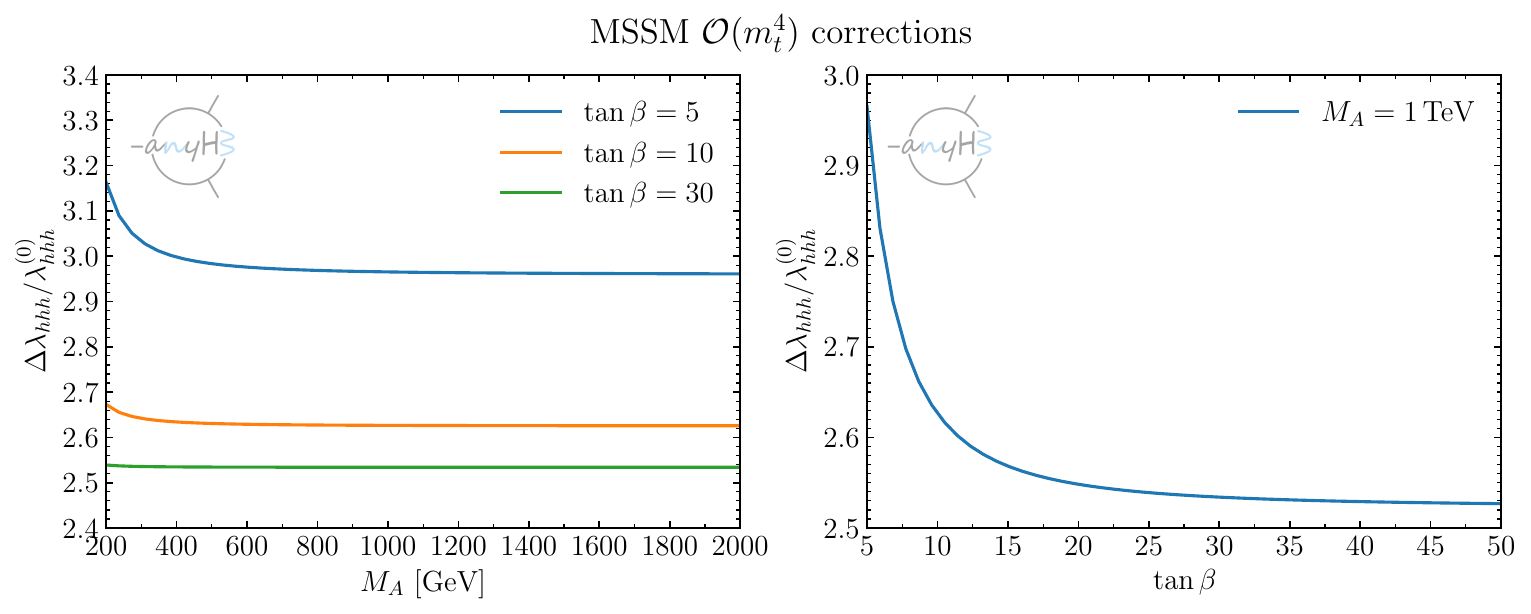}
    \caption{Reproduction of Fig.~2 of \ccite{Hollik:2001px} showing results for $\mathcal{O}(m_t^4)$ one-loop corrections $\Delta\lambda_{hhh}$ to the trilinear Higgs coupling (normalised to the tree-level coupling) in the MSSM. \textit{Left}: $\Delta\lambda_{hhh}$ as a function of $M_A$ for $\tan\beta = 5$ (blue curve), $\tan\beta = 10$ (orange curve), and $\tan\beta = 30$ (green curve). \textit{Right}: $\Delta\lambda_{hhh}$ as a function of $\tan\beta$ for $M_A = 1\tev$.}
    \label{fig:MSSM_mt4}
\end{figure}

As a cross-check of the MSSM implementation, we reproduced the results of \ccite{Hollik:2001px}. In this work, the leading $\mathcal{O}(m_t^4)$ corrections to the trilinear Higgs coupling originating from scalar top quarks were calculated in the limit of vanishing electroweak gauge couplings. Setting the SUSY (and SM) parameters as in \ccite{Hollik:2001px} (\ie, setting $M_{\tilde Q} = M_{\tilde U} = 15\tev$, $\mu = |A_t| = 1.5\tev$), we find very good agreement with their results (see our \cref{fig:MSSM_mt4} in comparison to Fig.~2 of \ccite{Hollik:2001px}).


\subsubsection{Recovering the SM result in the decoupling limit}
\label{sec:anydecoupling}

As an additional non-trivial cross-check of \anyH (and also of the model files distributed alongside it), we have verified that the BSM contributions decouple 
if the masses of the BSM scalars are increased in a uniform way (see below for details), so that the SM result for \lamhhh is recovered in this limit.

\begin{figure}
    \centering
    \includegraphics[width=0.9\textwidth]{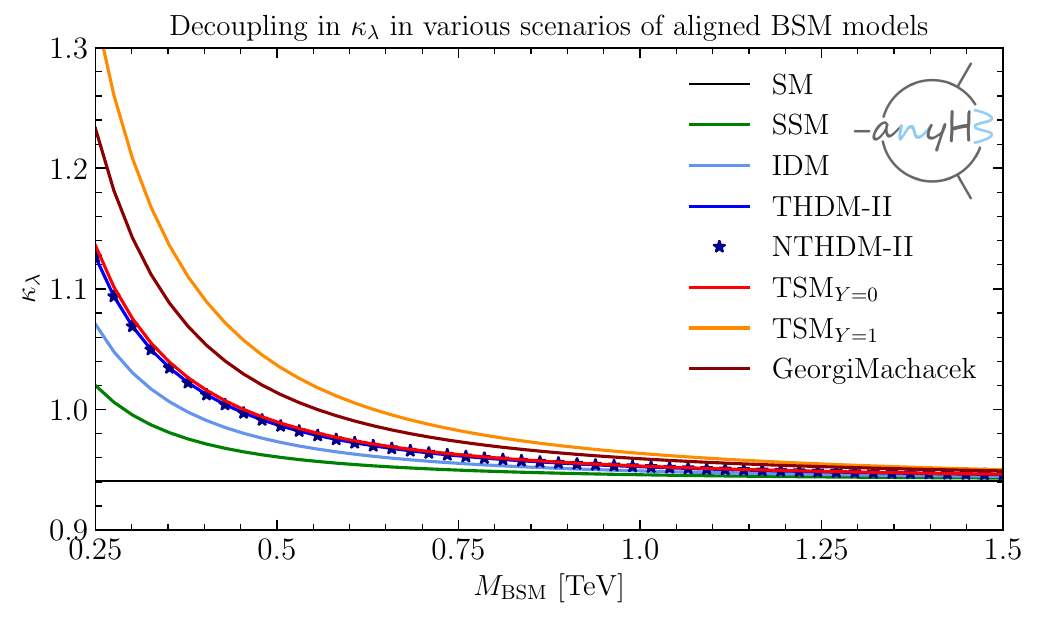}
    \caption{
    For the shown models all masses have been chosen to be degenerate with the value $M_\mathrm{BSM}$. Soft symmetry-breaking parameters are set to $\sqrt{M_\mathrm{BSM}-(\unit[250]{GeV})^2}$. Individually the relevant model parameters are fixed as follows. \textbf{SSM}: $\alpha=0,\, \kappa_S=\kappa_{SH}=\unit[-800]{GeV}$ and $v_S=\unit[300]{GeV}$. \textbf{IDM}: $\sqrt{M_\mathrm{BSM}^2-\mu_2^2}=\unit[250]{GeV}$ and $m_{A,H,H^+}=M_\mathrm{BSM}$. \textbf{THDM-II}: $\sqrt{M_\mathrm{BSM}^2-M^2}=\unit[250]{GeV}$, $\tan\beta=2,\,\sin(\beta-\alpha)=1$  and $m_{A,{h_2},{H^+}}=M_\mathrm{BSM}$. \textbf{NTHDM-II}: as in the \textbf{THDM-II} with $v_S=M_{\mathrm{BSM}}$, $\alpha_1+\alpha_3=\beta-\pi/2$, $\alpha_2=\pi/2$. \textbf{\TSM0}: $\lambda_{T\Phi}=2.5$ and $M_H^+=M_\mathrm{BSM}$. \textbf{\TSM1}: $\lambda_4=2.5$ and $m_{D^+}=m_{D^{++}}=M_{\mathrm{BSM}}$. \textbf{GeorgiMachacek}: $\sqrt{M_\mathrm{BSM}^2-M_\eta^2}=\unit[250]{GeV}$, $M_5=M_3=M_\mathrm{BSM}$ and $\sin(H)=0$. See \cref{app:models} for details about the various models.
    }
    \label{fig:decouplingall}
\end{figure}

This verification is shown in \cref{fig:decouplingall},  where \kaplam is displayed as a function of the BSM mass scale $M_\text{BSM}$. All BSM masses in each model have been chosen to be degenerate with each other with the mass value $M_\text{BSM}$. The results for \kaplam are shown
in the SSM (green), the IDM (light blue), the THDM-II (dark blue), the NTHDM-II (stars), the \TSM0 (red), the \TSM1 (orange), and the Georgi-Machacek model (brown). In all models we chose appropriate input parameters such that the lowest-order couplings of the SM-like Higgs boson to the other SM states are exactly as in the SM. This setting is referred to as \textit{aligned} scenario in the following.
Further details about the chosen parameters of the models are specified in the caption.
As a reference, the SM result for \kaplam is indicated as a black line. It is clearly visible that for increasing $M_\text{BSM}$ \kaplam quickly approaches the SM result for the chosen parameter settings. For $M_\text{BSM}\gtrsim 1\tev$, the deviations from the one-loop SM result are below $\sim 0.05$. We note that, as will be seen in \cref{sec:NTHDM}, values below the SM prediction are also possible in BSM models.
Additionally, the details of the decoupling patterns in the different models strongly depend on the chosen parameters. Therefore, from the shown example no general conclusions can be drawn about how quickly decoupling occurs with increasing BSM mass scale in the various models.
 

\section{Example applications}
\label{sec:applications}

After having discussed cross-checks for validation, we present here a series of example applications. 
We first discuss estimates of the remaining theoretical uncertainties and then provide examples of results that go beyond existing studies in the literature.


\subsection{Estimation of uncertainties in the computation of \texorpdfstring{\lamhhh}{lamhhh}}
\label{sec:applications:uncert}

When comparing predictions for an observable (or a pseudo-observable) with experimental measurements or limits, an important consideration to ensure the reliability of the comparison is to estimate the theoretical uncertainty associated with the obtained prediction. We devote this section to discussing different contributions to the theoretical uncertainty associated with computations of the trilinear Higgs coupling: on the one hand, uncertainties due to missing higher-order terms, and on the other hand, parametric uncertainties due to the limited precision with which quantities entering the calculation of \lamhhh are known experimentally.

\subsubsection{Uncertainty from missing higher-order corrections in the SM}

Focusing at first on the calculation of \lamhhh in the SM, we begin by investigating the possible size of missing higher-order contributions. While in the SM two-loop corrections of $\mathcal{O}(\alpha_t\alpha_s)$ and $\mathcal{O}(\alpha_t^2)$ are known~\cite{Senaha:2018xek,Braathen:2019pxr,Braathen:2019zoh}, we refer here to all higher-order corrections that go beyond the full one-loop result that is obtained with \anyH (see below for a discussion of the impact of the known two-loop corrections). Since calculations that are carried out at a given order in different renormalisation schemes differ by contributions that go beyond the calculated order, the comparison of different results of this kind for the same parameter point can be used as an estimate for the size of missing higher-order corrections --- provided that the perturbative behaviour in the different schemes is of similar quality. We employ this method and compare the result obtained with \anyH in the OS scheme with the one in the \MS scheme.

In the SM, we use for the quantities that enter the tree-level expression of \lamhhh~--- i.e., the Higgs-boson mass, the $W$- and $Z$-boson masses, and the electromagnetic coupling $\alpha_\text{em}$ (the latter three quantities are in turn used to compute the Higgs \vev, see the discussion in \cref{app:sm:vevren}) --- the following OS input values
\begin{align}
\label{EQ:SM_defaultOSinputs}
    &M_h=125.1\gev\,,\quad M_W=80.379\gev\,,\quad M_Z=91.187\gev\nonumber\\
    &\alpha_\text{em}^{-1}(0)=137.035999679\,,
\end{align}
where the notation $M_i$ indicates the OS mass of particle $i$ (we deviate here from the lower-case notation $m_h$ employed for the Higgs mass in the rest of the paper in order to avoid ambiguities between OS and \MS masses). This yields for the tree-level prediction of the trilinear coupling a value of $\lambda_{hhh}^{(0)}=187.3\gev$. For the full one-loop predictions of $\lamhhh$ in the on-shell scheme (where the tadpoles are renormalised in the OS scheme) we obtain for the two cases of vanishing external momenta and for the choice $p_1^2=(200\gev)^2$ and $p_2^2=p_3^2=M_h^2$
\begin{align}
    \lambda_{hhh}^\text{(1),\text{ OS}}(0,0,0)=176.2\gev\,,\nn\\
    \lambda_{hhh}^\text{(1),\text{ OS}}((200\gev)^2,M_h^2,M_h^2)=180.8\gev\,.
\end{align}
In order to compare these values with those in the \MS scheme, we must first convert the OS input parameters $M_h$, $M_W$, $M_Z$, $\alpha_\text{em}$ to the \MS scheme --- this conversion (and all other scheme conversions in this paper) will be performed at one-loop order. Working at $Q=172.5\gev$, and employing again an OS renormalisation of the tadpoles (see the discussion in \cref{app:sm:tadren} for further details), we find after the one-loop conversion 
\begin{align}
    m_h^\MS=&\ 121.4\gev\,,\nn\\
    m_W^\MS=&\ 80.1\gev\,,\nn\\
    m_Z^\MS=&\ 91.6\gev\,,\nn\\
    (\alpha_\text{em}^\MS)^{-1}=&\ 128.34\,.
\end{align}
Using now these values as inputs for the calculation of \lamhhh in the \MS scheme, we obtain at $Q=172.5\gev$ (again at full one-loop order, and with the same two choices of external momenta as above)
\begin{align}
    \lambda_{hhh}^\text{(1),\MS}(0,0,0)=175.8\gev\,,\nn\\
    \lambda_{hhh}^\text{(1),\MS}((200\gev)^2,M_h^2,M_h^2)=180.5\gev\,.
\end{align}
The difference of about $0.3-0.4\gev$ between the results obtained in the OS and the \MS schemes constitutes a first estimate of a part of the unknown higher-order corrections; in relative size, the obtained shifts correspond to a difference of less than 0.2\%. We note that if we had chosen to convert also the value of the squared Higgs mass used for the external momenta, the result for $\lambda_{hhh}^{(1),\MS}((200\gev)^2,(m_h^\MS)^2,(m_h^\MS)^2)$ would have decreased by $0.2\gev$, giving rise to only a slight change of our uncertainty estimate. 

Concerning the interpretation of the uncertainty estimates obtained so far, it should be emphasised that the scheme comparison done above in fact does not capture corrections to \lamhhh involving the strong coupling $\alpha_s$, because the performed scheme conversions of the quantities $M_h$, $M_W$, $M_Z$, $\alpha_\text{em}$ do not involve this coupling at one-loop order. Such effects can on the other hand be estimated by converting the input value used for the top-quark mass (in contrast to the quantities entering the prediction of the trilinear Higgs coupling at tree-level, converting the top-quark mass entering at the one-loop level from the OS to the \MS scheme directly gives rise to a two-loop effect in the prediction for 
\lamhhh). Starting from the OS value of $M_t=172.5\gev$, a conversion including the leading ${\cal O}(\alpha_s)$ and ${\cal O}(\alpha_t)$ contributions to the top-quark self-energy (involving the strong gauge and top Yukawa couplings) yields an \MS value of $m_t^\MS(Q=172.5\gev)=166.3\gev$. Using this \MS value in the computation of \lamhhh, we obtain
\begin{align}
   \lambda_{hhh}^\text{(1),\MS}(0,0,0)\big|^{m_t^\MS}=178.3\gev\,,\nn\\
    \lambda_{hhh}^\text{(1),\MS}((200\gev)^2,M_h^2,M_h^2)\big|^{m_t^\MS}=182.7\gev\,,
\end{align}
which corresponds to a total deviation of about $2\gev$ compared to the OS results above where $M_t$ was used as input value. For the sake of comparison, we note that employing the OS computation of \lamhhh using $M_h$, $M_W$, $M_Z$, and $\alpha_\text{em}^{-1}(0)$ from \cref{EQ:SM_defaultOSinputs} but with the \MS value of $m_t$, we obtain
\begin{align}
    \lambda_{hhh}^{(1),\text{ OS}}(0,0,0)\big|^{m_t^\MS}&=178.6\gev\,,\nn\\
    \lambda_{hhh}^{(1),\text{ OS}}((200\text{ GeV})^2,M_h^2,M_h^2)\big|^{m_t^\MS}&=182.8\gev\,.
\end{align}
As explained above, the dominant two-loop ${\cal O}(\alpha_t\alpha_s)$ and ${\cal O}(\alpha_t^2)$ corrections to \lamhhh are in fact known~\cite{Senaha:2018xek,Braathen:2019pxr,Braathen:2019zoh}, and amount to about $+3\gev$ (specifically, the ${\cal O}(\alpha_t\alpha_s)$ corrections amount to $+4.1$ GeV, while those of ${\cal O}(\alpha_t^2)$ amount to $-1.1$ GeV). Thus, our above estimates of higher-order corrections that are not included in the computation of \anyH\ turn out to be close to the actual size of the known higher-order corrections. While for the case of the SM those two-loop corrections could be incorporated into the \anyH prediction, in many other models for which \anyH can be employed the corresponding corrections are not fully known. For reasons of uniformity we also restrict the SM prediction for \lamhhh in \anyH to the full one-loop level. An extension of the code providing the incorporation of higher-order contributions is left for future work.

\subsubsection{Parametric uncertainties}
Another source of theoretical uncertainty in the prediction of \lamhhh arises from the experimental errors of the input parameters. In order to investigate the impact of these parametric uncertainties, we take into account the $1 \, \sigma$ ranges of the experimental input parameters as given in \ccite{ParticleDataGroup:2022pth},
\begin{align}
    \Delta M_h^\text{exp.}&=\pm 0.17\gev\,,\nn\\
    \Delta M_W^\text{exp.}&=\pm0.012\gev\,,\nn\\
    \Delta M_Z^\text{exp.}&=\pm0.0021\gev\,,\nn\\
    \Delta(\alpha^\text{exp.}_\text{em}(0))^{-1}&=\pm2.1\times10^{-8}\,,
\end{align}
while for the top-quark mass we use
\begin{align}
    \Delta M_t=\pm 1 \gev\,.
\end{align}
It should be noted that the variation of $\pm 1 \gev$ for $M_t$ is indicated for illustration. The parametric uncertainty of the top-quark mass receives a contribution both from the experimental error of the measured mass parameter of $\pm 0.30\gev$ at the $1\,\sigma$ level~\cite{ParticleDataGroup:2022pth} and from the systematic uncertainty arising from relating the measured quantity to a theoretically well-defined top-quark mass. 
The parametric uncertainties that are induced by the masses of the other quarks and the leptons are negligible. We furthermore note that we do not consider a parametric uncertainty from the strong gauge coupling because it does not enter the expression of \lamhhh at the one-loop level. 

Varying each of the indicated experimental errors independently, we find the theoretical uncertainties induced in \lamhhh shown in \cref{tab:parametricuncertainties}. As expected, the largest effect on \lamhhh, with an induced uncertainty of $\pm 0.5\gev$, originates from the experimental error of the mass of the detected Higgs boson. Indeed the Higgs-boson mass enters the prediction for \lamhhh already at the tree level, and while it is already known to a high level of accuracy its experimental error is still larger --- by more than an order of magnitude --- than the experimental errors of $M_W$ and $M_Z$ (and much larger than the parametric uncertainty associated with $\alpha_\text{em}$). The theoretical uncertainties that are induced by the gauge-boson masses have only effects at the level of some tens of MeV or less. On the other hand, the experimental uncertainty of the top-quark mass has a stronger impact on \lamhhh even though it only enters at the one-loop level. It should be noted that if in the future BSM parameters are 
measured, the parametric uncertainties in the prediction for \lamhhh induced by their experimental errors should also be taken into account.

\begin{table}\centering
\begin{tabular}{|c|c|c|}
   \hline
    Parameter & Exp.\ uncertainty & Impact on \lamhhh \\
   \hline
    $M_h$ & $0.17\gev$ & $0.53\gev$ \\
    $M_W$ & $0.012\gev$ & $0.048\gev$\\
    $M_Z$ & $0.0021\gev$ & $0.011\gev$\\
    $\alpha_\text{em}(0)$ & $2.1\times10^{-8}$ & $1.1\times10^{-8}$\\
    $M_t$ & $1\gev$ & $0.4\gev$\\
   \hline
\end{tabular}
\caption{Theoretical uncertainties in the calculation of \lamhhh\ arising from the experimental errors of the input parameters.}
\label{tab:parametricuncertainties}
\end{table}

\subsubsection{Uncertainty from missing higher-order corrections in a BSM model: the example of the IDM}
\label{sec:IDMuncertainties}

When considering the computation of \lamhhh in BSM theories, BSM parameters can enter the tree-level expressions. This is not the case in aligned scenarios, like the IDM, where the tree-level prediction for the trilinear coupling is the same as in the SM and only the mass of the detected Higgs boson and the associated vacuum expectation value enter the lowest-order prediction. It should be noted that also in this case a comparison between the OS and \MS results for \lamhhh computed at one-loop order with \anyH with a one-loop conversion of $M_h$ and $v$ would not be sensitive to the type of contributions that can give rise to the largest effects at the two-loop level. For the specific case of the IDM one can infer from simple arguments of dimensional analysis that the leading two-loop corrections to \lamhhh are of $\mathcal{O}(g_{h\Phi\Phi}^5/M_\Phi^4)$ and $\mathcal{O}(\lambda_2g_{h\Phi\Phi}^3/M_\Phi^2)$ (which were computed in  \ccite{Braathen:2019pxr}), where $\Phi$ denotes either of the BSM scalars of the IDM, $g_{h\Phi\Phi}$ is a coupling between the Higgs boson at 125~GeV and two BSM scalars, and $\lambda_2$ is the Lagrangian self-coupling of the inert doublet (c.f.~\cref{app:idm} for more details). These types of contributions are not generated by a one-loop conversion of $M_h$ or $v$.

Instead, the size of these contributions can be estimated via a one-loop conversion of the BSM scalar masses, which affect the size of the dominant one-loop corrections to \lamhhh of $\mathcal{O}(g_{h\Phi\Phi}^3/M_\Phi^2)$. In the following, we investigate for four different example benchmark points --- labelled BP1, BP2, BP3, and BP4 (defined in~\cref{tab:idmuncertainties}) --- the potential size of these leading two-loop effects. We choose BP1 and BP2 with small splittings between the BSM scalar masses and the BSM mass parameter $\mu_2$, so that the couplings $g_{h\Phi\Phi}$ (which are proportional to
the difference $M_\Phi^2-\mu_2^2$) remain small, while we choose larger splittings for BP3 and BP4. Additionally, we set $\lambda_2$ to zero in BP1 and BP3, in order to investigate only terms of the form $\mathcal{O}(g_{h\Phi\Phi}^5/M_\Phi^4)$, while for BP2 and BP4 we set $\lambda_2=2$ to also include effects of $\mathcal{O}(\lambda_2g_{h\Phi\Phi}^3/M_\Phi^2)$. We present the results obtained with the code \anyH for the one-loop conversions of the scalar masses and for \lamhhh in~\cref{tab:idmuncertainties} (note that for the computation of $(\lambda_{hhh}^{(1)})^\text{OS}$ and for the scheme conversion of the BSM scalar masses, the tadpole contributions are renormalised on-shell).

\begin{table}[]
    \centering
    \begin{tabular}{|c|c|c|c|c|c|c|c|c|}
    \hline
         &\multicolumn{2}{c|}{Inputs} & \multicolumn{3}{c|}{\MS masses} & \multicolumn{3}{c|}{\anyH results}\\
         & \multicolumn{2}{c|}{} & \multicolumn{3}{c|}{(at $Q=300\gev$)} & \multicolumn{3}{c|}{}\\
    \hline
        BP & $\mu_2$ & $\lambda_2$ & $m_H^\MS$ & $m_A^\MS$ & $m_{H^\pm}^\MS$ & $(\lambda_{hhh}^{(1)})^\text{OS}$ & $(\lambda_{hhh}^{(1)})^\MS$ & $\Delta$ \\
        & [GeV] & $-$ & [GeV] & [GeV] & [GeV] & [GeV] & [GeV] & [\%] \\
    \hline
        1 & 250 & 0 & 403.7 & 413.8 & 418.6 & 220.6 & 223.7 & 1.4\\ 
        2 & 250 & 2 & 406.7 & 416.7 & 421.4 & 220.6 & 226.2 & 2.5 \\
        3 & 0 & 0 & 409.9 & 419.9 & 424.6 & 356.1 & 373.9 & 4.8\\
        4 & 0 & 2 & 412.9 & 422.7&  427.4 & 356.1 & 379.4 & 6.1\\
    \hline
    \end{tabular}
    \caption{One-loop predictions for \lamhhh in the IDM for different example scenarios in the OS and the \MS scheme, as well as the relative difference $\Delta$. For the conversion of masses from the OS to the \MS scheme, as well as the \MS calculation of \lamhhh, the renormalisation scale is chosen to be $Q=300\gev$. The values of the \MS masses are also given. For all four benchmark scenarios we set the OS masses to $M_H=400\gev$, $M_A=410\gev$, $M_{H^\pm}=415\gev$, while the other free IDM parameters are given in the ``Inputs'' columns.}
    \label{tab:idmuncertainties}
\end{table}

As could be expected, we find that the OS and \MS results are in very good agreement --- differing only by 1.4\% and 2.5\% for the two choices of $\lambda_2$ --- for the scenarios with small mass splittings (BP1 and BP2). For BP3 and BP4 featuring larger splittings, the discrepancy between the two results increases to about $5-6\%$. This confirms the known fact that the inclusion of two-loop corrections to \lamhhh is increasingly important for parameter regions with larger splittings between the different BSM masses. Finally, we observe that the relative size of the $\mathcal{O}(\lambda_2g_{h\Phi\Phi}^3/M_\Phi^2)$ pieces compared to the $\mathcal{O}(g_{h\Phi\Phi}^5/M_\Phi^4)$ ones decreases for larger mass splittings, which simply follows from the lower power dependence on $g_{h\Phi\Phi}\propto (M_\Phi^2-\mu_2^2)$.


\subsection{Comparison of renormalisation scheme choices for the \texorpdfstring{\TSM0}{TSM with Y=0}}
\label{sec:TSM0}
\begin{figure}
    \centering
    \includegraphics[width=0.49\textwidth]{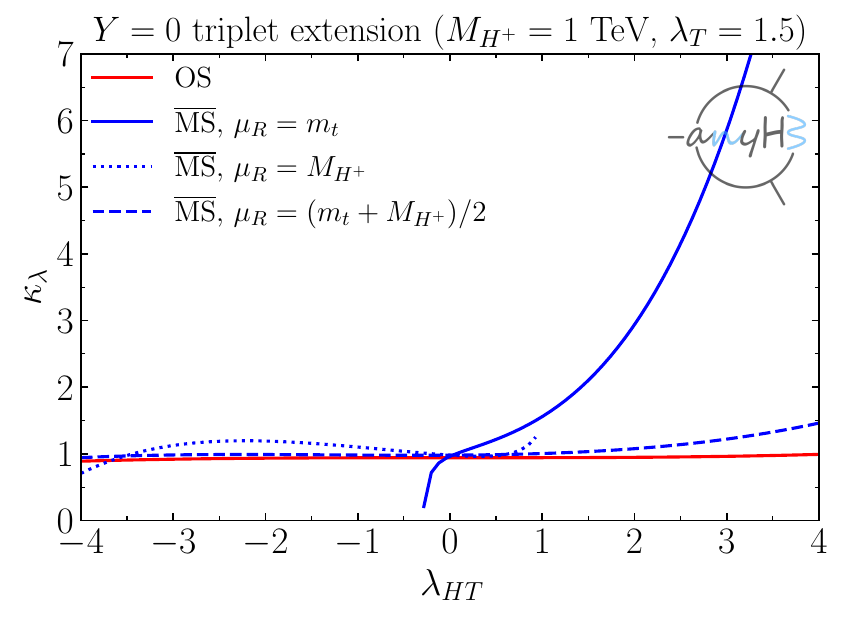}
    \includegraphics[width=0.49\textwidth]{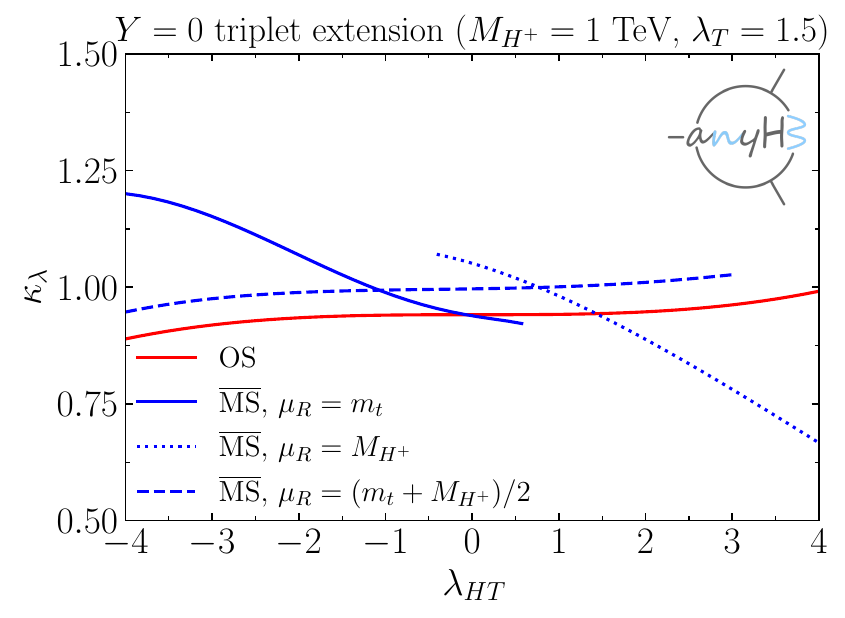}
    \caption{\kaplam in the $Y=0$ triplet extension of the SM as a function of $\lambda_{HT}$, comparing the OS (red curves) and \MS (blue curves) renormalisation schemes for the Higgs-boson mass and the EW VEV. For the \MS case, we consider different choices of the renormalisation scale, shown by the solid, dashed and dotted curves. The charged Higgs mass is set to $M_{H^+}=1\text{ TeV}$, while $\lambda_T=1.5$. \textit{Left}: Results using the FJ scheme for the tadpoles. \textit{Right}: Results using OS-renormalised tadpoles. 
    }
    \label{fig:Yeq0_triplet_tad}
\end{figure}

\begin{figure}
    \centering
    \includegraphics[height=5.5cm]{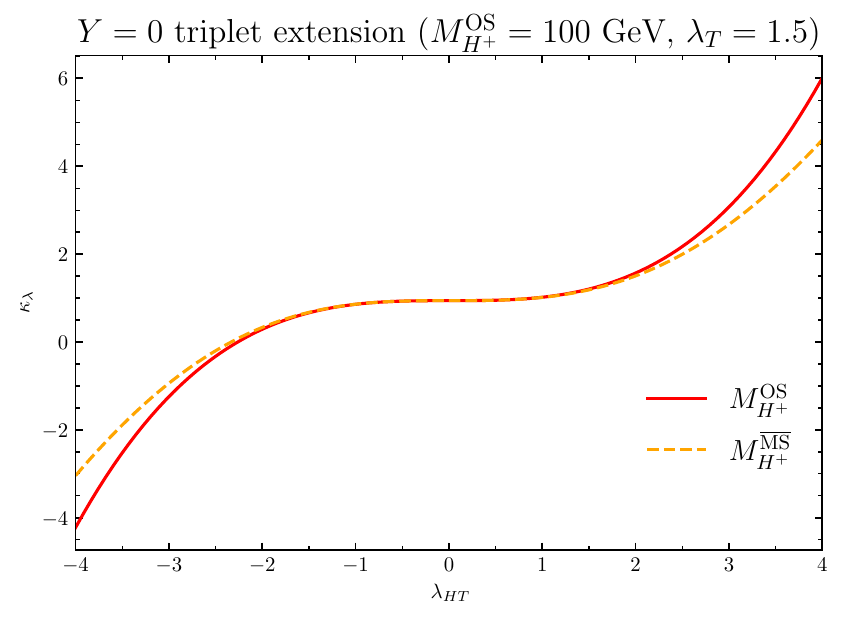}
    \includegraphics[height=5.5cm]{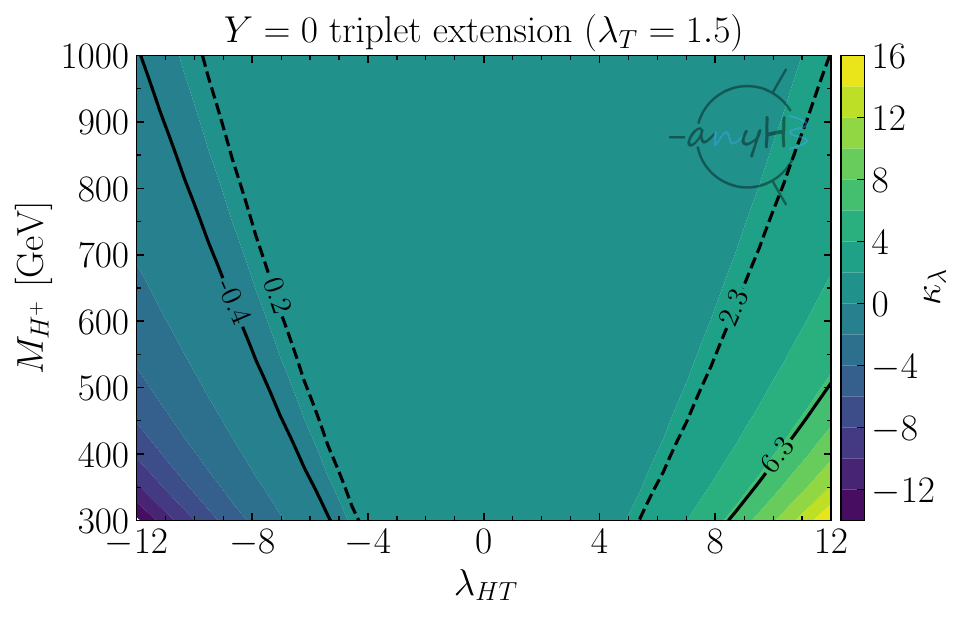}
    \caption{\kaplam in the $Y=0$ triplet extension of the SM. \textit{Left}: \kaplam as a function of $\lambda_{HT}$, comparing results employing the OS- and \MS-renormalised charged Higgs mass. The OS scheme is used for the Higgs mass, the EW VEV, and the tadpoles. 
    \textit{Right}: Results for \kaplam (calculated in the OS scheme) shown in the $(\lambda_{HT},M_{H^+})$ parameter plane.}
    \label{fig:Yeq0_triplet}
\end{figure}

Choosing a suitable renormalisation scheme is a crucial step for the calculation of \kaplam. We illustrate this in \cref{fig:Yeq0_triplet_tad,fig:Yeq0_triplet} for the $Y=0$ triplet extension of the SM (see \cref{app:tsm0} for details about the model and its implementation).

In \cref{fig:Yeq0_triplet_tad}, \kaplam is shown for different renormalisation schemes as a function of the quartic interaction between the SM Higgs doublet and the BSM triplet ($\lambda_{HT}$) with fixed $M_{H^+} = 1\tev$ and $\lambda_T = 1.5$. In the left plot, the tadpoles are treated in the FJ prescription, and therefore enter both the calculation of \lamhhh and the parameter conversion explicitly, while in the right plot an OS renormalisation is used for the tadpoles. If the mass of the SM-like Higgs boson and the EW \vev are renormalised in the on-shell scheme (red curve, identical in both plots), the dependence of \kaplam on $\lambda_{HT}$ is very small. This is expected since the BSM masses are chosen at the TeV scale implying that all BSM corrections should be small as a consequence of decoupling. If the mass of the SM-like Higgs boson and the EW \vev are renormalised in the \MS scheme (blue curves), the result for \kaplam depends strongly on the choice of the renormalisation scale, as well as on the chosen treatment of the tadpoles. 
For all curves, $m_h^\OS = 125.1 \gev$ and $v^\OS \simeq 250.7 \gev$ are used as input which are then converted in the first step to the \MS scheme. Then, these \MS quantities are used to calculate \kaplam. For the considered scenario, the conversion can lead to very large shifts between the OS and \MS quantities if the renormalisation scale is not chosen appropriately. If for example $\mu_R = m_t$ is chosen (solid blue curve), we encounter artificially large corrections to \kaplam for large positive $\lambda_{HT}$ when employing FJ tadpoles. We note that this is due exclusively to the impact of the tadpoles on the \MS parameters obtained from OS inputs, because the tadpole contributions in the calculation of \lamhhh itself are the same independently of the employed scheme, as shown in \cref{app:sm:tadren}. Moreover, for this scale choice the \MS\ mass of the SM-like Higgs boson quickly becomes tachyonic for negative $\lambda_{HT}$ in the case with FJ tadpoles, and for positive $\lambda_{HT}$ in the case with OS tadpoles. Similar issues appear for the choice of $\mu_R = M_{H^+}$ (blue dotted curve) for which the \MS mass of the SM-like Higgs boson becomes tachyonic for $\lambda_{HT} \gtrsim 1$ ($\lambda_{HT}\lesssim -1/2$) when using FJ (OS) tadpoles. For $\mu_R = (m_t + M_{H^+})/2$ (blue dashed curve), however, the corrections are quite well-behaved, and a result close to the OS curve is obtained --- although in the case of OS tadpoles, the Higgs mass once again becomes tachyonic for $\lambda_{HT}\gtrsim 3$. Overall, the choice of OS tadpoles leads to more moderate effects in $\kappa_\lambda$, and appears to be (if implemented) a preferable option for calculations in which scheme conversions of  parameters are performed. 

In the left plot of \cref{fig:Yeq0_triplet}, we next present results for $\kappa_\lambda$ using OS inputs for the Higgs mass and EW VEV (and OS-renormalised tadpoles), but with the charged Higgs boson mass in the OS (red curve) and the \MS scheme (orange curve). As in \cref{fig:Yeq0_triplet_tad}, we compare here results for the same parameter point, defined by $\lambda_T=1.5$ and an OS charged Higgs mass of $M_{H^+}^\text{OS}=100\gev$ (which for the \MS curve is converted to the \MS scheme). Because $M_{H^+}$ only enters the prediction for \lamhhh at the one-loop order, the difference between the red and orange curves is formally of two-loop order and, as was discussed for the IDM in \cref{sec:IDMuncertainties}, can serve as an estimate of the size of the unknown higher-order contributions to \lamhhh. We can observe, as expected, that the results in the two schemes remain very close for small values of $|\lambda_{HT}\lesssim 2|$, and only differ for larger couplings. Finally, additional results for \kaplam, calculated in the OS scheme, are shown the $(\lambda_{HT},M_{H^+})$ parameter plane (for fixed $\lambda_T = 1.5$) in the right panel of \cref{fig:Yeq0_triplet}.\footnote{We checked that perturbative unitarity is preserved at the tree level throughout the shown parameter plane.}  It is clearly visible that large corrections to \kaplam are obtained for low $M_{H^+}$ and large $|\lambda_{HT}|$. The solid black contour lines indicate the parameter region that is excluded by the current LHC bounds on \kaplam~\cite{ATLAS:2022kbf},\footnote{Since we work in the limit where the triplet VEV is zero, the couplings of the Higgs boson at $125\gev$ are SM-like. This limit also implies that no BSM scalars will contribute to the pair production of the SM-like Higgs boson at the tree level. This means that the existing experimental constraints on \kaplam can be applied for the $Y=0$ triplet model.} while the dashed black contour lines show the region that will be probed with the projected sensitivity based on the full HL-LHC dataset~\cite{Cepeda:2019klc}.


\subsection{Comparison of BSM effects arising from mass splittings}

In \cref{sec:anydecoupling} we showed for various models that the BSM contributions to \lamhhh vanish once \textit{all} BSM states are decoupled simultaneously in an \textit{appropriate} way. This behaviour is in accordance with the decoupling theorem \cite{Appelquist:1974tg} which states that, in the case of heavy new physics, all BSM contributions can be incorporated into coefficients $\mathcal{C}^{(n)}_i$ of higher-dimensional SM operators,
\begin{equation}
\mathcal{L}^{d>4} = \sum_{n=5}^\infty \frac{\mathcal{C}_i^{(n)}}{M_{\text{BSM}}^{n-4}}{\cal O}^{(n)}_i\,,
\end{equation}
such that all BSM effects vanish for $M_{\text{BSM}}\to~\infty$. A crucial requirement for this decoupling behaviour is that the $\mathcal{C}_i^{(n)}$ are small and do not increase with $M_{\text{BSM}}$. Mass splittings between the BSM particles can modify this behaviour. This can happen for instance if some of the masses of the heavy BSM states $\phi_\text{BSM}$ are mostly generated via the comparably small SM \vev,
\begin{equation}
      \frac{1}{2}\left(\lambda_{XH}\Phi_\text{SM}^\dagger\Phi_\text{SM}
           + M_L^2
           \right)\phi_\text{BSM}^2 
                   \xrightarrow{\text{SSB}} \frac{1}{2}\left(\frac{1}{2}\lambda_{XH}v^2 + M_L^2\right) \phi_\text{BSM}^2
                    = \frac{M_{\text{BSM}}^2}{2} \phi_\text{BSM}^2\,,
                    \label{eq:portalhiggs}
\end{equation}
where $\phi^2_\text{BSM}$ schematically stands for the quadratic term of some BSM  scalar, and $\Phi_\text{SM}$ is the SM-like doublet. Thus, $\phi_{\text{BSM}}$ receives mass contributions both from the quartic coupling $\lambda_{XH}$ and the mass parameter $M_L$. For the case of $M_L\sim v$ a large scalar mass $M_{\text{BSM}}~\gg~v$ can be realised via $\lambda_{XH}~\gg~1$. The quartic interaction of BSM scalars to SM-like Higgs bosons can lead to large contributions in this case,
\begin{equation}
    c_{hh \phi_{\text{BSM}} \phi_{\text{BSM}}}\propto\lambda_{XH} = 2\frac{M_{\text{BSM}}^2-M_L^2}{v^2}
    \,.
    \label{eq:nondec}
\end{equation}

Another way of understanding the origin of such large contributions is related to the symmetry argument of the decoupling theorem \ie\ that the decoupling of a heavy particle must not break any symmetries of the resulting effective theory (EFT). To demonstrate this we consider the states $X_i$ of some irreducible $SU(2)_L$ multiplet $\bm{X}$ with masses $M_{X_{i\neq L}}\equiv M_{\text{BSM}}$ and $M_{X_{L}}\equiv M_L\ll M_{\text{BSM}}$. Taking the limit $M_{\text{BSM}}\to\infty$ leads to an EFT which is not $SU(2)_L$-invariant anymore as $X_L$ cannot be incorporated into a smaller $SU(2)_L$ multiplet. As a consequence, portal couplings as in \cref{eq:portalhiggs} between the SM- and the BSM-Higgs bosons become large for a large mass splitting. In fact, for all models implemented in \anyH with additional states charged under $SU(2)_L$ we found the couplings $c_{hh X_i X_j}$ to behave as in \cref{eq:nondec}, provided that appropriate parameterisations for the input masses are assumed. Note that the above discussion applies not only to the $SU(2)_L$ gauge symmetry but more generally to global and gauge symmetries of the BSM theories.

\begin{figure}
    \centering
    \includegraphics[width=0.8\textwidth]{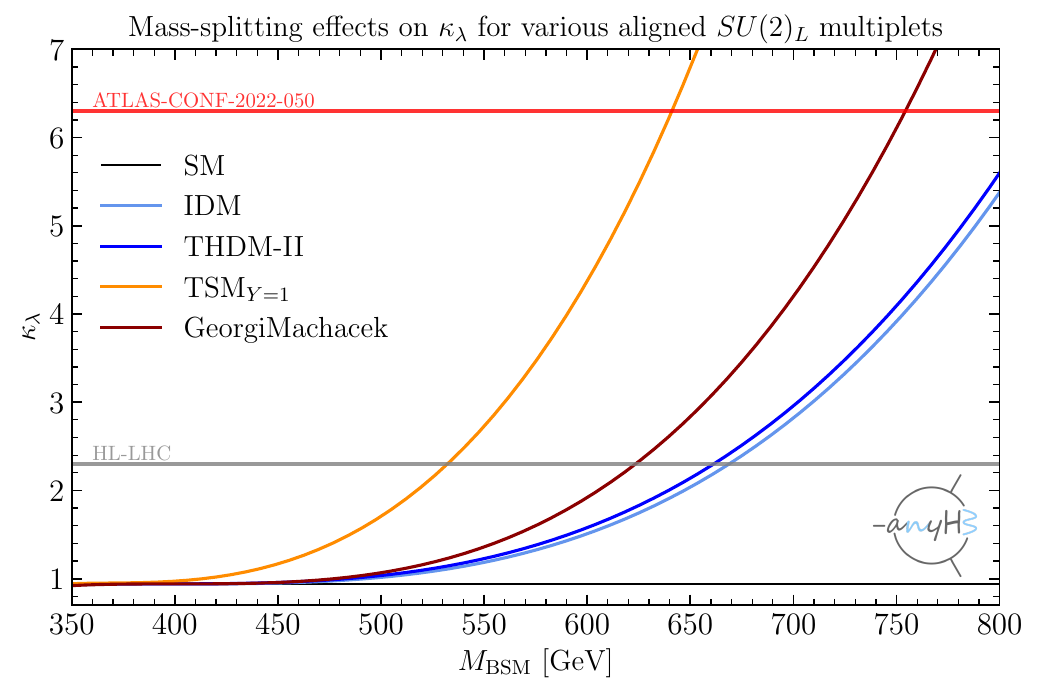}
    \caption{
    In all shown models the mass of the lightest BSM state which is charged under the $SU(2)_L$ gauge group is set to $M_L=\unit[400]{GeV}$. For the different models the following parameter choices have been made: \textbf{IDM}: $M_{H}=\mu_2=M_L$. \textbf{THDM-II}: $M=M_{H}=M_L$. \textbf{\TSM1}: $m_{D^{++}}=M_L$. \textbf{GeorgiMachacek}: $M_{h_2}=M_\eta=M_L$. All other parameters are chosen as in \cref{fig:decouplingall}. In particular, the other BSM masses are degenerate with a mass value of $M_{\text{BSM}}$.}
    \label{fig:nondec}
\end{figure}

It should be noted that there can be regions in parameter space where the splitting of the mass parameters $M_{\text{BSM}}^2-M_L^2$ is relatively large compared to the electroweak scale while the model can still be described perturbatively. From the THDM it is known that such large couplings can be constrained by the current experimental bounds on $\kaplam$ while being in agreement with all other experimental and theoretical constraints~\cite{Bahl:2022jnx}. With the help of \anyH\ one can easily go beyond the THDM and study the effect of couplings determined by \cref{eq:nondec} onto \kaplam in other SM extensions. To demonstrate this, we use the example of the different $SU(2)_L$ extensions from \cref{sec:anydecoupling} and fix one of the BSM scalar masses to $M_L=\unit[400]{GeV}$ rather than having all masses degenerate at $M_{\text{BSM}}$. \cref{fig:nondec} shows the resulting \kaplam prediction as a function of $M_{\text{BSM}}$ for the IDM (light blue), the THDM (blue), the \TSM1 (orange) and the Georgi-Machacek model (brown). We want to emphasise again that all shown parameter points are chosen to be in the alignment limit, \ie\ they have a tree-level prediction of $\kaplam^{(0)}=1$. In agreement with \cref{eq:nondec}, we observe in all models that $\kaplam\approx\kaplam^{\text{SM}}$ for $M_{\text{BSM}}\approx M_L=\unit[400]{GeV}$. For increasing values of $M_{\text{BSM}}$, corrections proportional to couplings of the form of \cref{eq:nondec} lead to a large increase of \kaplam so that it can become close to or even larger than the current experimental constraint (red horizontal line). The projection for the sensitivity on \kaplam at the HL-LHC (grey horizontal line) shows that it will be possible to probe mass splittings down to 150--200~GeV in the displayed examples. We explicitly verified, using the \texttt{anyPerturbativeUnitarity} module of \anyBSM, that all models fulfil the tree-level perturbative unitarity constraint in the high-energy limit for all shown values of \kaplam.

It is also important to stress that this discussion is not restricted to the $SU(2)_L$ gauge symmetry of the SM but also applies to any other symmetry within or beyond the SM.\footnote{See \eg\ \ccite{Gabelmann:2019jvz} for a discussion of contributions to trilinear Higgs couplings caused by BSM fermions in split-SUSY models.}

\subsection{Phenomenological results for \texorpdfstring{\lamhhh}{lamhhh} in the NTHDM}
\label{sec:NTHDM}
In this section we discuss an example of phenomenological results for \lamhhh in the NTHDM (other investigations of $\lambda_{hhh}$ in the NTHDM have also been performed with \texttt{BSMPT} in Ref.~\cite{Basler:2019iuu}). The scalar sector of this model contains three \cp-even scalars, which can mix, and in turn three mixing angles are required to diagonalise the $3\times 3$ \cp-even scalar mass matrix (see model definitions and details in \cref{app:nthdm}). It is of interest in this context to investigate the potential impact of mixing on the prediction for the trilinear Higgs coupling. 

\begin{figure}
    \centering
    \includegraphics[width=.9\textwidth]{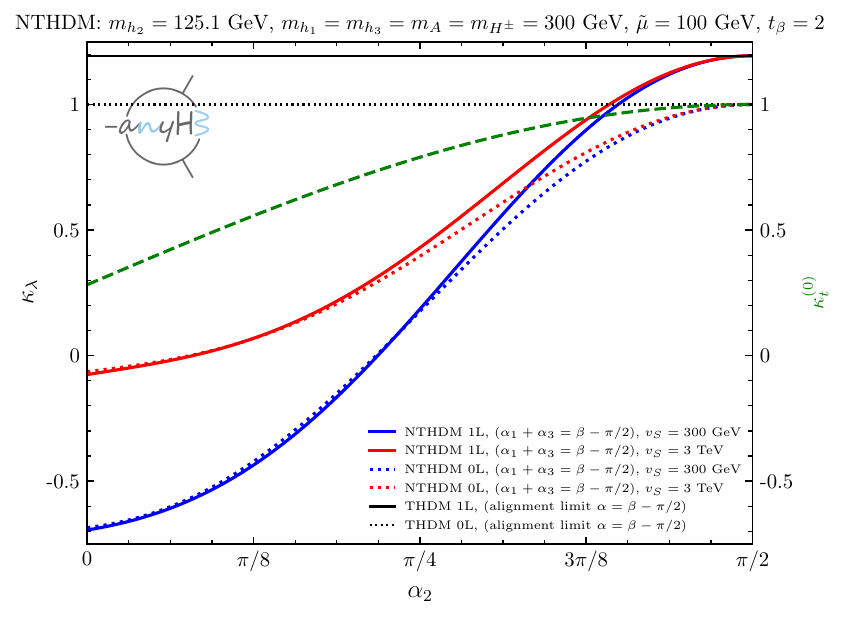}
    \caption{Predictions for $\kappa_\lambda$ (red, blue, and black lines) and for $\kappa_t$ (green line) in the NTHDM and THDM, as a function of the NTHDM mixing angle of the \cp-even scalar sector $\alpha_2$. Results for \kaplam are shown at tree level (dashed curves) and at one loop (solid lines). The masses of the BSM scalars are taken to be degenerate at 300 GeV, while the BSM mass scales --- $\tilde{\mu}$, defined by $\tilde{\mu}^2=m_{12}^2/(\cos\beta \sin\beta)$, for the NTHDM and $M$ for the THDM --- are chosen to be 100 GeV, and $\tan\beta=2$. For the singlet vev in the NTHDM we consider two scenarios: $v_S=300$ GeV in blue and $v_S=3$ TeV in red. }
    \label{fig:NTHDM}
\end{figure}

As a brief illustration of phenomenological studies made possible with \anyBSM, we present in \cref{fig:NTHDM} results for \kaplam as a function of the second \cp-even mixing angle $\alpha_2$. We choose a scenario of the NTHDM where $h_2$ is identified with the detected Higgs boson at 125~GeV, so that the alignment limit is reached for $\alpha_1+\alpha_3\to\beta-\pi/2$ and $\alpha_2\to\pi/2$. We consider in \cref{fig:NTHDM} scenarios of the NTHDM and THDM were the BSM scalars ($h_2$, $h_3$, $A$, $H^\pm$ for the NTHDM; $h_2$, $A$, $H^\pm$ for the THDM) are mass-degenerate with a mass value of 300~GeV. We fix the BSM mass scales of both models ($\tilde{\mu}$ for the NTHDM and $M$ for the THDM) to 100 GeV, resulting in a sizeable mass splitting giving rise to a significant contribution to \kaplam. Additionally, we set $\tan\beta=2$ and $\alpha_1+\alpha_3=\beta-\pi/2$, while for the singlet \vev, $v_S$, we adopt two values: $v_S=300$ GeV (blue curves) and $v_S=3$ TeV (red curves). Regarding our choice of renormalisation scheme, we employ here an OS renormalisation of all scalar masses and of the Higgs \vev, while the other parameters are renormalised in the \MS scheme. Note that this scenario is devised as a simple setting in which to demonstrate calculations that are made possible by \anyBSM. The range of parameters shown in \cref{fig:NTHDM} is not allowed in its entirety, however, we do not explicitly indicate the exclusion limits since we are not aiming at a thorough phenomenological analysis here. 

The solid curves in \cref{fig:NTHDM} show the full one-loop results for \kaplam, while the dashed lines correspond to the tree-level results. For $\alpha_2\to\pi/2$, we observe --- as expected --- that we recover the alignment limit, and for both possible values of $v_S$ the tree-level and one-loop predictions for \kaplam converge to the results in the THDM, indicated by the black horizontal lines. In this limit, the additional singlet decouples entirely, and the dependence of \kaplam on $v_S$ vanishes. A sizeable BSM contribution remains in this limit, yielding a value of $\kaplam\sim 1.25$, which arises from the corrections involving the THDM-like scalars ($h_2$, $A$, and $H^\pm$). On the other hand, away from the alignment limit, and for $\alpha_2\lesssim \pi/4$, the relative importance of the loop corrections to \lamhhh decreases significantly. It should be pointed out here, for completeness, that deviations of $\kappa_t$ from the SM are already constrained by experimental data to be below $\mathcal{O}(20\%)$ (see for instance \ccite{ATLAS:2022vkf}). This implies that values of $\alpha_2\lesssim \pi/4$ are already excluded in this scenario. 

Furthermore, we can observe the interesting feature that the prediction for \kaplam becomes negative as $\alpha_2$ decreases --- \ie\ as one departs from the alignment limit. At this point, it is however important to remark that the sign of \lamhhh is not a physical observable. A quantity that is of physical relevance is the relative sign between the trilinear Higgs coupling and other couplings of the Higgs boson, \eg\ its coupling to top quarks. For this reason, we also present in \cref{fig:NTHDM} results (green line) for the coupling modifier of the top Yukawa interaction, which we denote $\kappa_t$, at tree level. We find that for the entire range of $\alpha_2$, $\kappa_t^{(0)}$ remains positive, so that a change in the relative sign between the trilinear Higgs coupling and the top Yukawa does occur --- this can in principle lead to a significant increase in the Higgs pair production cross-section, as the destructive interference between the box and triangle diagrams occurring in the SM is avoided in this case. We leave a more thorough investigation of scenarios with negative trilinear Higgs couplings in the NTHDM for future work.

\subsection{Momentum-dependent effects in \texorpdfstring{\lamhhh}{lamhhh}}
\label{sec:momdep}

By default, \anyH evaluates the trilinear Higgs coupling setting the momentum of all external legs to zero. The code, however, also allows the evaluation of \lamhhh for finite momenta (via the argument \code{momenta} of the \code{lambdahhh} function).

\begin{figure}
    \centering
    \includegraphics[width=.8\textwidth]{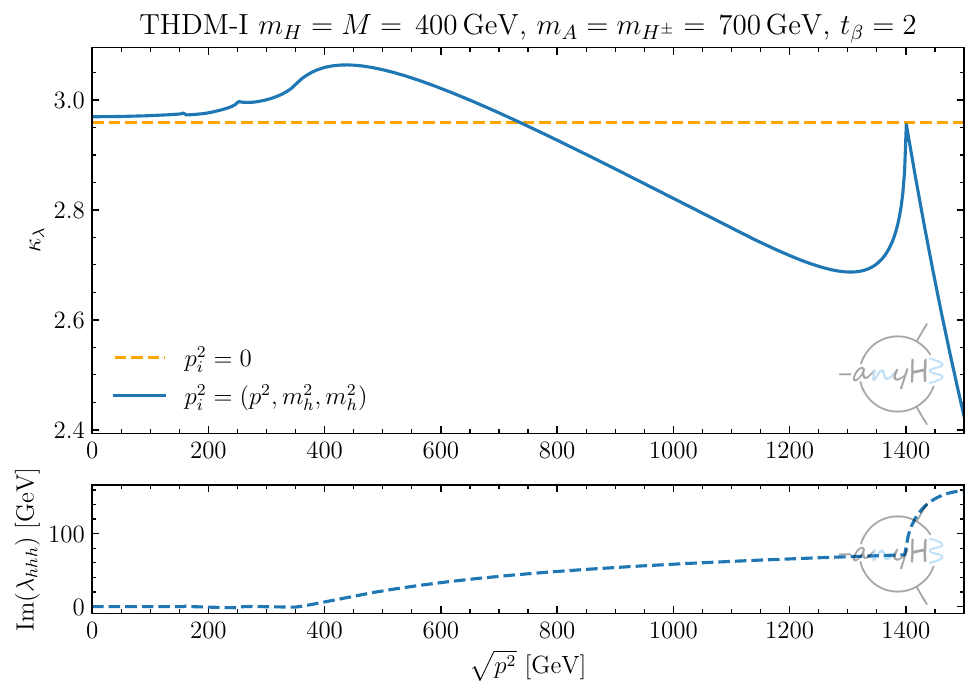}
    \caption{\textit{Upper panel}: momentum dependence of \kaplam in the THDM of type-I, with $M=m_{h_2}=400\gev$, $m_A=m_{H^\pm}=700\gev$, and $\tan\beta=2$. \textit{Lower panel}: same as upper panel but the imaginary part of \lamhhh is shown.}
    \label{fig:THDM-I_triple_p2_dep_BP2}
\end{figure}

\begin{figure}
    \centering
    \includegraphics[width=.8\textwidth]{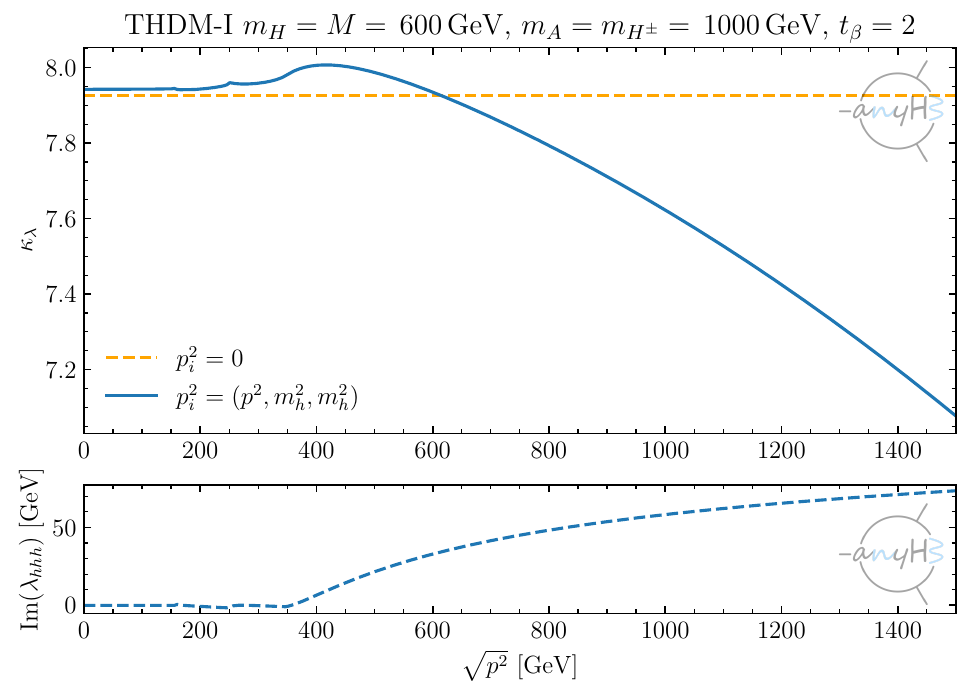}
    \caption{\textit{Upper panel}: momentum dependence of \kaplam in the THDM of type-I, with $M=m_{h_2}=600\gev$, $m_A=m_{H^\pm}=1000\gev$, and $\tan\beta=2$. \textit{Lower panel}: same as upper panel but the imaginary part of \lamhhh is shown.}
    \label{fig:THDM-I_triple_p2_dep_BP3}
\end{figure}

\begin{figure}
    \centering
    \includegraphics[width=.8\textwidth]{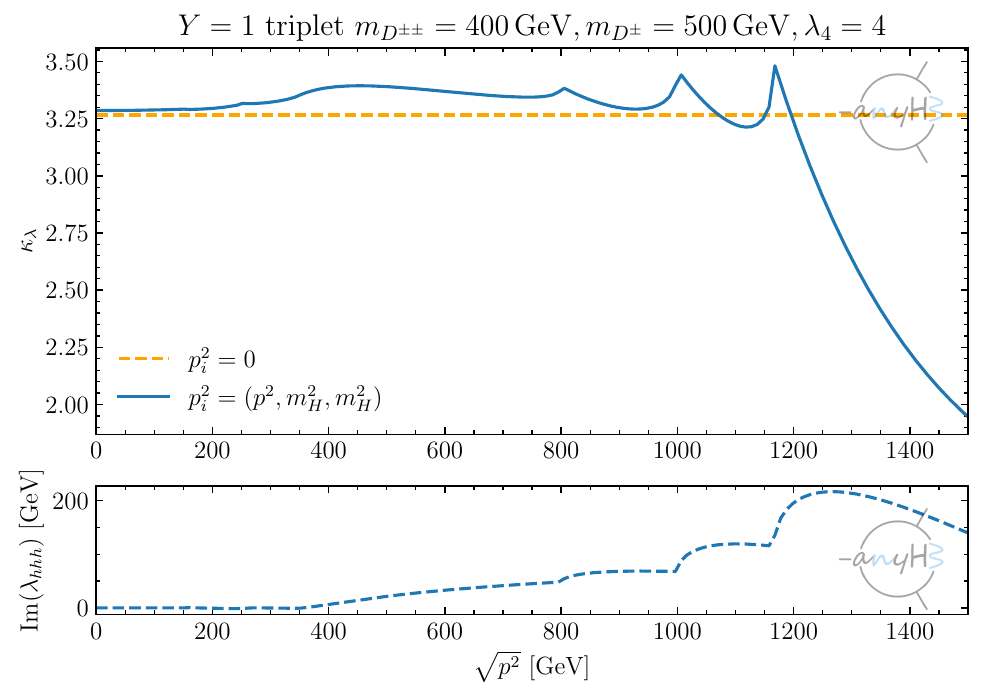}
    \caption{\textit{Upper panel}: momentum dependence of \kaplam evaluated in the \TSM1 with $m_{D^{\pm\pm}}=400\gev$, $m_{D^\pm} = 500\gev$, and $\lambda_4 = 4$. \textit{Lower panel}: same as upper panel but the imaginary part of \lamhhh is shown.}
    \label{fig:Yeq1_triple_p2_dep}
\end{figure}

We demonstrate this for the THDM of type I (see \cref{app:thdm}) and for the $Y=1$ triplet extension of the SM (\TSM1, see \cref{app:tsm1}) in \cref{fig:THDM-I_triple_p2_dep_BP2,fig:THDM-I_triple_p2_dep_BP3,fig:Yeq1_triple_p2_dep}. In all three scenarios, $\kappa_\lambda=1$ at the tree level. For the THDM-I (\cref{fig:THDM-I_triple_p2_dep_BP2,fig:THDM-I_triple_p2_dep_BP3}), we present results for two scenarios, with $M=m_{h_2}=400$ GeV, $m_A=m_{H^\pm}=700$ GeV for the first and $M=m_{h_2}=600$ GeV, $m_A=m_{H^\pm}=1000$ GeV for the second, and with $\tan\beta=2$ for both. Next, for the \TSM1 (\cref{fig:Yeq1_triple_p2_dep}), we set $m_{D^{\pm\pm}}=400\gev$, $m_{D^\pm} = 500\gev$, and $\lambda_4 = 4$. All three scenarios are chosen such that significant BSM effects occur in the trilinear Higgs coupling, and additionally the second THDM-I scenario is devised specifically to obtain a value of \kaplam larger than the current upper experimental bound of 6.3. 

In the upper panels of \cref{fig:THDM-I_triple_p2_dep_BP2,fig:THDM-I_triple_p2_dep_BP3,fig:Yeq1_triple_p2_dep}, we show \kaplam as a function of a varying external momentum scale $\sqrt{p^2}$. The orange dashed line denotes the \kaplam value if all squared external momenta $p_i^{2}$ are set to zero. For the blue curve, two of the external momenta are set equal to the mass of the SM-like Higgs boson, while the momentum of the third leg is kept general and set to $p^2$. For this off-shell leg, we also do not include an external wave-function renormalisation, see also \cref{sec:extleg}. This means that the result shown in blue corresponds precisely to the quantity that enters the evaluation of the triangle diagram contributing to di-Higgs production. Comparing the solid blue and the orange-dashed curves, we observe that, for low to intermediate ranges of $\sqrt{p^2}$, the momentum effects are small in comparison to the overall size of the BSM effects, which shift \kaplam to about $2.95$, $7.9$, and $3.3$ respectively for the three figures. It is only for larger values of $\sqrt{p^2}$ --- namely $\sqrt{p^2}\gtrsim 600-700$ GeV for the THDM-I scenarios and $\sqrt{p^2} \gtrsim 1.2\tev$ for the \TSM1 --- that the momentum effects become sizeable and cause a significant decrease in \kaplam.

Finite external momenta can also induce imaginary parts for \lamhhh (for the calculation of \kaplam, we take the real part of \lamhhh). These are shown in the lower panels of \cref{fig:THDM-I_triple_p2_dep_BP2,fig:THDM-I_triple_p2_dep_BP3,fig:Yeq1_triple_p2_dep}. Several particle thresholds are visible (corresponding to what can be seen in the upper panels of the corresponding figures): \eg, the di-Higgs threshold around $\sqrt{p^2}\sim 250\gev$, the di-top threshold around $\sqrt{p^2}\sim 350\gev$, and for instance for the \TSM1 (\cref{fig:Yeq1_triple_p2_dep}), also the $D^{\pm\pm}D^{\mp\mp}$ threshold ($\sqrt{p^2}\sim 800\gev$) and the $D^{\pm}D^{\mp}$ threshold ($\sqrt{p^2}\sim 1000\gev$). Note that for the THDM-I (\cref{fig:THDM-I_triple_p2_dep_BP2,fig:THDM-I_triple_p2_dep_BP3}), there are no $h_2h_2$ thresholds, because the $\lambda_{h_1h_2h_2}$ coupling vanishes (due to the equality $M=m_{h_2}$) and hence diagrams with internal \cp-even scalars $h_2$ do not contribute to $\lambda_{hhh}$. In \cref{fig:THDM-I_triple_p2_dep_BP2}, the $AA/H^+H^-$ threshold is visible, as expected, at $\sqrt{p^2}=1.4\tev$.

The results shown in \cref{fig:THDM-I_triple_p2_dep_BP2,fig:THDM-I_triple_p2_dep_BP3,fig:Yeq1_triple_p2_dep} can directly be applied to di-Higgs boson production by treating \lamhhh as a momentum-dependent quantity entering the cross-section calculation. In this context, it is important to note that the di-Higgs invariant mass distribution typically peaks around $\sqrt{p^2}\sim 400\gev$ (see \eg~\ccite{LHCHiggsCrossSectionWorkingGroup:2016ypw}) and then quickly falls off (by several orders of magnitude) as $\sqrt{p^2}$ increases. This implies that the sizeable momentum dependence found for larger values of $\sqrt{p^2}$ only has a small impact on the total di-Higgs boson production cross section. For the representative value of $\sqrt{p^2}=400\gev$, \ie\ around the peak of the di-Higgs differential cross-section, the momentum-dependence contributes positive shifts of 3.4\% for the first THDM-I scenario, 1.0\% for the second THDM-I case, and 3.6\% for the \TSM1 scenario. Furthermore, we observe that, in all three considered scenarios, the imaginary part of \lamhhh remains minute until $\sqrt{p^2}\gtrsim 350-400\gev$ (\ie\ until the di-top threshold), and only reach sizeable values for $\sqrt{p^2}$ well above $400\gev$, and thus far from the peak of the di-Higgs invariant mass distribution. Consequently, we find that evaluating \lamhhh at zero external momenta is a good approximation of the full result for the scenarios considered here. We leave a more detailed investigation of the impact of non-zero momenta in di-Higgs production for future work. Moreover, we observe that for the points with the most sizeable BSM deviations in \kaplam~--- such as the second THDM-I scenario --- the relative magnitude of the momentum-dependent effects are the smallest, compared to the overall value of \kaplam. Similarly, the magnitude of the imaginary part of \lamhhh, which originates dominantly from the SM-like loop contributions, remains approximately the same in \cref{fig:THDM-I_triple_p2_dep_BP2,fig:THDM-I_triple_p2_dep_BP3} (apart from the additional threshold at $1.4\tev$ in \cref{fig:THDM-I_triple_p2_dep_BP2}). Thus, its relative size compared to its real part diminishes for scenarios with larger $\kappa_\lambda$. 


\subsection{Use of \anyH together with a spectrum generator: an example in the MSSM}
\label{sec:applications_MSSM}

\begin{figure}
    \centering
    \includegraphics[width=\textwidth]{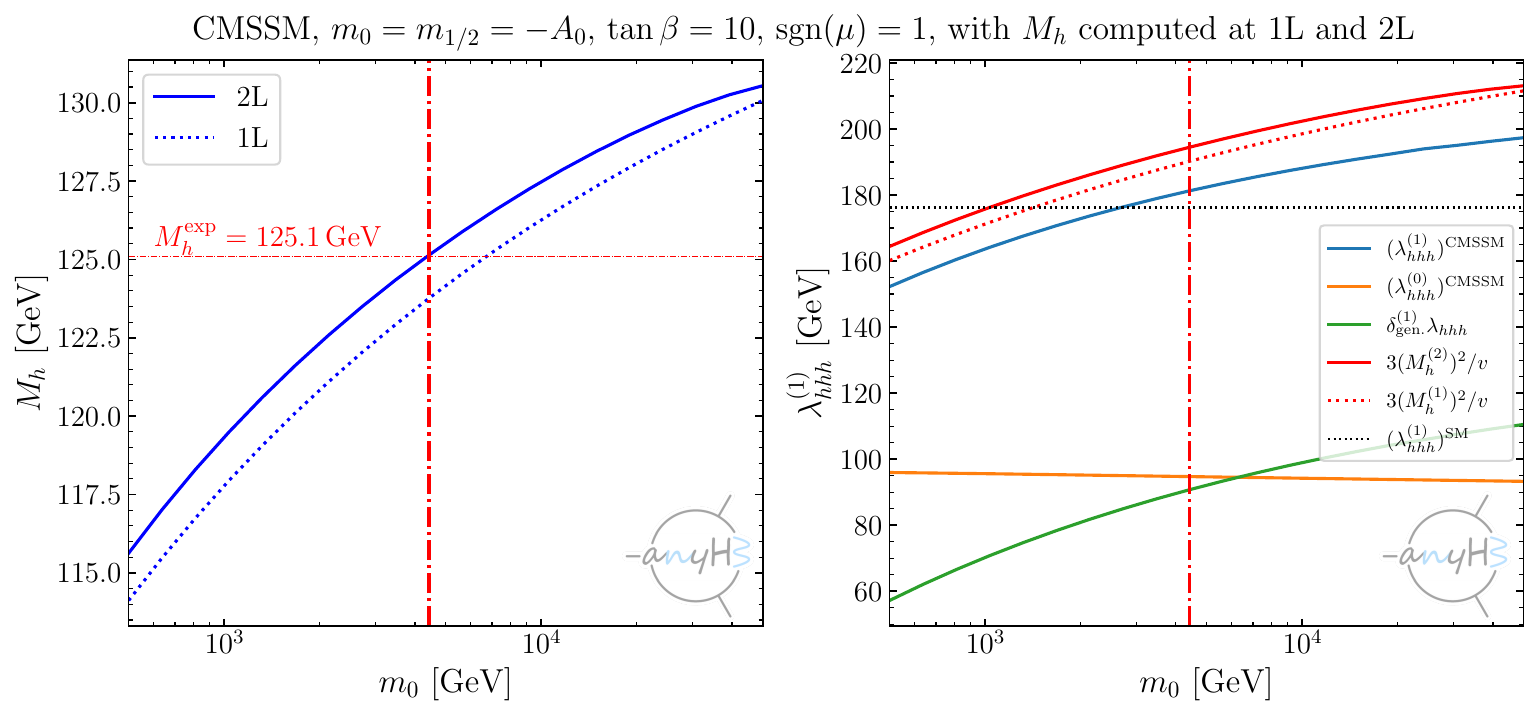}
    \caption{\textit{Left}: Mass of the SM-like Higgs boson computed with \texttt{SARAH}/\texttt{SPheno}~\cite{Staub:2013tta, Porod:2011nf} at the one- and two-loop levels in the CMSSM as a function of the SUSY scale parameter $m_0$. 
    \textit{Right}: Same as the left panel, but results for the trilinear Higgs coupling are shown.
    }
    \label{fig:MSSM}
\end{figure}

\anyBSM can easily be interfaced with spectrum generator tools. This is demonstrated in \cref{fig:MSSM} for the constrained Minimal Supersymmetric Standard Model (CMSSM). For this example, we use \texttt{SARAH}/\texttt{SPheno} to generate the mass spectrum for different values of the SUSY scale parameter $m_0$ (fixing the other BSM parameters via $m_0 = m_{1/2} = - A_0$, $\tan\beta = 10$, $\text{sgn}(\mu)=1$). The resulting predictions for the mass of the SM-like Higgs boson are shown in the left panel of \cref{fig:MSSM}: the solid curve shows the two-loop results, while the dotted curves corresponds to the one-loop result. The experimental value of $\sim 125\gev$ is reached for $m_0 \sim 4.4\tev$ in this example.

The loop-corrected mass spectrum computed with \texttt{SARAH}/\texttt{SPheno} is then passed to \anyH using the \texttt{SLHA} interface (see \cref{sec:tutorial:setparams}). The resulting prediction for the trilinear Higgs coupling is shown in the right panel of \cref{fig:MSSM}. The orange curve shows the tree-level prediction, which is given in terms of the tree-level Higgs mass ($M_{h,\text{tree}}^2\simeq M_Z^2 \cos^2{2\beta}$) divided by the electroweak VEV. The finite part of the genuine one-loop corrections to the trilinear Higgs coupling, represented by the green curve, is roughly of the same size as the tree-level prediction. Adding both contributions (together with the additional counterterm, external-leg, and tadpole contributions), the full one-loop result is obtained (blue curve). For $m_0\sim 4.4\tev$, for which $M_h^{(2)} \sim 125\gev$, we find $\lamhhh \simeq 180\gev$. This result is very close to the one-loop SM value of $\sim 176\gev$. For comparison, we also show in red the result of 
the effective lowest-order contribution $3 M_h^2/v$, where $M_h$ incorporates the corrections to the mass of the SM-like Higgs boson up to the one-loop (red dotted curve) or the two-loop level (red solid curve). These results, which are quite close to the full one-loop result (within $\sim 10\gev$), indicate that the bulk of the corrections to $\lambda_{hhh}$ in the CMSSM enters via the loop-corrected prediction to the mass of the SM-like Higgs boson.


\subsection{Non standard couplings}
While \cref{sec:genericcalc} discusses specifically the calculation of the trilinear coupling of the SM-like Higgs boson, the program \anyH is in general able to calculate any trilinear self-coupling $\lambda_{h_i h_i h_i}$ 
where the three external Higgs bosons are the same. To demonstrate this feature we again consider the real singlet extension of the SM, the SSM, and compute both the SM-like Higgs coupling $\lamhhh$ and the singlet coupling $\lambda_{sss}$. For simplicity, we set the singlet--doublet mixing angle to zero, $\alpha=0$, which leads to the tree-level expressions
\begin{subequations}
    \begin{align}
        \lamhhh^{(0)}       &=  \frac{3 m_h^2}{v}\,, \quad \text{and} \\
        \lambda_{sss}^{(0)} &= \frac{3 m_s^2}{v_S} - \kappa_S + \frac{3}{2}\frac{v^2}{v_S^2}\kappa_{SH}\,,
    \end{align}
\end{subequations}
where we fix the mass of the SM-like Higgs boson to $m_h=\unit[125]{GeV}$ and allow the singlet mass $m_s$ to be either larger or smaller than $m_h$ (see \cref{app:ssm} for more details about the model). In this case the renormalisation of $\lamhhh$ at the one-loop order is identical to the SM. For the trilinear singlet coupling, we choose to renormalise $\kappa_S$ and $\kappa_{SH}$ in the \MS and $m_s$ in the OS scheme. The renormalisation of the singlet \vev $v_S$ involves contributions from the singlet tadpole $t_s$. \anyH provides the necessary ingredients to compute the one-loop prediction for $\lambda_{sss}$ in the considered scenario. 

\begin{figure}
    \centering
    \includegraphics[width=\textwidth]{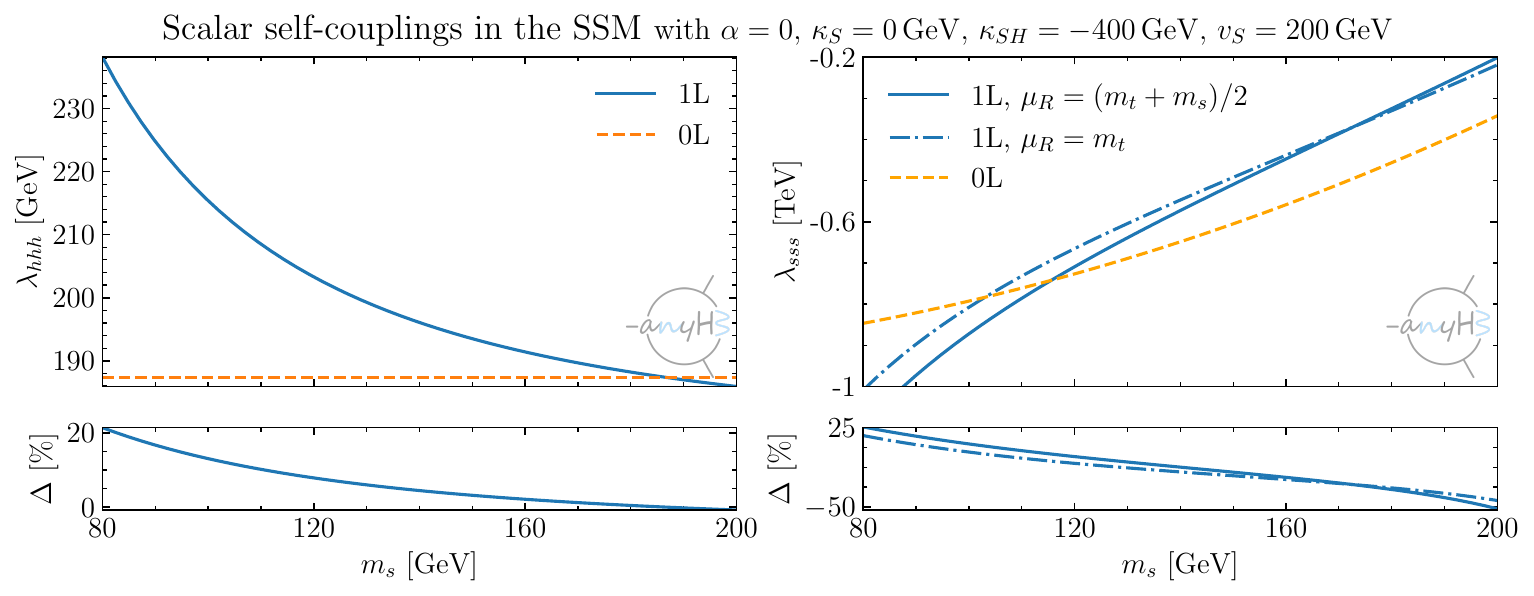}
    \caption{One-loop corrections to trilinear scalar self-couplings in the SSM as a function of the tree-level singlet mass. \textit{Left:} the SM-like Higgs self-coupling. \textit{Right:} the scalar singlet coupling for a fixed renomalisation scale, $\mu_R=m_t$ (dash-dotted), and a dynamically chosen scale, $\mu_R=(m_t+m_s)/2$ (solid). Tree-level couplings are shown in orange (dashed). \textit{Lower panels:} relative difference between one-loop and tree-level predictions in percent.}
    \label{fig:SSMlamsss}
\end{figure}

\cref{fig:SSMlamsss} shows the prediction for \lamhhh (left) and $\lambda_{sss}$ (right) for $\kappa_S=0$, $\kappa_{SH}=\unit[-400]{GeV}$ and $v_S=\unit[200]{GeV}$ as a function of the singlet mass $m_s$ in the interval $80\,\text{GeV}\leq m_s \leq 200\,\text{GeV}$. This corresponds to the scenario $\lambda_{\Phi S}=1$ that has been studied in \ccite{He:2016sqr}.\footnote{In contrast to \ccite{He:2016sqr}, we use vanishing external momenta in this example.
The theoretical constraints checked in \ccite{He:2016sqr} are not affected by this choice.} The corrections to $\lamhhh$ 
reach up to 20\% for small values of $m_s$ in this scenario. For increasing $m_s$ the prediction for $\lamhhh$ approaches the one of the one-loop SM result. We find corrections to $\lambda_{sss}$ between $-50$ to $+25\%$ in the considered singlet mass range. For non-zero soft-$\Ztwo$-breaking parameters the result for $\lambda_{sss}$ depends on the chosen renormalisation scale (while \lamhhh does not depend on the renormalisation scale since the \MS parameters $\kappa_S$ and $\kappa_{SH}$ do not enter the prediction for \lamhhh at the tree-level). In order to demonstrate the dependence on the renormalisation scale we calculate $\lambda_{sss}$ for the two options of using a fixed scale $\mu_R=m_t$ and a dynamical scale $\mu_R=(m_t+m_s)/2$. We find that the difference in the $\lambda_{sss}$ prediction for these two scale choices is at least about a factor 3 smaller than the overall size of the corrections for most of the considered $m_s$ range. We have explicitly verified (besides the UV-finiteness of the result obtained with \anyH) that $\lambda_{sss}$ is independent of the renormalisation scale for $\kappa_{SH}=\kappa_S=0$.


\section{Conclusions}
\label{sec:conclusions}

Obtaining information about the trilinear Higgs coupling \lamhhh is crucial for determining the shape of the Higgs potential and for gaining a better understanding of the nature of the electroweak phase transition. BSM contributions to \lamhhh can be large even for cases where all the couplings of the detected Higgs boson to gauge bosons and fermions are very close to the SM predictions. Thus, the comparison of the theoretical predictions for \lamhhh in different models --- taking into account contributions at the quantum level --- with the available experimental constraints on \lamhhh\ plays an important role for discriminating between the SM and extensions or alternatives of it.  

It is therefore the main purpose of the public \py code \anyH, which we have presented in this paper, to provide precise predictions for \lamhhh in a wide variety of models. \anyH, which is part of the broader \anyBSM framework, calculates the trilinear Higgs coupling in the SM and renormalisable BSM extensions of it at the one-loop level. For the model input, the code supports the widely used \UFO format. This allows the user to easily extend the library of models shipped alongside the code. Already 14 models are provided in this library.

The code implements generic one-loop corrections which are mapped to the respective \UFO model. For renormalisation, semi-automatic routines allow the user to easily implement different renormalisation schemes. Besides calculating the trilinear Higgs coupling, \anyBSM also supports the calculation of other quantities like scalar and vector boson self-energies.

We have validated the results of \anyH by explicit analytical cross-checks, various consistency checks (such as cancellation of UV divergences and decoupling of BSM contributions), and numerical cross-checks against known results in the literature. Besides comparing with known results, we have also presented new results for various models. In this context we have investigated different aspects like renormalisation scheme dependence, non-decoupling effects, momentum dependence, and negative trilinear Higgs couplings.

\anyH can be used in the form of a \py module, called from the command line, or accessed via a \mat interface. All output quantities can either be evaluated analytically or numerically. To evaluate the required loop functions, \anyH employs a link to the \texttt{COLLIER} library, which is available as the independent \py module \pyCollier. Besides detailed \examples (including scripts to reproduce all plots in this paper), we also provide an extensive \docs. 

The code base of \anyBSM is not restricted to the calculation of \lamhhh, but can be easily extended to support the calculation of other observables like trilinear Higgs couplings with different external scalars ($i.e.$ of the form $\lambda_{ijk}$) or electroweak precision observables. We leave this for future work.


\section*{Acknowledgements}
\sloppy{We thank M.\ Bosse, M.\ Hannig, K.\ Radchenko Serdula, J.\ Scheibler and A.\ Verduras for interesting discussions and valuable contributions to testing the program, as well as N.\ Nomachi for designing the logo. H.B.\ acknowledges support from the Alexander von Humboldt foundation. J.B., M.G., and G.W.\ acknowledge support by the Deutsche Forschungsgemeinschaft (DFG, German Research Foundation) under Germany's Excellence Strategy --- EXC 2121 ``Quantum Universe'' --- 390833306. 
This work has been partially funded by the Deutsche Forschungsgemeinschaft 
(DFG, German Research Foundation) --- 491245950. 
}


\appendix


\section{Definitions and conventions for the generic results implemented in \anyBSM}
\label{app:generic}
This appendix discusses the conventions employed in the generic expressions within \anyBSM (and \anyH) and how to convert the conventions of an already existing \UFO model to match these. For generic vertices, which are used to obtain general results for Feynman diagrams, the program closely follows the conventions introduced in \SARAH/\SPheno \cite{Goodsell:2017pdq}, which are equivalent to the definitions used by the option \texttt{InsertionLevel->\{Generic\}} of \FeynArts \cite{Hahn:2000kx}. The \UFO format, however, is much more general in the sense that every vertex $\boldsymbol{V}$ defined by a list of fields consists of two concurrent lists: a set of arbitrary Lorentz structures $\boldsymbol{L}_i$ and a set of arbitrary couplings $\boldsymbol{C}_i$. The decomposition of $\boldsymbol{V}=\boldsymbol{L}\cdot\boldsymbol{C}$, however, is not unique and can differ between different specific \UFO models. The explicit choice of Lorentz structures $\boldsymbol{L}_i$ (which in turn fixes $\boldsymbol{C}$ for a given generic $\boldsymbol{V}$) for all renormalisable couplings used within \anyBSM is shown in \cref{tab:ufolorentz}. If one wants to use \anyBSM with a \UFO model which does not obey the conventions of \cref{tab:ufolorentz}, the command line tool \code{anyBSM_import} (\cf next section) can be used to re-write all vertices of a given \UFO model in terms of the Lorentz structures in \cref{tab:ufolorentz}.
\begin{table}[t]
    \centering
\begin{tabular}{c|c|c}
\makecell{\UFO \\ vertex type $\boldsymbol{V}$}   & \makecell{Lorentz\\ structure(s) $\boldsymbol{L_i}$} & \makecell{expected \UFO Lorentz\\ structure by \anyBSM} \\ \hline
$S_1S_2S_3$      & 1                 & \code{'1'}  \\
$S_1S_2S_3S_4$   & 1                 & \code{'1'}  \\
$F_1F_2S_3$      & $P_L$             & \code{'ProjM(2,1)'} \\
                 & $P_R$             & \code{'ProjP(2,1)'} \\
$F_1F_2V_3^{\mu}$& $\gamma^{\mu} P_L$& \code{'Gamma(3,2,-1)*ProjM(-1,1)'} \\
                 & $\gamma^{\mu} P_R$ & \code{'Gamma(3,2,-1)*ProjP(-1,1)'} \\
$S_1V_2^{\mu}V_3^{\nu}$  & $g^{\mu\nu}$      & \code{'Metric(2,3)'}\\
$S_1S_2V_3^{\mu}V_4^\nu$ & $g^{\mu\nu}$      & \code{'Metric(3,4)'}\\
$S_1S_2V_3^\mu$          & $p_{1}^\mu-p_{2}^\nu$ & \code{'P(3,1)-P(3,2)'}\\
\makecell{$V_1^{\mu_1} V_2^{\mu_2}V_3^{\mu_3}$ \vspace{1cm}}     & \makecell{$\big[\,\, g^{\mu_2\mu_3}(p_3^{\mu_1}- p_2^{\mu_1} )$\\ $ + g^{\mu_1\mu_3}( p_1^{\mu_2} - p_3^{\mu_2})$\\ $\,\,\,\,+ g^{\mu_1\mu_2}( p_2^{\mu_3} - p_1^{\mu_3} )\,\,\big]$} &
\makecell{
{\code{'}\color[HTML]{BA2121}\texttt{-Metric(2,3)*P(1,2)+Metric(2,3)*P(1,3)}}\\
$\,\,\,${\color[HTML]{BA2121}\texttt{+Metric(1,3)*P(2,1)-Metric(1,3)*P(2,3)}}\\
$\,\,\,\,\,${\color[HTML]{BA2121}\texttt{-Metric(1,2)*P(3,1)+Metric(1,2)*P(3,2)}}\code{'}
}
\\
$V_1^{\mu_1}V_2^{\mu_2}V_3^{\mu_3}V_4^{\mu_4}$  
                 & $g^{\mu_1\mu_2}g^{\mu_3\mu_4}$ & \code{'Metric(1,2)*Metric(3,4)'}\\
                 & $g^{\mu_1\mu_3}g^{\mu_2\mu_4}$ & \code{'Metric(1,3)*Metric(2,4)'}\\
                 & $g^{\mu_1\mu_4}g^{\mu_2\mu_3}$ & \code{'Metric(1,4)*Metric(2,4)'}\\
$S_1U_2U_3$        & 1                              & \code{'1'}\\
$U_1U_2V_3^{\mu}$  & $p_1^\mu$                      & \code{'P(3,1)'}\\
                   & $p_2^\mu$                      & \code{'P(3,2)'}
\end{tabular}
    \caption{\UFO Lorentz structures used in \anyBSM for vertices involving scalars $S_i$, fermions $F_i$, vectors $V_i^\mu$, and ghosts $U_i$. The four-momentum $p_i^{\mu_j}=\code{'P(j,i)'}$     in the second and third columns is carried by the field with the label $i$ in the first column. In this table, $\gamma^\mu$ denotes Gamma matrices, $g^{\mu\nu}$ the metric tensor, and $P_{L,R}$ the left-/right-handed projectors defined as $P_{L,R}=(1\mp \gamma_5)/2$.
    }
    \label{tab:ufolorentz}
\end{table}

\subsection{Conversion between different conventions}
\label{app:generic:converter}
As an illustrative example of how to handle different conventions, we consider the \UFO implementation of the Inert-Doublet-Model (IDM) that is available at the \FeynRules \href{https://feynrules.irmp.ucl.ac.be/wiki/2HDM}{webpage} as well as the \UFO implementation which is included in \anyBSM and was generated with the help of \SARAH. The coupling between the neutral Goldstone boson and the top/anti-top quarks is defined in the \FeynRules model as $\boldsymbol{V} = \boldsymbol{L}\cdot\boldsymbol{C} = [ \gamma_5\cdot \frac{-m_t}{v}]$. This is obviously not compatible with the \anyBSM conventions, in which \textit{all} fermion couplings are written in terms of left/right-handed projectors, and \anyBSM therefore would expect the vertex in the form $\boldsymbol{V} = \boldsymbol{L}\cdot\boldsymbol{C} = [P_L\cdot \frac{m_t}{v}, P_R\cdot \frac{-m_t}{v}]$. Thus, running \anyBSM from within the directory of the \FeynRules-generated \UFO model will show the error message:
\begin{minted}[bgcolor=bg]{text}
anyBSM ./InertDoublet_UFO
>> Take model from absolute path.
>> ERROR:unexpected lorentz structure for vertex V_112 ([T, t, G0])!
\end{minted}
In order to still make use of that \UFO model, one can run the converter:
\begin{minted}[bgcolor=bg]{text}
anyBSM_import ./InertDoublet_UFO -s -o ./IDMconverted
\end{minted}
which takes the path of the \UFO model that should be converted as first argument. The option \code{-o ./IDMconverted} specifies the path the converted \UFO model should be saved to. The option \code{-s} skips the conversion of vertices that are not supported by \anyBSM such as \eg effective gluon-Higgs couplings or other non-renormalisable couplings. The tool will also write a tabular file into the new model directory containing a mapping of vertices/couplings between the original and the newly created \UFO model, which can be printed to the command line using the option \code{-v}. In the example of the Goldstone-top coupling the output looks like like:
\begin{center}
\begin{minted}[bgcolor=bg,fontsize={\fontsize{7}{7}\selectfont}]{text}
+-------------+--------------------------------------------------+-----------------------------------------------------------+
| Vertex-Type |          Old model [couplings(lorentz)]          |              New model [couplings(lorentz)]               |
+-------------+--------------------------------------------------+-----------------------------------------------------------+                                                                                                                 
|     ...     |                    ...                           |                         ...                               |   
+-------------+--------------------------------------------------+-----------------------------------------------------------+
|   (T,t,G0)  | V_112 ['GC_92(-ProjM(F2, F1) + ProjP(F2, F1))']  | V_109 ['GC_101(ProjP(F2, F1))', 'GC_102(ProjM(F2, F1))']  |                                                                                                                 
|             |                 GC_92 = -(MT/v)                  |                      GC_102 = MT/v                        |                                                                                                                 
|             |                                                  |                      GC_101 = -MT/v                       |                                                                                                                 
+-------------+--------------------------------------------------+-----------------------------------------------------------+ 
\end{minted}
\end{center}
Note that in the original model the vertex (\code{V_112}) was defined by only one coupling/Lorentz structure, while in the new model it consists of a list of length two defining the left-/right-handed couplings. A similar procedure is also applied for couplings involving \eg vector bosons (not shown here).

For a variety of models, we checked explicitly that the prediction for \lamhhh obtained with \anyH is numerically identical when using the \UFO\ model generated with \SARAH (which can directly be used with \anyH) as well as with \FeynRules after the use of the importer.\footnote{In most cases one still needs to adjust the (SM) input parameters accordingly since both tools use different default values/relations for them.} For instance, the $U(1)_{B-L}$ extended SM from the \FeynRules model database can be converted in the following way:
\begin{minted}[bgcolor=bg,fontsize=\small]{bash}
wget https://feynrules.irmp.ucl.ac.be/raw-attachment/wiki/B-L-SM/B-L-N-4_UFO.2.zip
unzip B-L-N-4_UFO.2.zip
anyBSM_import B-L-N-4_UFO -o BmL -vv
anyBSM ./BmL
\end{minted}
We have verified that the result obtained for $\lamhhh$ with this ``converted'' \UFO model agrees with the one obtained with a version of the \UFO~model that was generated with the help of \SARAH and built into \anyBSM (\cf \cref{app:bmlsm}), after adjusting all input parameters.\footnote{A detailed discussion on this comparison is contained in the \href{https://anybsm.gitlab.io/anybsm/models/BmLSM.html\#alternative-ufo-model}{online documentation of the $U(1)_{B-L}$ model.}}

\subsection{Available topologies and generic diagrams}
\label{app:diagrams}

Diagrammatically, the calculation of \lamhhh (see \cref{eq:lambdahhh}) can be expressed in the form
\begin{align}
\lamhhh = &
\underbrace{\vcenter{\hbox{\includegraphics{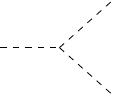}}}
}_{\text{tree-level: } \lambda_{hhh}^{(0)}}
+ \,\,\,
\underbrace{
    \vcenter{\hbox{\includegraphics{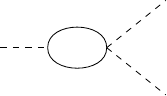}}}
+ \,\,\,
\vcenter{\hbox{\includegraphics{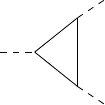}}}%
}_{\text{one-particle irreducible: } \delta^{(1)}_{\text{genuine}}\lambda_{hhh}} \nn
\\
&
+ \,\,\,
\underbrace{
\vcenter{\hbox{\includegraphics{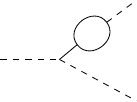}}}
+ \,\,\,
\vcenter{\hbox{\includegraphics{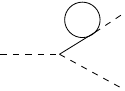}}}
\phantom{\hspace{2.6cm}}
}_{\text{external leg corretions: } \delta^{(1)}_{\text{WFR}}\lambda_{hhh}}
\hspace{-2.6cm}
+ \,\,\,
\overbrace{
\vcenter{\hbox{\includegraphics{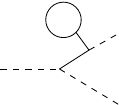}}}
+ \,\,\,
\vcenter{\hbox{\includegraphics{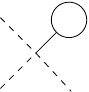}}}
}^{\text{tadpoles: } \delta^{(1)}_\text{tad. WFR}\lambda_{hhh}+\delta^{(1)}_{\text{tadpoles}}\lambda_{hhh}} \nn
\\
&
+ \,\,\,
\underbrace{
\vcenter{\hbox{\includegraphics{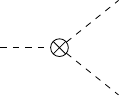}}}
}_{\text{renormalisation: } \delta^{(1)}_{\text{CT}}\lambda_{hhh}}
\label{eq:lamhhhdiag}
\end{align}
where the solid internal lines are meant to be placeholders that are populated with all possible field-insertions (spin 0, $\nicefrac{1}{2}$ and 1) from a given \UFO model upon running \anyBSM. The model-specific results are obtained in \anyBSM by inserting the Feynman rules into the result of the corresponding \textit{generic} diagram. It should be noted that the tadpole contribution appears in both $\delta^{(1)}_{\text{tadpoles}}\lambda_{hhh}$ and $\delta^{(1)}_{\text{tad. WFR}}\lambda_{hhh}$, and that the tadpoles as well as the external-leg corrections can optionally be turned off separately and instead be included in $\delta^{(1)}_{\text{CT}}\lambda_{hhh}$ (see \cref{app:sm:tadren,app:ssm:ren} for detailed discussions).
Finite contributions from non-minimal counterterms, $\delta^{(1)}_{\text{CT}}\lambda_{hhh}$, have been discussed in \cref{sec:genericcalc:ren} and will be discussed in more detail in \cref{app:sm:tadren,app:sm:vevren}. One crucial input for the renormalisation are scalar one-point functions,
\begin{equation}
t_{S} =
\vcenter{\hbox{\includegraphics{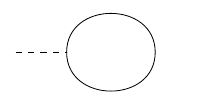}}}
,
\end{equation}
as well as all bosonic two-point functions,
\begin{equation}
\Sigma_{XY}(p^2) = 
\vcenter{\hbox{\includegraphics{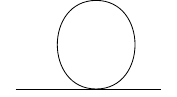}}}
+
\vcenter{\hbox{\includegraphics{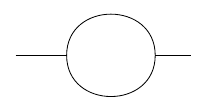}}}
+
\vcenter{\hbox{\includegraphics{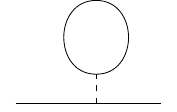}}}
    \label{fig:onetwopoint}%
    ,
\end{equation}
where $S$ can be any scalar, and $X,\,Y$ any scalar or vector boson. The internal solid lines are analogous to those in \cref{eq:lamhhhdiag}. 

Thus, \anyBSM is at the moment able to calculate the scalar one-, bosonic two-point, and scalar three-point functions of any renormalisable QFT. 


\section{Implementing new models}
\label{app:newmodels}
In this Appendix, we briefly describe the different possibilities and general strategies to create \anyBSM-compatible \UFO models. For detailed instructions and examples we refer to the \docs and the \examplesrepo, respectively.

Models in \anyBSM are fully described using the \UFO standard~\cite{Degrande:2011ua,Darme:2023jdn}. The code requires the following \UFO files to be present in the model directory:
\begin{itemize}
    \item \texttt{particles.py} --- specifying the fields present in the model,
    \item \texttt{parameters.py} --- specifying all input parameters (called ``external'' parameters) and ``internal'' parameters which are computed in terms of the input parameters,
    \item \texttt{lorentz.py} --- specifying all Lorentz structures appearing in the model,
    \item \texttt{couplings.py} --- specifying all couplings in terms of the model parameters,
    \item \texttt{vertices.py} --- specifying all vertices along with their corresponding Lorentz structures and couplings.
\end{itemize}
Additional \py files which are normally distributed for \UFO models (\eg\ \texttt{function\_library.py}) are directly incorporated into \anyBSM and are therefore not needed/ignored, \cf\ \cref{app:additionalufo}.

\UFO models for \anyBSM can be created using \eg\ \texttt{SARAH}~\cite{Staub:2008uz,Staub:2009bi,Staub:2010jh,Staub:2012pb,Staub:2013tta} or \texttt{FeynRules}~\cite{Christensen:2008py,Degrande:2011ua,Alloul:2013bka,Alloul:2013fw}. However, the \anyBSM package uses a specific convention for the Lorentz structures that are used in the derivation of generic results, \cf\ \cref{app:generic}. These conventions are identical to those of \SARAH. For \UFO models from other sources it is, however, not guaranteed that all vertices are expressed in terms of the Lorentz structures used in \anyBSM. 
In such cases the tool \texttt{anyBSM\_import}, which is discussed in \cref{app:generic:converter}, can be used to test a given \UFO model for compatibility with \anyBSM and to perform a conversion, if required.

Nevertheless, there are currently some limitations on the models that can be used with \anyBSM. The model
\begin{itemize}
    \item must not contain any non-renormalisable couplings (\ie, they are ignored),
    \item needs to obey the conventions for Lorentz structures listed in \cref{app:generic} (otherwise the model-converter should be used), and
    \item should define all SM-like particles, either via their PDG code as defined in~\ccite{Boos:2001cv,Alwall:2006yp,ParticleDataGroup:2022pth} or by defining at least one renormalisation scheme in the \code{schemes.yml} file (see~\cref{sec:documentation:ren}). 
\end{itemize}
Additional current limitations are:
\begin{itemize}    
    \item no external coloured states for the calculation of $\lambda_{h_ih_ih_i}$ are supported (\ie, the scalar state under consideration must not be charged under $SU(3)_C$).
    \item No colour representations beyond $\bm{1}, \bm{3}, \bm{\bar{3}}, \bm{8}$ are supported for particles on internal lines.
    \item The automatic \OS renormalisation of the electroweak VEV   (\code{VEV_counterterm: OS} in the \texttt{schemes.yml}) is only implemented for models with an electroweak $\rho$ parameter that does not differ from one at lowest order, see the discussion in \cref{app:sm:vevren}.
    \item For the caching (\cf\ \cref{app:cache}) to work, the couplings defined in the file \texttt{couplings.py} need to be in ascending order (\code{GC_1}, \code{GC_2}, \dots, \code{GC_23}, \code{GC_24}). Otherwise only the caching of insertions should be used (set \code{anyBSM.caching = 1}).
\end{itemize}
New models can be added simply by including the corresponding \UFO model files (run through the converter described above if necessary) in the models repository of \anyBSM. The location of this directory is specified in the \anyBSM configuration file which is written at the first start of \anyBSM after the installation, \cf \cref{sec:tutorial:installation}. Both locations, the config-path and the models-path, can be issued as follows
\begin{minted}[bgcolor=bg]{python}
import anyBSM
from os import path
print('anyBSM config file:      ', path.join(
        anyBSM.config.appdirs.user_config_dir('anyBSM'),
        'anyBSM_config.yaml'))
print('anyBSM model directory:  ', anyBSM.anyBSM.models_dir)
\end{minted}
To get an up-to-date list of known models contained in the model directory one can issue  \code{anyBSM -l} within the command-line. For a list of built-in models, one can also consult the \docs or visit the \modelsrepo which contains more information about the specific models and their implementation details. We also note that all underlying \SARAH or \FeynRules models, which have been used to generate the built-in \UFO models, are publicly available in the \examplesrepo.

Alternatively, new models that are unknown to \anyBSM (\ie stored outside of the models directory) can be used by providing the absolute/relative path to the \UFO directory rather than just providing the model name (which is determined by the name of the model directory) at the initialisation step or as the first argument of the command-line tool.

For more information on how to implement a new model, we refer to the \docs.

\section{Models currently provided in \anyH}
\label{app:models}

In the following we review the models that are shipped alongside \anyBSM and \anyH in the \UFO format. However, it should be stressed that the program is \textit{not} restricted to this set of BSM models (and the SM) but works with all \UFO models that fulfill the requirements described in \cref{app:newmodels}. 

This appendix is organised as follows: In the first part addressing the SM, \cref{app:sm}, we in particular discuss the renormalisation of the trilinear Higgs coupling. In this context we describe the renormalisation of the tadpole and the VEV as well as the OS renormalisation of the mass of the Higgs boson. An OS mass renormalisation is also used as the default scheme in most of the BSM models discussed below (explicit examples on how to implement other schemes will also be given). In \cref{app:sm,app:ssm,app:thdm,app:idm,app:nthdm,app:tsm0,app:tsm1,app:gm,app:bmlsm,app:mssm} we briefly describe the models in terms of their Lagrangian densities and chosen parametrisations as well as necessary conditions to achieve the alignment limit (\ie, the limit in which the the tree-level couplings of the SM-like Higgs boson are identical to the respective SM couplings). More detailed information about the individual models can be found in the \docs and in the references therein. We emphasise once again that throughout this appendix the quantity \lamhhh refers to the trilinear Higgs coupling of the detected Higgs boson with a mass of about 125~GeV. 

\subsection{The Standard Model (SM)}
\label{app:sm}
The SM Lagrangian used to generate the \UFO model with \SARAH is given by:
\begin{align}
    \mathcal{L}_{\text{SM}} &= \boldsymbol{Y}_d\, \Phi^\dagger\cdot Q_L d_R + \boldsymbol{Y}_e\, \Phi^\dagger\cdot L_L e_R + \boldsymbol{Y}_u\, \Phi\cdot Q_L u_R + h.c - V^{(0)}_{\text{SM}}\,, \\
              V^{(0)}_{\text{SM}} &= \mu^2 \Phi^\dagger\Phi + \frac{\lambda}{2} |\Phi^\dagger\Phi|^2\,.
\end{align}
The Yukawa matrices $\boldsymbol{Y}_i$ are assumed to be diagonal for simplicity (i.e.\ the CKM matrix is unity). Their entries are traded for the measured lepton and quark masses used as inputs in the \UFO model.

We continue with the discussion of the tree-level scalar potential and its renormalisation. Assuming $\mu^2<0$, the Higgs doublet of the SM obtains a VEV and is parametrised at the minimum of the potential in terms of the physical Higgs field $h$, the neutral (charged) Goldstone(s) $G$ ($G^\pm$) and the VEV $v$,
\begin{align}
    \Phi=\frac{1}{\sqrt{2}}\begin{pmatrix}\sqrt{2}G^+\\ v+h+i G
    \end{pmatrix}\,.
\end{align}
Dropping all terms involving $G^\pm$ or $G$, we can expand the tree-level potential as
\begin{align}
    V^{(0)}_{\text{SM}}\supset&\ \frac{1}{2}\mu^2(v+h)^2+\frac{1}{8}\lambda (v+h)^4\nn\\
    =&\ \frac{1}{2}\mu^2v^2+\frac18 \lambda v^4+\left(\mu^2+\frac{1}{2}\lambda v^2\right)vh+\frac{1}{2}\left(\mu^2+\frac32 \lambda v^2\right)h^2+\frac{1}{2}\lambda v h^3+\frac18 \lambda h^4\,.
\end{align}
We choose to replace $\mu^2$ and $\lambda$ by defining the tree-level minimum $t_h$ as well as the squared tree-level mass $m_h^2$ as follows 
\begin{equation}
      t_h \equiv \left.\frac{\partial V^{(0)}_{\text{SM}}}{\partial h}\right|_{h=0} = \left(\mu^2+\frac{1}{2}\lambda v^2\right)v\,,\qquad
     m_h^2 \equiv \left.\frac{\partial^2 V^{(0)}_{\text{SM}}}{\partial h^2}\right|_{h=0} = \mu^2+\frac{3}{2}\lambda v^2\,.
\end{equation}
In terms of $m_h^2$ and $t_h$, the potential becomes
\begin{align}
    V^{(0)}_{\text{SM}}\supset t_h h+\frac12 m_h^2 h^2+\frac{m_h^2-t_h/v}{2v}h^3+\frac{m_h^2-t_h/v}{8v^2}h^4\,.%
\end{align}
We introduce counterterms at one-loop order for the different parameters and the Higgs field entering the tree-level scalar potential as
\begin{align}
    t_h&\to t_h+\delta^{(1)}_\text{CT}t_h\,,\nn\\
    m_h^2&\to m_h^2+\delta^{(1)}_\text{CT}m_h^2\,,\nn\\
    v&\to v+\delta^{(1)}_\text{CT}v\,,\nn\\
    h&\to Z_h^{1/2}h= h\left(1+\frac12\delta^{(1)}_\text{CT}Z_h\right) \, .
\end{align}
Turning now to the trilinear Higgs coupling, we find at the tree level that
\begin{align}
    \label{eq:sm:lamhhh0}
    \lambda_{hhh}^{(0)}
    =\left.\frac{\partial^3 V^{(0)}_\text{SM}}{\partial h^3}\right|_{h=0}
    =\frac{3(m_h^2-t_h/v)}{v}\,.
\end{align}
Correspondingly, the vertex counterterm (including field renormalisation) is given by
\begin{align}
   \label{eq:sm:lamhhhct}
    \delta^{(1)}_\text{CT, vertex}\lambda_{hhh}=
    \frac{3}{v}\delta^{(1)}_\text{CT}m_h^2
    -\frac{3}{v^2}\delta^{(1)}_\text{CT}t_h
    -3\left(\frac{m_h^2}{v^2}-\frac{2t_h}{v^3}\right)\delta^{(1)}_\text{CT}v
    +\frac{3}{2}\frac{3(m_h^2-t_h/v)}{v}\delta^{(1)}_\text{CT}Z_h\,.
\end{align}
At the one-loop order we use the parametrisation
\begin{align}
    \label{eq:sm:lamhhh}
    \lambda_{hhh}=\frac{3(m_h^2-t_h/v)}{v}+\delta^{(1)}_\text{diag.}\lambda_{hhh}+\delta^{(1)}_\text{CT}\lambda_{hhh}\,,
\end{align}
where $\delta^{(1)}_\text{diag.}\lambda_{hhh}$ contains the one-loop diagrammatic corrections to the trilinear coupling (see the first two lines of \cref{eq:lamhhhdiag}), and 
\begin{align} 
\label{eq:sm:lamhhhCT}
\delta^{(1)}_\text{CT}\lambda_{hhh}\equiv \delta^{(1)}_\text{CT, vertex}\lambda_{hhh}-\frac{3}{v^2}\delta^{(1)}_\text{CT}t_h\,.
\end{align}
The second term in \cref{eq:sm:lamhhhCT} arises from the counterterms that are associated with the diagrammatic tadpole contributions as described in \cref{eq:sm:lamhhhtadpolesnonpi} below. It turns out to have the same form as the tadpole counterterm contribution in \cref{eq:sm:lamhhhct} arising from the parametric dependence of $\lambda_{hhh}^{(0)}$ on $t_h$. 

The diagrammatic one-loop corrections in \cref{eq:sm:lamhhh} are decomposed as
\begin{align}
    \delta^{(1)}_\text{diag.}\lambda_{hhh}=\delta^{(1)}_\text{genuine}\lambda_{hhh}+\delta^{(1)}_\text{WFR}\lambda_{hhh}+\delta^{(1)}_\text{tadpoles}\lambda_{hhh}\,,
\end{align}
where $\delta^{(1)}_\text{tadpoles}\lambda_{hhh}$ refers to diagrams where a one-loop tadpole is attached to a Higgs-quartic coupling, \ie
\begin{equation}
    \label{eq:sm:lamhhhtadpolesnonpi}
     \delta^{(1)}_{\text{tadpoles}} \lamhhh = \lambda_{hhhh}^{(0)}\frac{(-1)}{m_h^2} \delta^{(1)}t_h =  -\frac{3}{v^2}\delta^{(1)}t_h\,.
\end{equation}
Rather than expressing $\lambda_{hhh}$ in terms of the tree-level Higgs-boson mass and the VEV, we want to express it in terms of physical inputs --- namely the pole masses of the Higgs, $W$, and $Z$ bosons ($M_h$, $M_W$ and $M_Z$) as well as the fine-structure constant $\alpha_\text{em}(0)$ (and $\Delta\alpha$, see \cref{eq:dalpha} below). In practice, this can be done, either by performing conversions of mass parameters or by applying the OS scheme to fix the counterterms, with pole-mass relations of the form (using here the sign conventions of \texttt{anyBSM}) 
\begin{align}
    M_h^2&=m_h^2+\mathrm{Re}\Sigma^{(1),\text{no tad.}}_{hh}(p^2=m_h^2)+\delta^{(1)}_\text{CT,no tad.} m_h^2-\frac{3}{v}(\delta^{(1)}t_h+\delta^{(1)}_\text{CT}t_h) \nn  \\
         &\equiv m_h^2+\mathrm{Re}\Sigma^{(1)}_{hh}(p^2=m_h^2)+\delta^{(1)}_\text{CT}m_h^2 \label{eq:sm:polemasses}\\
    M_V^2&=m_V^2 - \mathrm{Re}\Sigma_{VV}^{T,(1)}(p^2=m_V^2) +\delta^{(1)}_\text{CT} m_V^2\,, \quad (V=W,Z)
\end{align} 
where we distinguish between a self-energy $\Sigma^\text{no tad.}$, which does not contain one-loop tadpole insertions, and the full one-particle-irreducible (1PI) self-energy $\Sigma$. 
The transverse part of the gauge-boson self-energies is defined from the general decomposition
\begin{align}
\label{EQ:gaugeboson_self_TL}
    \Sigma_{VV^\prime}^{\mu\nu}(p)=\left(g^{\mu\nu}-\frac{p^\mu p^\nu}{p^2}\right)\Sigma^T_{VV^\prime}(p^2)+\frac{p^\mu p^\nu}{p^2}\Sigma^L_{VV^\prime}(p^2)\,.
\end{align}
Demanding that the tree-level input masses equal the pole masses, $m_i^2=M_i^2$, fixes the mass counterterms entering $\delta^{(1)}_\text{CT}\lamhhh$ and $\delta^{(1)}_\text{CT}v$ in the OS scheme (see \cref{app:sm:vevren} below). However, it is important to stress that, until this point, we have not yet specified the renormalisation of the tree-level minimum $t_h$ and of the electroweak VEV. In the following two sections we describe different possible treatments of the tadpoles and the VEV.

\subsubsection{Equivalence of different tadpole renormalisation schemes}
\label{app:sm:tadren}

We now investigate different approaches for  treating the tadpoles in the SM. The general tadpole contribution to $\lamhhh$ including all possible sources of tadpoles in \cref{eq:sm:lamhhh} reads:
\begin{align}
\label{eq:sm:lamhhalltadpoles}
    \lamhhh^\text{tadpoles} =& 
    - \frac{3t_h}{v^2}
    - \frac{6}{v^2}\delta^{(1)}_\text{CT}t_h 
    + \delta^{(1)}_\text{tadpoles}\lamhhh
    +\frac{3}{v}\delta^{(1)}_\text{CT, tadpoles}m_h^2  \nn \\
   &- \frac{3 m_h^2}{v^2}\delta^{(1)}_\text{CT, tadpoles}v
    - \frac{3}{2}\frac{3 m_h^2}{v}\delta^{(1)}_\text{CT, tadpoles} Z_h\,.
\end{align} 
The first term in \cref{eq:sm:lamhhalltadpoles} originates from the tree-level tadpole contribution, \cf \cref{eq:sm:lamhhh0}, the second term arises from the vertex counterterm and the counterterms of the tadpole diagrams as described above, the third term contains the one-loop diagrammatic contributions to \lamhhh of tadpoles (see the last diagram in the second line of \cref{eq:lamhhhdiag}), \cf \cref{eq:sm:lamhhhtadpolesnonpi}, and the fourth term arises from possible tadpole contributions in the mass counterterm. The next-to-last term denotes tadpole contributions to the VEV counterterm, while the last term vanishes since tadpole contributions to the field renormalisation (which is purely diagonal in the SM) drop out in the derivative w.r.t.\ the squared momentum. With \cref{eq:sm:lamhhalltadpoles} at hand, we can now discuss different choices of renormalisation schemes which amount to different relations/identities of/for the individual parts in \cref{eq:sm:lamhhalltadpoles}. In the following, we restrict the discussion to the UV-finite parts of the various counterterms since UV-divergences are universal, and the UV-finiteness is not affected by the following discussions.\footnote{We explicitly checked for UV finiteness separately.}

\paragraph{Tadpole-free \MS scheme (T\MS)}

A popular convention in \eg\ \ccite{Martin:2003it,Martin:2016xsp,Martin:2019lqd} (this is for instance the default choice for loop calculations in \SARAH/\SPheno) is to employ \MS renormalisation for the tadpoles, $\delta^{(1)}_\text{CT} t_h=0$, and to work at the minimum of the loop-corrected potential, which is realised by demanding that the total tadpole at one-loop order must vanish, \ie
\begin{align}
    T_h\equiv t_h+\delta^{(1)}t_h+\delta^{(1)}_\text{CT}t_h= t_h+\delta^{(1)}t_h = 0\,,
\end{align}
from which we can write
\begin{align}
\label{eq:sm:tmsscheme}
    t_h=-\delta^{(1)}t_h=-\frac{\partial V^{(1)}}{\partial h}\bigg|_\text{min.}\,,
\end{align}
where $V^{(1)}$ denotes the one-loop contributions to the effective potential. We note that this means that $t_h$ is formally of one-loop order. In addition, the VEV $v$ is taken to be the true minimum of the loop-corrected potential. Using this in \cref{eq:sm:lamhhalltadpoles}, we find at one-loop order the following total tadpole contribution to $\lamhhh$,
\begin{align}
\label{eq:sm:SPMtadpoles}
    \left. \lamhhh^\text{tadpoles}\right|_\text{T\MS}(m_h^2,v^2) 
    = -\frac{3t_h}{v^2}+\delta^{(1)}_\text{tadpoles}\lambda_{hhh}
    = 0\,,
\end{align}
where we made use of \cref{eq:sm:lamhhhtadpolesnonpi} and \cref{eq:sm:tmsscheme} in the last step. The absence of the explicit dependence on any of the tadpoles is why this scheme is frequently called the \textit{tadpole-free} \MS renormalisation scheme (T\MS).

However, to compare the result in this prescription with the results obtained in a different renormalisation scheme, we need to express $\lambda_{hhh}(m_h^2)$ --- especially the tree-level piece --- in terms of physical observables (\ie, $\lambda_{hhh}(M_h^2)$), since $m_h^2$ is not at the pole of the propagator in this scheme. The relation between the tree-level Higgs mass $m_h^2$ and the pole mass $M_h^2$ in this scheme reads 
\begin{align}
    \label{eq:sm:SPMpolemass}
    m_h^2=M_h^2+\frac{\delta^{(1)}t_h}{v}-\Sigma^\text{(1), no tad.}_{hh}(m_h^2)\,.
\end{align}
Once inserted into the tree-level expression of $\lambda_{hhh}$ this gives an additional shift
\begin{align}
    &\frac{3m_h^2}{v}=\frac{3M_h^2}{v}+\frac{3}{v^2}\delta^{(1)}t_h-\frac{3}{v}\Sigma_{hh}^{(1),\text{ no tad.}}(m_h^2)\nn\\
    &\Rightarrow \left. \lamhhh^\text{tadpoles}\right|_\text{T\MS} (M_h^2,v^2) = \underbrace{\left.\lamhhh^\text{tadpoles}\right|_\text{T\MS} (m_h^2,v^2)}_{=0}+\frac{3}{v^2}\delta^{(1)}t_h\,.\label{eq:sm:SPMlamhhh}
\end{align}
It should be noted that the counterterm contribution that is associated with the Higgs VEV, which is assumed to be the loop-corrected VEV in this scheme, does not explicitly introduce any tadpole contribution to \lamhhh~because no self-energy diagrams with tadpole insertions are included in this scheme. 

\paragraph{OS tadpole renormalisation (tOS)}

An alternative treatment of the tadpoles was \eg\ proposed in \ccite{Aoki:1982ed,Bohm:1986rj,Bohm:2001yx}, where the one-loop tadpole corrections to the minimum of the tree-level potential (\ie, $t_h=0$) are required to be canceled by the tadpole counterterm such that the one-loop corrected minimum corresponds to the tree-level minimum (this is also often referred to as ``OS tadpole condition'' or ``parameter renormalised tadpole scheme'' (PRTS)),
\begin{align}
    \delta^{(1)}_\text{CT}t_h=-\delta^{(1)}t_h\,. 
\end{align}
This has \eg\ the advantage that no tadpole contributions explicitly contribute to the conversion between the \MS and OS pole masses and that in general all diagrammatic tadpole contributions to any process are cancelled by the corresponding tadpole-counterterm diagrams (but tadpole counterterms furthermore appear in some other counterterms).

Using these ingredients in \cref{eq:sm:lamhhalltadpoles} (together with \cref{eq:sm:lamhhhtadpolesnonpi}) we find
\begin{equation}
    \left. \lamhhh^\text{tadpoles}\right|_{\text{t}\text{OS}}(M_h^2,v^2) 
     = \delta^{(1)}_\text{tadpoles}\lamhhh -\frac{6}{v^2}\delta^{(1)}_\text{CT}t_h
     = +\frac{3}{v^2}\delta^{(1)}t_h \,,
\end{equation}
which is identical to the result obtained in the T\MS scheme, \cf\ \cref{eq:sm:SPMlamhhh}.

Note that we did not specify the scheme used for $\delta^\text{CT}v$. However, a cancellation of the one-loop genuine and counterterm tadpole contributions will occur separately therein due to the on-shell renormalisation of the tadpoles.

\paragraph{\MS tadpole renormalisation at the tree-level minimum (FJ)}

Finally, we consider another scheme in which we again set $t_h=0$, but now renormalise the tadpoles in the \MS scheme. Concretely, we require that $\delta^\text{CT}t_h$ cancels (only) the divergent part of $\delta^{(1)}t_h$ but is zero otherwise. This scheme is equivalent to the one known as the Fleischer-Jegerlehner (FJ) scheme~\cite{Fleischer:1980ub}. Returning to the master expression, \cref{eq:sm:lamhhalltadpoles}, of $\lambda_{hhh}$ we have in this scheme
\begin{align}
    \left. \lamhhh^\text{tadpoles}\right|_{\text{FJ}}(m_h^2,v^2) 
    = 
    \delta^{(1)}_\text{tadpoles}\lamhhh 
    + \frac{3 }{v^2}\delta^{(1)}_\text{CT, tadpoles}m_h^2
    - \frac{3 m_h^2}{v^2}\delta^{(1)}_\text{CT, tadpoles}v\,.
\end{align}
In this scheme we need to properly extract $m_h^2$ as well as the VEV from their relation to physical observables including tadpole contributions (since they are not cancelled by their OS counterterms), which are then finally inserted into the tree-level expression for \lamhhh. For simplicity, let us assume now that we extract the VEV from its relation to the OS pole mass of the $Z$ boson $M_Z$ using \MS values of the EW gauge couplings,\footnote{This is just for demonstration purposes. The default treatment of the electroweak VEV in \anyH is discussed in \cref{app:sm:vevren}.}
\begin{align}
    v(M_Z)\equiv \frac{2}{\sqrt{g_2^2+g_Y^2}}M_Z \,.
\end{align}
With the FJ treatment of tadpoles, this results in a tadpole contribution to the VEV counterterm that reads 
\begin{align}
\label{eq:sm:vevshiftMZ}
    \delta^{(1)}_\text{CT, tadpoles}v = \frac{v(M_Z)}{2} \left.\frac{\delta^{(1)}M_Z^2}{M_Z^2}\right|_\text{tadpoles}  \,.
\end{align}
with 
\begin{align}
    \delta^{(1)}M_Z^2=\text{Re}\Sigma^{T, (1)}_{ZZ}(M_Z)\supset \frac{2}{v}\frac{M_Z^2}{m_h^2}\delta^{(1)}t_h\,.
\end{align}
This yields for the tadpole contribution from the VEV counterterm to the trilinear Higgs coupling 
\begin{align}
    -\frac{3 m_h^2}{v^2} \delta^{(1)}_\text{CT, tadpoles}v
    = -\frac{3}{v^2}\delta^{(1)}t_h\,.
\end{align}
In addition, we again need to make sure to express $m_h^2$ in terms of the pole mass $M_h^2$, which we achieve by renormalising it on-shell. With \MS-renormalised tadpoles, the finite shift relating the lowest-order Higgs boson mass to the pole mass contains a tadpole contribution of the form $+3/v\delta^{(1)}t_h$. Thus we find a shift to $\lambda_{hhh}$ of the form $+9/v^2 \delta^{(1)}t_h$.

Summing all contributions involving the tadpoles, we find a total of 
\begin{align}
    \left. \lamhhh^\text{tadpoles}\right|_{\text{FJ}}(M_h^2, v(M_Z)) = \frac{(-3-3+9)\delta^{(1)} t_h}{v^2}= +\frac{3}{v^2}\delta^{(1)} t_h\,,
\end{align}
which is again in agreement with the two previous schemes.

In conclusion, the three tadpole schemes discussed here yield the same finite contribution (up to higher-orders) to the renormalised $\lambda_{hhh}$ of the form $+3/v^2\delta^{(1)}t_h\big|_\text{UV-finite}$, once we take into account that for a proper comparison we need to express all parameters in terms of physical observables. This result is expected since the relation between the input parameters that are expressed in terms of physical observables and the process of Higgs pair production corresponds to a relation between physical observables for which the actual treatment of the tadpoles must not matter.

\paragraph{Comparison of tadpole schemes and implementation in \anyH}

Given the equivalence of the tadpole schemes demonstrated above, we are free to choose the most convenient treatment. Since the \UFO format in general does not contain any information about the (tree-level) tadpoles, we choose the FJ treatment per default, since here we have $t_h=0$. While in the equivalence proof above we re-wrote all tadpole-inserted contributions in terms of tree-level couplings and propagators multiplied by the one-loop one-point function, in the actual code implementation we cannot automatically perform this re-organisation of the calculation but instead we generate and calculate all tadpole-inserted diagrams separately.

However, the aim of \anyH is to be very flexible regarding the choice of renormalisation schemes. Therefore, the FJ scheme is only the default choice but the program easily allows the user to restrict to genuine loop diagrams and to add the tadpole contributions using a custom user-defined counterterm.

For example, in the SM the default FJ and the alternative tOS scheme can be defined in the following way in the \texttt{schemes.yml} file:
\begin{minted}[bgcolor=bg]{yaml}
renormalization_schemes:
  OS: # corresponds to FJ scheme
    mass_counterterms:
      h: OS
    VEV_counterterm: OS

  OStadpoles: # corresponds to tOS scheme
    mass_counterterms:
      h: OS
    VEV_counterterm: OS
    tadpoles: False # do not compute tadpole-inserted diagrams
    custom_CT_hhh: |
      dTad = Tadpole('h')
      self.custom_CT_hhh = f'-3*({dTad})/(vvSM**2)'
\end{minted}
where the ``\code{OS}'' scheme corresponds to the standard FJ scheme while the ``\code{OStadpoles}'' scheme corresponds to the tOS scheme and makes use of a custom counterterm ``\code{custom_CT_hhh}'' which is precisely the tadpole contribution that was derived above.\footnote{Note that the function \code{Tadpole('h')} computes the 1-point function, $-\delta^{(1)}t_h$, in the notations of the discussion above. }
In addition ``\code{tadpoles: False}'' is used in the ``\code{OStadpoles}'' scheme to switch off all tadpole contributions in the calculation of the other counterterms, of self-energies, and of the loop corrections to \lamhhh. The switch ``\code{VEV_counterterm: OS}'' refers to the calculation of the VEV counterterm contribution as described in the next section. It should be stressed that both, the ``\code{VEV_counterterm}'' and the ``\code{mass_counterterms}'' options, are compatible with the ``\code{tadpoles: False/True}'' option. In the example above this means that all 1PI  self-energy diagrams, including tree-level propagators with one-loop tadpole insertions, are automatically taken into account in the scheme ``\code{OS}'' but not in the scheme ``\code{OStadpoles}'' where, because of the setting ``\code{tadpoles: False}'', $\Sigma^{\text{no tad.}}$ is used for all self-energies. Therefore, all tadpole contributions have to be added manually when using ``\code{tadpoles: False}'' via the ``\code{custom_CT_hhh}'' directive.

The two different schemes can easily be compared numerically as follows:
\begin{minted}[bgcolor=bg]{python}
from anyBSM import anyH3
SM = anyH3('SM', scheme_name='OS')
lam_tMS = SM.lambdahhh()['total']
SM.load_renormalization_scheme('OStadpoles')
lam_tOS = SM.lambdahhh()['total']
print(lam_tMS-lam_tOS)
>>(-2.842170943040401e-14+0j)
\end{minted}
which shows perfect agreement within the numerical accuracy. We note that all other parameters in this example are renormalised in the OS scheme. If this is not the case, then a conversion has to be performed, resulting in numerical differences --- see the discussion in \cref{sec:applications:uncert}.

For numerical studies, a possible drawback of the FJ scheme is that is can suffer from numerical instabilities due to large numerical cancellations. In \cref{app:thdm}, we show with a THDM example that the code is not affected by this issue in the phenomenologically relevant parameter region. However, as discussed in \cref{sec:TSM0}, if conversions of parameters are performed, we recommend to use an OS scheme for the tadpoles, whenever this is technically feasible, in view of the better behaviour of the perturbative series that this choice exhibits. 

\subsubsection{Vacuum expectation value renormalisation from OS quantities}
\label{app:sm:vevren}
In the previous discussion we used \MS gauge couplings in the context of the extraction of the VEV within the FJ scheme. However, for practical reasons a direct determination of $v$ in terms of measured quantities is desirable. Solving the tree-level relation between the electromagnetic charge $e$, the vector boson masses $M_{Z/W}$,  the
weak mixing angle $\sws\equiv \sw$ and the VEV $v$,
\begin{equation}
    \label{eq:sm:evev}
    e = \sqrt{4 \pi \alpha_{\text{em}}} = \frac{g_Y g_2}{\sqrt{g_Y^2+g_2^2}} = \frac{2 M_W}{v} \sws = \frac{2 M_W}{v} \sqrt{1-\frac{M_W^2}{M_Z^2}}\,,
\end{equation}
for the electroweak VEV $v$, we can define the \vev in terms of the measured OS values of $\alpha_\text{em}(0)$ (with $e = \sqrt{4\pi\alpha_\text{em}(0)}$), $M_W$, and $M_Z$ as
\begin{equation}
    v^{\OS} \equiv  \frac{2 M_W}{e} \sqrt{1-\frac{M_W^2}{M_Z^2}}\, .
\end{equation}
The counterterms for the parameters entering this relation are defined as
\begin{subequations}
\begin{align}
    \frac{\delta^{(1)}M_{V}^2}{M_V^2} &=\frac{\mathrm{Re}\Sigma_{VV}^{T,(1)}}{M_V^2}(p^2=M_{V}^2)\,,\  (V=W,Z) \, , \\
    \frac{\delta^{(1)} e}{e} &=\left( \frac{1}{2}\Pi^{(1)}_\gamma(p^2=0) + \sing(\sws) \frac{\sws}{M_Z^2 \cws}\Sigma^{T,(1)}_{\gamma Z}(p^2=0)\right) ,\,\label{eq:sm:deltae}
\end{align}
\end{subequations}
where the term $\sing(\sws)$ ensures that the sign convention in the covariant derivative is taken into account correctly --- indeed a change in the sign with which the covariant derivatives are defined results in a sign flip in the sign of the weak mixing angle, and both sign conventions exist in the literature and can be employed when creating \UFO model files. The transverse part of massive vector self-energies is defined in \cref{EQ:gaugeboson_self_TL}, while for the photon self-energy we have
\begin{align}
  \Sigma_{\gamma}^{\mu\nu}(p^2)=(p^2g^{\mu\nu}-p^\mu p^\nu)\Pi_{\gamma}(p^2)=\left(g^{\mu\nu}-\frac{p^\mu p^\nu}{p^2}\right)\Sigma^T_\gamma(p^2)+\frac{p^\mu p^\nu}{p^2}\Sigma^L_\gamma(p^2)\,.
\end{align}
The resulting \vev counterterm in the OS scheme reads
\begin{equation}
    \label{eq:sm:deltav}
    \frac{\delta^{(1)} v^{\OS}}{v^{\OS}} = 
    \frac{\delta^{(1)}M_{W}^2}{2M_W^2} + \frac{\cos^2\theta_w}{2\sin^2\theta_w} \left( \frac{\delta^{(1)}M_{Z}^2}{M_Z^2} -   \frac{\delta^{(1)}M_{W}^2}{M_W^2}\right) - \frac{\delta^{(1)} e}{e}\,,
\end{equation}
from which we can also relate $v^{\OS}$ and $v^{\MS}$. In \cref{eq:sm:deltae}, the photon vacuum polarisation $\Pi_\gamma(0)$ is split into three contributions:
\begin{align}
    \left.\frac{\partial \Sigma_\gamma^{T}(p^2)}{\partial p^2}\right|_{p^2=0}\equiv \Pi_\gamma(0) = & 
    \left.\Pi_\gamma(0)\right|_{\text{heavy}} + \left.\Pi_\gamma(0)\right|_{\text{light}}
    \nn \\
    =& \left.\Pi_\gamma(0)\right|_{\text{heavy}} + \underbrace{\left.\Pi_\gamma(0)\right|_{\text{light}} -\frac{ \left.\mathrm{Re}\Sigma^T_\gamma(M_Z^2)\right|_{\text{light}}}{M_Z^2}}_{\equiv \Delta \alpha} + \frac{ \left.\mathrm{Re}\Sigma^T_\gamma(M_Z^2)\right|_{\text{light}}}{M_Z^2} \label{eq:sm:vacuumpol} \\
    = & \left.\Pi_\gamma(0)\right|_{\text{heavy}} \,+ \frac{ \left.\mathrm{Re}\Sigma^T_\gamma(M_Z^2)\right|_{\text{light}}}{M_Z^2} + \Delta\alpha\nn\,,
\end{align}
where $\left.\Pi_\gamma(0)\right|_\text{heavy}$ contains contributions from heavy fermions as well as all bosonic contributions and where $\left.\mathrm{Re}\Sigma^T_\gamma(M_Z^2)\right|_{\text{light}}$ is
the transverse part of the photon self-energy considering only the \textit{light} degrees of freedom of the SM, \ie\ all leptons and the five light quarks.
The contribution of the light fermions to the vacuum polarisation, $\Pi_\gamma(0)|_{\text{light}}$, 
would develop infra-red (IR) divergences in the limit of vanishing fermion masses. Thus, its contribution is absorbed in the quantity $\Delta\alpha$, defined as
\begin{equation}\label{eq:dalpha}
    \Delta \alpha=\Delta \alpha^{(5)}_\text{had.} + \Delta \alpha_\text{lep.}= 0.02766 + 0.031497687\,,
\end{equation}
where $\Delta \alpha_{\mathrm{had}}^{(5)}$ is extracted experimentally~\cite{ParticleDataGroup:2020ssz}, while $\Delta \alpha_\text{lep.}$ was computed in \ccite{Steinhauser:1998rq}. Note that this value can also be changed using the \UFO format, \cf \cref{app:additionalufo}. The evaluation of $\left.\mathrm{Re}\Sigma^T_\gamma(M_Z^2)\right|_{\text{light}}$ as well as of the heavy field contributions to the photon vacuum polarisation, $\Pi_\gamma(0)|_{\text{heavy}}$, is straightforward and numerically stable. 

From \cref{eq:sm:deltav}, it can also be seen that the tadpole contribution in this treatment of the \vev is identical to the one chosen in the discussion of the FJ scheme in \cref{app:sm:tadren}: the tree-level couplings between the Higgs boson and the massive vector bosons normalised to their squared masses are universal (\ie\ the same for $W$ and $Z$) such that the tadpole contribution to $\delta^{(1)}M_{W}^2/M_{W}^2$ and $\delta^{(1)}M_{Z}^2/M_{Z}^2$ are identical. 
Thus the tadpole shift is only caused by $v\Sigma^{T}_{WW}(M_W^2)/(2M_W^2)$ in this scheme, which is identical to the one in \cref{eq:sm:vevshiftMZ} up to higher-order corrections.

\subsection{The SM with a real singlet (SSM)}
\label{app:ssm}
The most general potential that couples a real scalar gauge singlet $S$ to the SM reads
\begin{equation}
    V(\Phi,S) = 
    \underbrace{\mu^2|\Phi|^2 + 
    \frac{\lambda_H}{2} |\Phi|^4}_{V_{\Phi}} + 
    \underbrace{\frac{m_S^2}{2} S^2 + 
    \frac{\kappa_S}{3} S^3 +
    \frac{\lambda_S}{2} S^4}_{V_S} +
    \underbrace{\kappa_{SH} S|\Phi|^2 + 
    \frac{\lambda_{SH}}{2}S^2|\Phi|^2}_{V_{S\Phi}}  \,.
\end{equation}
This model is implemented in the \SARAH package under the name \texttt{SSM}. This implementation was used to create the \UFO files after adapting the conventions for the input parameters of the SM sector, as well as those discussed in the following, to the \anyH conventions. After spontaneous symmetry breaking, the \cp-even component of the Higgs doublet and the singlet are assumed to acquire VEVs, 
\begin{equation}
    S = s + v_S \qquad\text{and}\qquad
    \Phi = \frac{1}{\sqrt{2}} \begin{pmatrix} \sqrt{2}G^+ \\ v+h+i G \end{pmatrix}.
\end{equation}
The fields $s$ and $h$ mix to two \cp-even eigenstates $h_{1,2}$ with masses $m_{h_1}<m_{h_2}$:
\begin{equation}
    \left(\begin{array}{cc} m_{h_1}^2 & 0 \\ 0 & m_{h_2}^2 \end{array}\right) = R^{-1}(\alpha) \textbf{m}^2 R(\alpha)\,,\quad R(\alpha)=\left(\begin{array}{cc} c_\alpha & s_\alpha \\ -s_\alpha & c_\alpha \end{array}\right) \,,
\end{equation}
where $c_x\equiv \cos x$ and $s_x\equiv \sin x$.
We eliminate the parameters $\mu^2$ and $m_S^2$ using the tree-level tadpole equations
\begin{subequations}
\begin{align}
    0=\frac{\partial V}{\partial h} &= \mu^2 v + v v_S\kappa_{SH}  + \frac{\lambda_{SH}}{2} v v_S^2 + \frac{\lambda_H}{2}v^3\,,\\
    0=\frac{\partial V}{\partial s} &= \frac{v^2}{2}\kappa_{SH} + m_S^2 v_S + \frac{v^2 v_S}{2}\lambda_{SH} + 2v_S^3 \lambda_S + v_S^2\kappa_S,
\end{align}
\end{subequations}
which yields the following squared mass matrix:
\begin{equation}
\textbf{m}^2 = \left(
    \begin{array}{cc}
        v^2 \lambda_H &  v(\kappa_S+\lambda_{SH}v_S) \\
        v(\kappa_S+\lambda_{SH}v_S) & v_S(4\lambda_S v_S + \kappa_S)-\frac{\kappa_S v^2}{2v_S} 
    \end{array}
    \right)\,.
\end{equation}
Furthermore, we choose to eliminate the three dimensionless parameters $\lambda_H,\ \lambda_S,\ \lambda_{SH}$ in favour of the two masses $m_{h_{1,2}}^2$ and the mixing angle $\alpha$:
\begin{subequations}
\begin{align}
    \lambda_H & = \frac{1}{2v^2}\left(m_{h_1}^2+m_{h_2}^2+(m_{h_1}^2-m_{h_2}^2)\cos(2\alpha) \right),\\
    \lambda_S & =\frac{1}{8v_S^3}\left( \kappa_{SH} v^2 + 
 v_S (m_{h_1}^2 + m_{h_2}^2 - 2 v_S \kappa_S) - (m_{h_1}^2 - m_{h_2}^2) v_S \cos(2\alpha) \right), \\
    \lambda_{SH} & = \frac{1}{2 v v_S}\left( (m_{h_1}^2-m_{h_2}^2)\sin(2\alpha) - 2v \kappa_{SH} \right),\\
    \tan(2\alpha) & = \frac{2 (\textbf{m}^2)_{12}}{(\textbf{m}^2)_{11}-(\textbf{m}^2)_{22}} = \frac{4 v v_S(\kappa_{SH} + v_S \lambda_{SH})}{v^2\kappa_{SH}-2v_S(4v_S^2\lambda_S + v_S\kappa_S - v^2\lambda_H)}.
\end{align}
\end{subequations}
Consequently, the scalar sector of the model is determined by the following input parameters:
\begin{equation}
    m_{h_1},\, m_{h_2},\, \kappa_{S},\, \kappa_{SH},\, \alpha,\, v_S,\,  M_W,\, M_Z,\, \alpha_\text{em}\,.
\end{equation}
At tree level, the expressions of the trilinear self-couplings of the two \cp-even states read
\begin{align}
\label{eq:SSM_lamh1h1h1}
    \lambda_{h_1h_1h_1}&=-\kappa_Ss_\alpha^3+\kappa_{SH}\frac{3 v s_\alpha^2 ( vs_\alpha-2 v_S c_\alpha )}{2 v_S^2}+3m_{h_1}^2\left(\frac{c_\alpha^3}{v}+\frac{s_\alpha^3}{v_S}\right)\,,\nn\\
    \lambda_{h_2h_2h_2}&=-\kappa_S c_\alpha^3+\kappa_{SH}\frac{3 v c_\alpha^2 (v c_\alpha + 2 v_S s_\alpha)}{2 v_S^2}+3m_{h_2}^2\left(\frac{c_\alpha^3}{v_S}-\frac{s_\alpha^3}{v}\right)\,.
\end{align}

In the limit $\kappa_{S},\kappa_{SH} \to 0$, the model obeys a spontaneously broken $\Ztwo$ symmetry and only adds three BSM parameters (the $\Ztwo$ breaking VEV, one mass as well as one mixing angle) to the SM parameters. For more detailed information about the model file generation we refer to the \examplesrepo.

\subsubsection{Alignment}
The alignment limit, in which all tree-level couplings of $h_1$ to the SM sector (including the trilinear Higgs coupling) become identical to those of the SM, is achieved by choosing:
\begin{equation}
    \alpha=0\,.
\end{equation}
However, in this limit the coupling $\lambda_{SH}$ is still non-zero and thus the one-loop prediction for $\lamhhh=\lambda_{h_ih_ih_i}^{(1)}$ can differ from the SM prediction. Note that the SM-like Higgs boson can in principle be either $h_1$ or $h_2$, see for instance the discussion of \cref{fig:SSMlamsss}. In the following discussion we assume without loss of generality that $h_1=h$ is the SM-like and $h_2=s$ is the singlet-like Higgs boson.

\subsubsection{Renormalisation}
\label{app:ssm:ren}
In the following we briefly describe the default renormalisation conditions in the (BSM) Higgs sector that have been implemented in the \texttt{schemes.yml} file. It should be emphasised that this choice is not fixed and can be changed by the user. Note also that in the following (as well as for the discussion of renormalisation in other models), we omit the superscript $(1)$ and subscript $\text{CT}$ on counterterms, as there should be no risk of confusion. 

\paragraph{$\bm{\delta v}$ and $\bm{\delta m_{h_1}^2}$:}
The mass of the SM-like Higgs boson as well as the electroweak VEV are renormalised as in the SM. Thus the user can choose to renormalise them \OS or \MS.

\paragraph{$\bm{\delta m_{h_2}^2}$:}
The mass of the singlet-like Higgs boson, $m_{h_2}^2$, does not enter the tree-level prediction of $\lambda_{h_1 h_1 h_1}$ (see \cref{eq:SSM_lamh1h1h1}), and therefore is not required to be renormalised at the considered loop order.

\paragraph{$\bm{\delta v_S}$:}
From the one-loop RGEs one can infer that the singlet VEV does not receive a UV-divergent counterterm at the one-loop order. As a simple cross-check of \anyH, we diagrammatically verified this fact. Therefore the singlet VEV counterterm does not introduce a renormalisation scale dependence in $\lambda_{h_1 h_1 h_1}^{(1)}$. Thus, we could choose not to renormalise $v_S$. However, due to the chosen default FJ renormalisation scheme in \anyH for all appearing tadpoles (meaning that $v_S$ is the VEV of the tree-level potential) we can also consider a finite shift to the singlet VEV which ensures that it corresponds to the minimum of the loop-corrected potential:
\begin{equation}
    \delta v_S = \cos\alpha \frac{\delta^{(1)} t_{h_2}}{m_{h_2}^2} + \sin\alpha \frac{\delta^{(1)} t_{h_1}}{m_{h_1}^2}\,.
    \label{eq:ssm:dvs}
\end{equation}

\paragraph{$\bm{\delta\alpha}$ and $\bm{\delta Z_{h_1 h_2}}$:}
The default choice in \anyH is to calculate \textit{all} external leg corrections at the given external momenta, which are set to zero by default. Thus, per default a fully OS wave-function renormalisation which properly removes the mixing between external states is not used. This is the default ``OS'' scheme (denoted ``OS'' because the SM-sector is still renormalised on-shell) defined in the \texttt{schemes.yml} file:
\begin{minted}[bgcolor=bg]{yaml}
OS:
  description: OS masses and singlet VEV
  SM_names:
    Higgs-Boson: h1
  VEV_counterterm: OS
  mass_counterterms:
    h1: OS
  custom_CT_hhh: |
    # Tadpole contribution to singlet vev
    dvS = f"-( cos(alphaH)*({Tadpole('h2')})/Mh2**2 \
        + sin(alphaH)*({Tadpole('h1')})/Mh1**2 )"
    dlambda_dvS = Derivative(lambdahhh_tree, 'vS')
    self.custom_CT_hhh = f'+({dlambda_dvS})*({dvS})'
\end{minted}
In the example above the counterterm \code{dvS} corresponds to \cref{eq:ssm:dvs}, while \code{Derivative(lambdahhh_tree, 'vS')} takes the derivative of the tree-level prediction for $\lambda_{h_1h_1h_1}$ w.r.t.\ $v_S$. The final result of the counterterm contribution again needs to be saved in the variable \code{self.custom_CT_hhh}, since this variable is internally used by \anyH. In this default scheme, $\alpha$ is renormalised in the \MS scheme. 

For demonstration purposes, we also implemented a non-minimal renormalisation scheme based on the OS scheme used in \ccite{Bojarski:2015kra} for heavy Higgs decays. This version of the OS scheme fixes the mixing angle counterterm to
\begin{equation}
    \delta \alpha = \frac{1}{2}\frac{\Sigma_{h_1 h_2}(p^2=m_{h_1}^2) + \Sigma_{h_1 h_2}(p^2=m_{h_2}^2)}{m_{h_2}^2-m_{h_1}^2}\, .
    \label{eq:ssm:dalpha}%
\end{equation}
The corresponding renormalisation scheme in the \texttt{schemes.yml} file is called ``OSmixing'':
\begin{minted}[bgcolor=bg]{yaml}
OSmixing:
  description: OS masses, singlet VEV and mixing angle/external legs
  SM_names:
    Higgs-Boson: h1
  VEV_counterterm: OS
  wfrs: False # turn off all WFR diagrams
  mass_counterterms:
    h1: OS
  custom_CT_hhh: |
    # OS counterterm contribution of the mixing angle
    dalphaH = f"({Sigma('h1','h2',momentum='Mh1**2')}\
               + {Sigma('h2','h1',momentum='Mh2**2')})/(2*(Mh2**2-Mh1**2))"
    dlambda_dalpha = Derivative(lambdahhh_tree, 'alphaH')
    self.custom_CT_hhh = f'({dalphaH})*({dlambda_dalpha})'

    # OS Z-factor contribution
    lam112 = self.getcoupling('h1','h1','h2')['c'].value
    dZ21 = f"+3*({Sigma('h2','h1',momentum='Mh1**2')})/((Mh1**2 - Mh2**2))"
    self.custom_CT_hhh += f'+complex(0,1)*({dZ21})*({lam112})'
    self.custom_CT_hhh += f"+3/2*({lambdahhh_tree})*({Sigmaprime('h1')})"

    # Tadpole contribution to singlet vev
    dvS = f"-(cos(alphaH)*({Tadpole('h2')})/Mh2**2 \
        + sin(alphaH)*({Tadpole('h1')})/Mh1**2 )"
    dlambda_dvS = Derivative(lambdahhh_tree, 'vS')
    self.custom_CT_hhh += f'+({dlambda_dvS})*({dvS})'
\end{minted}
Note the additional entry \code{wfrs: False} which turns off the automatic calculation of all external-leg corrections (which defaults to $p^2=0$). Instead they have been added in the \code{custom_CT_hhh}-section using on-shell momenta for all three external legs.

\paragraph{$\bm{\delta \kappa_S}$ and $\bm{\delta \kappa_{SH}}$:}
The soft-$\Ztwo$-breaking parameters are renormalised in the $\MS$ scheme. Thus, the trilinear Higgs self coupling of the SM-like Higgs boson is given purely in terms of $\OS$ parameters if we work in the $\Ztwo$-symmetric limit where $\kappa_{S},\kappa_{SH}\to 0$.

We have explicitly verified that in this limit the result obtained in the ``OSmixing'' scheme is manifestly renormalisation-scale independent. We also provide a \mat notebook in the \examplesrepo which explicitly demonstrates the UV finiteness of this scheme using the \mat interface of the \anyBSM code. 

\subsection{The Two-Higgs-Doublet Model (THDM)}
\label{app:thdm}

For the description of our implementation of the THDM 
we proceed along similar lines as for the SSM. The tree-level scalar potential of the built-in \cp-conserving THDMs reads 
\begin{align}
    V_{\text{THDM}} =
      &\ \mu_1^2 |\Phi_1|^2 + \mu_2^2 |\Phi_2|^2 +\lambda_1|\Phi_1|^4+\lambda_2|\Phi_2|^4 \nonumber \\
      &\quad+\lambda_3|\Phi_1|^2|\Phi_2|^2+\lambda_4|\Phi_1^\dagger\Phi_2|^2+\bigg(\frac{1}{2}\lambda_5(\Phi_1^\dagger\Phi_2)^2  + m_{12}^2\Phi_1^\dagger\Phi_2 +\text{h.c.}\bigg)\, .
\end{align}
This potential obeys a $\mathbb{Z}_2$ symmetry ($\Phi_1\to\Phi_1, \Phi_2\to -\Phi_2$), which is softly broken by the $m_{12}^2$ term. Even though the different types in the Yukawa sector~\cite{Barger:1989fj,Grossman:1994jb,Aoki:2009ha} play only a minor role in the prediction of $\lambda_{hhh}$, we implemented for convenience four distinct models corresponding to type-I, type-II, type-X and type-Y (the latter two types are also often called type~III and type~IV, or flipped and lepton-specific) --- see \ccite{Branco:2011iw} for a review. 

After spontaneous symmetry breaking, the two doublets obtain the VEVs $v_{1,2}$ with $\nicefrac{v_2}{v_1}=\tan\beta$. The mixing between the two \cp-even states is described by the parameter $\alpha$. We eliminate $\mu_1^2$ and $\mu_2^2$ using the tadpole equations and we trade $\lambda_{1,2,3,4,5}$ for the masses of the two \cp-even Higgs masses $m_{h_{1,2}}$, the \cp-odd Higgs mass $m_A$, the charged Higgs mass $m_{H^\pm}$, as well as the mixing angle $\alpha$. Thus, the chosen (BSM) input parameters are:
\begin{equation}
    m_{h_1},\, m_{h_2},\, m_{A},\, m_{H^{\pm}},\, M,\,\tan\beta,\, \sin(\beta-\alpha),\quad \text{with}\,\, M^2=-\frac{m_{12}^2}{\cos\beta\sin\beta}.
\end{equation}
For more detailed information we refer to the \docs and the \examplesrepo.

\subsubsection{Alignment}
\label{sec:THDM_alignment}
The alignment limit of the THDM is achieved if the electroweak vacuum expectation value is aligned with the SM-like Higgs boson. If the lightest \cp-even state is SM-like, this is achieved by setting $\sin(\beta-\alpha) = 1$; if the heavier \cp-even state is SM-like, $\cos(\beta-\alpha) = 1$ corresponds to the alignment limit.

\subsubsection{Renormalisation}
The renormalisation schemes provided for the THDM are similar to the SSM. The default scheme only renormalises the SM sector OS, while for the mixing angles the \MS scheme is used. As an additional option, which we employed for checking UV finiteness, we provide a scheme which uses an OS mixing angle as in the SSM, \cf\ \cref{eq:ssm:dalpha}. Furthermore, an OS counterterm for $\tan\beta$ can be either defined via the charged Higgs/Goldstone self-energies or via the self-energies of the pseudoscalar Higgs/Goldstone bosons. Since the two schemes differ by finite contributions, we have implemented both of them. The parameter $M$ is always renormalised in the \MS scheme.

A complete list of all renormalisation schemes defined in the \texttt{schemes.yml} file can be accessed
by \eg\ calling the \code{list_renormalization_schemes()} function:
\begin{minted}[bgcolor=bg]{python}
from anyBSM import anyBSM
THDM = anyBSM('THDMII')
THDM.list_renormalization_schemes()
\end{minted}
or by consulting the \docs.

In the alignment limit (see \cref{sec:THDM_alignment}), the tree-level prediction for $\lamhhh^{(0)}$ is identical to the SM (\cf\ \cref{app:sm}), which means that \lamhhh can be fully expressed in terms of SM OS quantities in this case. Thus the tOS and the FJ schemes will formally yield the same result, as discussed in \cref{app:sm:tadren}. However, the FJ scheme is known to be perturbatively unstable due to large cancellations between different tadpole-inserted diagrams. To assess potential instabilities, we also implemented the tOS scheme in the alignment limit.

\begin{figure}[t]
    \centering
    \includegraphics[width=1.0\textwidth]{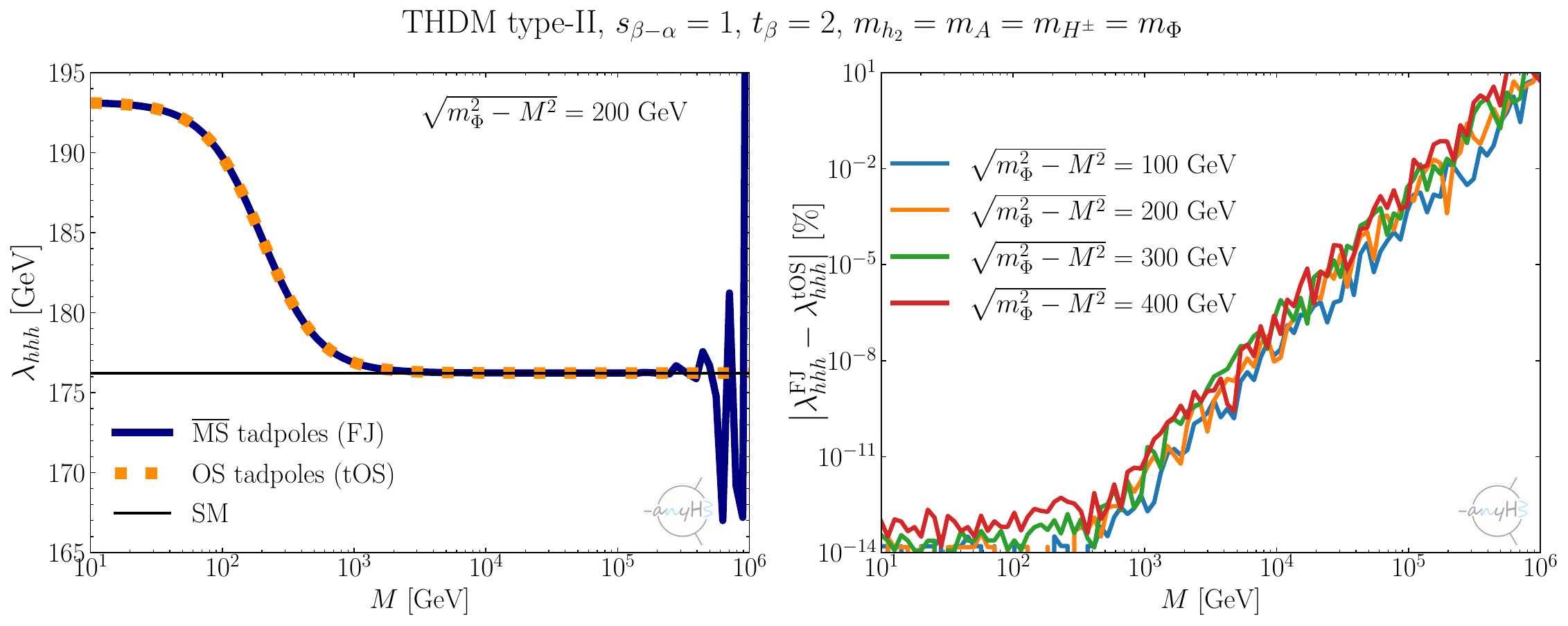}
    \caption{Comparison of the \lamhhh calculations using the tOS and FJ schemes in the THDM. 
    Left: \lamhhh as a function of $M$ for $\sqrt{m_\Phi^2-M^2}=200\gev$. Right: Relative difference in percent of \lamhhh calculated in the FJ and the OS-tadpole scheme for different values of $\sqrt{m_\Phi^2-M^2}$.}
    \label{fig:app:TadComp}
\end{figure}

In the left panel of \cref{fig:app:TadComp}, we compare the numerical prediction for \lamhhh in the THDM-II using the tOS and FJ schemes for a specific scenario with $\sin(\beta-\alpha)=1$, $\tan\beta=2$, and $m_{H^\pm}=m_A=m_{h_2}\equiv m_{\Phi}=\sqrt{M^2+\Delta^2}$ (with $\Delta=\unit[200]{GeV}$) as a function of $M$. In this parametrisation, it is expected that the BSM contributions decouple in the limit $M\gg v$, where the BSM states become heavy and mass-degenerate, and that the THDM result converges towards the SM result (black line). This is indeed observed for both schemes near the scale $M=\unit[10^3-10^4]{GeV}$. One can see that as long as $M\lesssim \unit[10^5]{GeV}$ --- well within the decoupling region --- the tOS scheme (orange-dotted) agrees very well with the FJ scheme (blue-solid line). For $M\gtrsim \unit[10^5]{GeV}$, numerical instabilities start to develop in the FJ scheme.

In the right panel of \cref{fig:app:TadComp}, we show the relative difference in \lamhhh between the two schemes for various values of $\Delta$. Even for very large couplings close to the unitarity limit (where the latter is reached for $\Delta\gtrsim 400\gev$),
implying very large loop corrections for $M\sim v$, the behaviour is very similar. 
Hence, we can conclude in general that the use of the (default) FJ scheme is not affected by numerical instabilities as long as all scalar masses are lighter than \unit[$100-1000$]{TeV}. As discussed above, numerical instabilities can still appear if a scheme conversion of the input parameters is performed.

\subsection{The Inert-Doublet Model (IDM)}
\label{app:idm}
The IDM is a variant of the THDM in which the $\mathbb{Z}_2$ symmetry is imposed exactly. Its potential is for this reason identical to that of the THDM, with the exception that $m_{12}^2$ vanishes:
\begin{equation}
    V_{\text{IDM}} = \left. V_{\text{THDM}}\right|_{m_{12}^2=0}\, .
\end{equation}
After spontaneous symmetry breaking we only allow the SM-like Higgs doublet to obtain a VEV, $\langle \Phi_1\rangle = \nicefrac{v}{\sqrt{2}}$, but not the second doublet, $\langle \Phi_2\rangle = 0$, because the $\Ztwo$ symmetry is exact. We again eliminate $\mu_1^2$ using the tadpole equation as well as write $\lambda_{1,3,4,5}$ in terms of the scalar masses and $\mu_2^2$. Thus, the input parameters of the scalar potential are:
\begin{equation}
    m_{h_1},\, m_{h_2},\, m_A,\, m_{H^\pm},\, \mu_2^2\, \,\text{and}\,\, \lambda_2,
\end{equation}
with
\begin{align}
  m_{h_1}^2&=\lambda_1v^2, \nn\\
  m_{h_2}^2&=\mu_2^2+\frac{1}{2}(\lambda_3+\lambda_4+\lambda_5) v^2,\nn\\
  m_A^2&=\mu_2^2+\frac{1}{2}(\lambda_3+\lambda_4-\lambda_5) v^2,\nn \\
  m_{H^\pm}^2&=\mu_2^2+\frac{1}{2}\lambda_3 v^2.
\end{align}
For more information about the model we refer to the \docs.

\subsubsection{Alignment}
Since the \cp-even Higgs bosons do not mix in this model, the tree-level couplings of the SM-like Higgs boson are automatically equal to the respective SM couplings.

\subsubsection{Renormalisation}
The tree-level prediction of $\lambda_{hhh}$ in this model is always identical to the SM. Hence, we only need to fix the renormalisation conditions for the SM parameters at the one-loop order as discussed in \cref{app:sm}.

\subsection{The THDM with a real singlet (NTHDM)}
\label{app:nthdm}
The Higgs sector of the NTHDM consists out of two $SU(2)_L$ doublets $\Phi_{1,2}$ and a real singlet $S$. In the model file, the discrete symmetries
\begin{alignat}{4}
   &\mathbb{Z}_2:\hspace{.5cm} &&\Phi_1 \rightarrow \Phi_1, \hspace{.3cm} &&\Phi_2\to - \Phi_2, \hspace{.3cm} &&S\to S, \\
   &\mathbb{Z}_2^\prime:\hspace{.5cm} &&\Phi_1 \rightarrow \Phi_1, \hspace{.3cm} &&\Phi_2\to \Phi_2, \hspace{.3cm} &&S\to -S
\end{alignat}
are imposed. Then, the most general potential reads\footnote{We note that the off-diagonal mass term $m_{12}^2$ is defined here with the opposite sign compared to its counterpart in the THDM (the latter follows the convention in the THDM \SARAH model files).}
\begin{align}
    V(\Phi_1,\Phi_2,S) =&\ m_{11}^2\Phi_1^\dagger\Phi_1 + m_{22}^2\Phi_2^\dagger\Phi_{2} - (m_{12}^2\Phi_1^\dagger\Phi_2 + \text{h.c.}) \nonumber \\
    & + \frac{\lambda_1}{2}(\Phi_1^\dagger\Phi_1)^2 + \frac{\lambda_2}{2}(\Phi_2^\dagger\Phi_2)^2 + \lambda_3(\Phi_1^\dagger\Phi_1)(\Phi_2^\dagger\Phi_2) + \lambda_4(\Phi_1^\dagger\Phi_2)(\Phi_2^\dagger\Phi_1) \nonumber\\
    & + \frac{\lambda_5}{2}\left[(\Phi_1^\dagger\Phi_2)^2 + \text{h.c.}\right]\nonumber\\
    & + \frac{1}{2}m_S^2 S^2  + \frac{\lambda_6}{8}S^4 + \frac{\lambda_7}{2}(\Phi_1^\dagger\Phi_1)S^2 + \frac{\lambda_8}{2}(\Phi_2^\dagger\Phi_2)S^2\,.
\end{align}
After spontaneous symmetry breaking, the \cp-even components of the Higgs doublet $\rho_{1,2}$ and the singlet $\rho_S$ mix with each other. The mixing matrix is defined as
\begin{align}
R = 
\begin{pmatrix}
c_{\alpha_1} c_{\alpha_2} & s_{\alpha_1} c_{\alpha_2} & s_{\alpha_2} \\ - s_{\alpha_1} c_{\alpha_3} - c_{\alpha_1} s_{\alpha_2} s_{\alpha_3} & c_{\alpha_1} c_{\alpha_3} - s_{\alpha_1}s_{\alpha_2}s_{\alpha_3} & c_{\alpha_2} s_{\alpha_3} \\ s_{\alpha_1} s_{\alpha_3} - c_{\alpha_1} s_{\alpha_2} c_{\alpha_3} & - c_{\alpha_1} s_{\alpha_3} - s_{\alpha_1}s_{\alpha_2}c_{\alpha_3} & c_{\alpha_2} c_{\alpha_3}
\end{pmatrix},%
\end{align}
where $\alpha_{1,2,3}$ are the three mixing angles. Then,
\begin{align}
\begin{pmatrix} h_1 \\ h_2 \\ h_3 \end{pmatrix} = R \begin{pmatrix} \rho_1 \\ \rho_2 \\ \rho_S \end{pmatrix}
\end{align}
The mixing angle of the \cp-odd and charged components of the Higgs doublets is $\beta$ (with $\tan\beta = v_2/v_1$), like in the THDM.
The Yukawa sector of the model provided with \anyH is defined as a type-II Yukawa sector.
For more details about the NTHDM, see \eg\ \ccite{Heinemeyer:2021msz}.

\subsubsection{Alignment}
One way to achieve SM-like couplings for $h_1$ is by choosing~\cite{Krause:2017mal}
\begin{equation}
    \alpha_1=\beta\,,\quad\alpha_2=0\,,\quad\text{and}\quad \alpha_3=0\,,
\end{equation}
while the same limit can be obtained for $h_2$ with
\begin{equation}
    \alpha_1+\alpha_3=\beta-\pi/2,\quad\text{and}\quad \alpha_2=\pi/2\,.
\end{equation}

\subsubsection{Renormalisation}
For the NTHDM, only the default renormalisation (\ie, OS or \MS conditions for the boson masses and the VEV as well as \MS for the mixing angle), is available in the in-built model so far. However, a generalisation of the OS scheme to the mixing angles in the NTHDM has been performed in \eg\ \ccite{Krause:2017mal} and could in principle be implemented straightforwardly in the \texttt{schemes.yml} of the model if needed.

\subsection{The SM with a real triplet (\texorpdfstring{\TSM0}{TSM with Y=0})}
\label{app:tsm0}
The most general potential which couples a real scalar $SU(2)_L$ triplet $T$ with hypercharge $Y_T=0$ to the SM reads
\begin{equation}
    V(\Phi,T) = 
    \mu^2|\Phi|^2 + 
    \frac{\lambda}{2} |\Phi|^4 + 
    \frac{M_T^2}{2} |T|^2 + 
    \frac{\lambda_T}{2} |T|^4 +
    \kappa_{T\Phi} \Phi^\dagger \hat{T} \Phi + 
    \frac{\lambda_{T\Phi}}{2}|T|^2|\Phi|^2  \,,
\end{equation}
with
\begin{align}
    \hat{T} = \sigma_a T^a = \left(\begin{array}{cc}
        \frac{t^0}{\sqrt2} & t^+  \\
         t^- & -\frac{t^0}{\sqrt2}
    \end{array}\right)\,,
\end{align}
where $\sigma_a$ are the Pauli matrices.
After spontaneous symmetry breaking, the neutral triplet component may acquire a VEV $t^0\to t^0 + 2^{-1/2}v_T$ analogous to the neutral doublet component $(\Phi)_0\to 2^{-1/2}(h + v)$.
The tree-level mixing between the fields $t^0$ and $h$ is solely controlled by the parameters $v_T$ and $\kappa_{T\Phi}$. However, since the triplet is charged under the $SU(2)_{L}$ group, it contributes to the $W^\pm$-boson mass via gauge-kinetic terms,
\begin{equation}
    M_{W^\pm}^2 = g_2^2\left(\frac{v^2}{4}+ \frac{v_T^2}{2}\right),
\end{equation}
but not to the $Z$-boson mass. Therefore the $\rho$ parameter already differs from one at the tree level, which tightly constraints the ratio $\nicefrac{v_T}{v}$. For these small values of $\nicefrac{v_T}{v}$, the dependence of \lamhhh on $v_T$ is negligible. For this reason, we consider a VEV-less triplet: $v_T =0$.

Considering the tadpole equations in the VEV-less limit, 
\begin{subequations}
\begin{align}
    2 M_T^2 v_T + \lambda_{T\Phi} v^2 v_T + \lambda_T v_T^3 -\kappa_{T\Phi} v^2 = 0\,,  \\
    \mu^2 v + \frac{\lambda v^3}{2} - \frac{\kappa_{T\Phi} v v_T}{2} + \frac{\lambda_{T\Phi} v v_T^2}{4} = 0\,,
\end{align}
\end{subequations}
we recover the SM tadpole condition and $\kappa_{T\Phi}=0$. Thus we have
\begin{equation}
    \label{eq:trip:vevless}
    \kappa_{T\Phi} = v_T = 0 \,.
\end{equation}
The charged Higgs boson mass matrix has a particular simple form in the limit of \cref{eq:trip:vevless},
\begin{equation}
\bm{m^\pm} = \left(\begin{array}{cc}
    M_W^2 \xi_W & 0 \\
        0 & M_T^2 + \frac{\lambda_{T\Phi}v^2}{2}
\end{array}\right) =
\left(\begin{array}{cc}
    m_{G^+}^2 & 0 \\
    0 & M_{H^+}^2
\end{array}\right)\,,
\end{equation}
such that we can eliminate $M_T^2$ in favour of the charged Higgs boson mass
\begin{equation}
    \label{eq:trip:mhp}
    M_T^2 = M_{H^{+}}^2-\frac{v^2\lambda_{T\Phi}}{2}\,.
\end{equation}
This gives an implicit bound on the input parameters $\lambda_{T\Phi} < \nicefrac{2 M_{H^+}^2}{v^2}$ in order to preserve the structure of the tree-level vacuum.

The neutral Higgs boson mass matrix is also already diagonal in the limit of \cref{eq:trip:vevless},
\begin{equation}
    \bm{m}^0 = \left(\begin{array}{cc}
         \lambda v^2 & 0  \\
         0           & M_T^2 + \frac{\lambda_{T\Phi}v^2}{2} 
    \end{array}\right)
    \equiv\left(\begin{array}{cc}
        m_h^2 & 0 \\
        0 & M_{H^+}^2
    \end{array}\right)\,.
\end{equation}
This means that the additional neutral Higgs boson is degenerate in mass with the charged Higgs boson. Additionally, there is no mixing between the SM and the triplet field implying that the trilinear Higgs coupling is identical to the SM at tree level. However, at the one-loop order the parameters $\lambda_{T\Phi}$ and $M_{H^+}$ induce BSM corrections to \lamhhh.

\subsubsection{Alignment}
Similar to the IDM, the VEV-less version of all triplet extensions are also automatically aligned with the SM concerning all SM-like Higgs boson interactions.

\subsubsection{Renormalisation}
All couplings of the SM-like Higgs boson in this model are as in the SM at tree-level.
Consequently, the renormalisation follows the description in \cref{app:sm}.

\subsection{The SM with a complex triplet (\texorpdfstring{\TSM1}{TSM with Y=1})}
\label{app:tsm1}
The discussion of the complex triplet extension closely follows that for the real triplet extension. Instead of a hypercharge-less triplet, we add a triplet $\Delta$ with $Y=1$, which leads to the following potential:
\begin{align}
    V(\Phi, \Delta) ={}& m^2 \Phi^\dagger\Phi + M^2 \text{Tr}(\Delta^\dagger\Delta) \nonumber\\
        & + \lambda_1 (\Phi^\dagger\Phi)^2 + \lambda_2 \left[\text{Tr}(\Delta^\dagger\Delta)\right]^2 + \lambda_3 \left[\text{Tr}(\Delta^\dagger\Delta)^2\right] \nonumber\\
        & + \lambda_4 (\Phi^\dagger\Phi) \text{Tr}(\Delta^\dagger\Delta) + \lambda_5 \Phi^\dagger\Delta\Delta^\dagger\Phi\, .
\end{align}
We again demand that the triplet does not take part in electroweak symmetry breaking. Thus, there is no mixing in the \cp-even sector. The mass parameter $m^2$ is eliminated using the SM-like tadpole condition. Then, the tree-level Higgs boson masses are given by
\begin{align}
    m_h^2 &= 2\lambda_1 v^2, \\
    m_\chi^2 &= m_{D^0}^2 = M^2 + \frac{1}{2}(\lambda_4 + \lambda_5)v^2, \\
    m_{D^\pm}^2 &= M^2 + \frac{1}{4} (2\lambda_4 + \lambda_5)v^2, \\
    m_{D^{\pm\pm}}^2 &= M^2 + \frac{1}{2} \lambda_4v^2.
\end{align}
We use $m_{D^\pm}$, $m_{D^{\pm\pm}}$, and the couplings $\lambda_{2,3,4}$ as BSM input parameters. All couplings of the SM-like Higgs boson in this model are as in the SM at tree-level. Therefore the renormalisation is treated as in the \TSM0 and the IDM.
For more details about the $Y=1$ inert triplet model, see \eg\ \ccite{Bahl:2022gqg}.

\subsection{The Georgi-Machacek Model (GM)}
\label{app:gm}
The Georgi-Machacek model is a triplet extension of the SM, which --- in contrast to the previously discussed triplet extensions --- respects the custodial symmetry of the SM at tree-level and therefore has a tree-level $\rho$ parameter of one. The model adds one complex and one real triplet, respectively denoted $\xi$ and $\eta$, to the SM scalar sector. The doublet and the triplets are effectively re-written in terms of one bi-doublet $\Phi$ and one bi-triplet $X$ transforming under the custodial group $SU(2)_L\times SU(2)_R$:
\begin{equation}
\Phi = \left( \begin{array}{cc}
    \phi^{0*} & \phi^+ \\
    -\phi^- & \phi^0 \end{array}\right), \,\,\,\,
  X = \left(\begin{array}{ccc}
    \xi^{0\ast} & \eta^+ & \xi^{++} \\
    -\xi^{-} & \eta^0 & \xi^{+} \\
   -\xi^{--} & -\eta^- & \xi^{0}
\end{array}\right)\, .%
\end{equation}%
The most general potential that is manifestly invariant under $SU(2)_L\times SU(2)_R$ reads~\cite{Hartling:2014zca}
\begin{align}
V_{\text{GM}} = &\ 
    \frac{\mu^2}{2} \text{Tr}\, \Phi^\dagger\Phi +
    \frac{M_X^2}{2} \text{Tr}\, X^\dagger X +
    \frac{\lambda_1}{2} \left(\text{Tr}\, \Phi^\dagger\Phi\right)^2 +
    \lambda_2 \left(\text{Tr}\, \Phi^\dagger\Phi\right) \left(\text{Tr}\, X^\dagger X\right)\nn \\
    & + \lambda_3 \text{Tr}\, \left(X^\dagger X\right)^2  +
    \lambda_4 \left(\text{Tr}\, X^\dagger X\right)^2  
   - 
   \lambda_5 \left(\text{Tr}\, \Phi^\dagger\sigma^a\Phi\sigma^b\right)\left(\text{Tr}\, X^\dagger t_a X t_b\right)  \\
   &
    -\left[
        M_1 \left(\text{Tr}\, \Phi^\dagger\sigma^a\Phi\sigma^b\right)
      + M_2 \left(\text{Tr}\, X^\dagger t^a X t^b\right)
    \right] \left(U X U^\dagger\right)_{ab}\nn
\end{align}
with the rotation matrix $U$
\begin{equation}
U = \left( \begin{array}{ccc}
    -\frac{1}{\sqrt{2}} & 0 & \frac{1}{\sqrt{2}} \\
    -\frac{i}{\sqrt{2}} & 0 & -\frac{i}{\sqrt{2}} \\
                      0 & 1 & 0
\end{array}\right)\,,
\end{equation}
which transforms the bi-triplet $X$ in terms of Cartesian field-coordinates. The $\sigma_i$ ($t_i$) are the $SU(2)$ generators in the adjoint (fundamental) representation. We provide two \UFO models which implement the two scenarios where either $M_X^2<0$ or $M_X^2>0$.

In the first case the bi-triplet obtains a VEV $v_X$ that triggers mixing between the doublet and the triplet states, \ie, the SM-like Higgs will have a non-zero triplet component. We solve the tadpole equation for the SM-like doublet to eliminate $\mu^2$. The tadpole equation for the bi-triplet is used to eliminate $M_1$ rather than $M_X^2$,  as 
\begin{align}
    M_1=\frac{4v_X}{v_\text{SM}^2} \big[M_X^2 + (2\lambda_2 - \lambda_5) v_\text{SM}^2 - 6M_2 v_X + 4(\lambda_3 + 3\lambda_4) v_X^2\big]\,,
\end{align}
which conveniently allows to take the alignment limit $M_1, v_X\to 0$ (while keeping $\nicefrac{v_X}{M_1}$ finite). After spontaneous symmetry breaking we trade $\lambda_1$, $\lambda_2$ and $\lambda_5$ for the SM-like Higgs mass, the triplet mass $M_3$ and the fiveplet mass $M_5$. Thus, the input parameters of the BSM sector are
\[
\sin {\theta_H},\, M_2,\, M_3,\, M_5,\, M_X,\, \lambda_3\,\text{ and }\, \lambda_4\,,
\]
with $\sin {\theta_H}=\frac{2 \sqrt{2} v_X}{v_{\text{SM}}}$.

\medskip

In the case of $M_X^2>0$, we have one tadpole equation less. As a trade-off we automatically have $\sin\theta_H=0$ (and thus $v_X=M_1=0$) as well as $M_3$ not being independent anymore: $M_3^2=\frac{1}{3} \left(M_5^2+2 m_{h_2}^2\right)$. Hence in the VEV-less case we have
\[
M_2,\, M_5,\, M_X,\,\lambda_2,\, \lambda_3\,\text{ and }\, \lambda_4\,.
\]
as BSM input parameters. We do not trade $\lambda_2$ for $m_{h_2}$, but keep $\lambda_2$ as input since the parametrisation which uses $m_{h_2}$ as input suffers from numerical instabilities in the decoupling limit (see also \ccite{Braathen:2017izn} for related discussions).

For more information on the model we refer to \eg\ \ccite{Hartling:2014zca,Krauss:2017xpj} and the \docs.

\subsubsection{Alignment}
Alignment in this model can only be achived with a vanishing triplet VEV (\ie, $\sin {\theta_H}=0$). Since this also significantly simplifies many tree-level relations, we provide a separate model (\texttt{GeorgiMachacekAligned}) without a triplet VEV (\ie, $M_X^2>0$) in addition to the \UFO~model implementing the general (\ie $M_X^2<0$) case (\texttt{GeorgiMachacek}).

\subsubsection{Renormalisation}
For the \texttt{GeorgiMachacekAligned} \UFO model the usual SM-like OS/\MS renormalisation procedure is sufficient. In the general \texttt{GeorgiMachacek}-implementation all BSM parameters are assumed to be \MS~parameters for simplicity. Non-minimal renormalisation conditions in this model have been proposed \eg\ in \ccite{Chiang:2018xpl}.

\subsection{A simple \texorpdfstring{$U(1)_{B-L}$}{U(1)\_(B-L)} extension of the SM (BmLSM)}
\label{app:bmlsm}
We also provide a \UFO model which extends the SM by a gauged $U(1)_{B-L}$ symmetry that is spontaneously broken by the VEV $v_X$ of a complex scalar $S$. The scalar potential reads
\begin{equation}
    V_{\text{B-L}}(\Phi,S) =  \mu^2 \Phi^\dagger\Phi
        + \frac{\lambda_1}{2} |\Phi^\dagger\Phi|^2
        + \mu_S^2 S^* S
        + \frac{\lambda_2}{2} |S^* S|^2
        + \lambda_3 S^* S \Phi^\dagger\Phi\,.
\end{equation}
We use the tadpole equations to eliminate $\mu$ and $\mu_S$ as well as trade $v_x$, $\lambda_1$, $\lambda_2$ and $\lambda_3$ in favor of the $Z^{\prime}$ mass $M_{Z^\prime}$, the two scalar masses $m_{h_{1,2}}$, and the mixing angle $\alpha$. 

Right-handed neutrinos are added in order to cancel the additional gauge anomalies:
\begin{equation}
    \lag_{\text{B-L}}^{\text{Yukawa}} = -\boldsymbol{Y}_{\nu}\, \Phi^\dagger\cdot L_L \nu_R - \boldsymbol{Y}_{x}\, S\nu_R \nu_R +   h.c.
\end{equation}
For simplicity we assume that the Yukawa matrices $\boldsymbol{Y}_\nu$ and $\boldsymbol{Y}_x$ are diagonal. Thus, we can diagonalise the neutrino mass matrices, resulting in 6 eigenvalues which scale for $v_{SM}\ll v_X$ to first non-vanishing order as
\begin{align}
    m_{{\nu}_i} = &\frac{v_{SM}^2 (\boldsymbol{Y}_v^2)_{ii}
    }{v_x (\boldsymbol{Y}_x^2)_{ii}} + \mathcal{O}\left(\frac{v_{SM}^3}{v_x^2}\right),\\
    m_{{\nu}_{i+3}} = & \sqrt{2}v_x (\boldsymbol{Y}_x)_{ii} + \mathcal{O}\left(\frac{v_{SM}^2}{v_x}\right)\,.\,\, (i=1,2,3)%
\end{align} \noindent%
We choose to trade the components of $\boldsymbol{Y}_v$ and $\boldsymbol{Y}_x$ for the neutrino masses. In consequence, the BSM input parameters of the model are
\begin{equation}
    m_{h_1},\, m_{h_2},\, \alpha,\, M_{Z^\prime},\, m_{\nu_{1\dots6}}
\end{equation}
Kinetic mixing between the $U(1)_{Y}$ and $U(1)_{B-L}$ is not considered (\ie, set to zero at tree level) in this model implementation. For more information on the model see \eg\ \ccite{Basso:2010yz}.

\subsubsection{Alignment}
All couplings of the SM-like Higgs boson to the SM sector are rescaled by $\cos\alpha$. Thus, $\alpha=0$ is the proper alignment limit in the Higgs sector. However, it should be noted that in this limit the bosonic BSM sector completely decouples from the SM and that the one-loop corrections to \lamhhh are identical to the SM. Furthermore, in this limit charged lepton-currents are still not SM-like as they get modified by factors of $\mathcal{O}(m_{\nu_i}/m_{\nu_{i+3}}),\, i=1,2,3$. However, those contributions are numerically negligible. 

\subsubsection{Renormalisation}
All SM parameters are renormalised using the automatic OS/\MS procedures described above. The BSM sector is renormalised in the \MS scheme. Alternatively, we provide an \OS scheme for the $Z^{\prime}$ mass as well as a scheme with an \OS counterterm for $\alpha$ along the same lines as in the SSM and THDM.

\subsection{Minimal Supersymmetry (MSSM)}
\label{app:mssm}
A minimal supersymmetric version of the SM is also provided in terms of a \UFO model. The \UFO files were generated for the \cp-conserving but non-minimal flavor-violating MSSM using \SARAH. Therefore it is recommended to make use of the corresponding \SPheno spectrum generator in order to calculate parameter points for this model (see \eg the example in \cref{sec:applications_MSSM}). An example SLHA file produced with \SPheno is contained in the MSSM model directory and can be used \eg\ in the following way:
\begin{minted}[bgcolor=bg]{bash}
anyBSM MSSM \
  -f ~/.config/anyBSM/models/MSSM/SPheno.spc.MSSM --non-interactive
\end{minted}
using the command-line mode or 
\begin{minted}[bgcolor=bg]{python}
from anyBSM import anyH3
MSSM = anyH3('MSSM')
MSSM.setparameters(anyH3.built_in_models['MSSM'] + '/SPheno.spc.MSSM')
MSSM.lambdahhh()
\end{minted}
using the \py library.

\subsubsection{Renormalisation}
In contrast to the previously discussed models, the tree-level expression for $\lambda_{hhh}^{(0)}$ in the MSSM does not depend on the Higgs boson mass, but on $m_Z$, $v_{SM}$, $\alpha$, and $\tan\beta$. We are again able to automatically renormalise all SM-parameters OS. $\tan\beta$ and $\alpha$ are renormalised in the \DR scheme. Since \anyBSM by default uses dimensional regularisation (\ie, the \MS scheme), one needs to switch to dimensional reduction (\ie, the \DR scheme) in the \texttt{schemes.yml} of the MSSM model file in the following way
\begin{minted}[bgcolor=bg]{yaml}
dimensional_reduction: True
\end{minted}

\section{Additions to the \UFO standard}
\label{app:additionalufo}
While the \UFO standard itself provides a very general strategy to store detailed information about a particular QFT, we found that several additions and adjustments in its actual implementation are useful in order to overcome difficulties in the renormalisation procedure as well as to improve computing performance.

As a first step, \anyBSM expects various input parameter definitions to be present in the file \texttt{parameters.py} that are required for the renormalisation procedure. If the parameters do not exist in the \UFO~files, the program creates them using the \UFO object library with the according default values:
\begin{itemize}
        \item \code{Qren} is a parameter used for setting the renormalisation scale and defaults to $Q_{ren}=\unit[172.5]{GeV}$ (\ie~the top-quark mass).
        \item \code{GFermi} is used to set the Fermi constant. Alternative parameter names that are also searched for are \code{Gfermi, gFermi, gF, GF} and \code{Gf}. The default value is $\unit[G_F=1.16637~\cdot~10^{-5}]{GeV^{-2}}$.
        \item \code{Deltaalpha} determines the value for $\Delta \alpha$, defined in \cref{eq:sm:vacuumpol}. Alternative names are \code{dalpha, Dalpha} and \code{deltaalpha}. The default value, $\Delta \alpha=\Delta \alpha_{had.} + \Delta \alpha_{lep.}= 0.02766 + 0.031497687$, is taken from \ccite{Steinhauser:1998rq,ParticleDataGroup:2022pth}.
        \item \code{Zsignfac} is a parameter which fixes $\sing(\sin \theta_w)$ (\ie, the sign of the weak mixing angle, \cf\ \cref{eq:sm:deltae}). If it is not provided by the user, it is obtained automatically upon run-time using the method \code{getSignSinThetaW()} which determines the sign from the relative sign of the Z-top-top and photon-top-top couplings.
\end{itemize}
In addition, models are loaded with an \anyBSM-custom \texttt{object\_library.py} that adds some convenience and performance features:
\begin{itemize}
    \item The global \texttt{all\_<ufo object>} variables are dictionaries instead of lists.
    \item Warnings appear if \UFO objects with the same name (based on the \code{.name}-attribute) are initialised.
    \item A \code{nvalue}-attribute was introduced, containing the numerical value obtained with the current set of inputs.
    \item Similar to \code{nvalue}, a \code{nmass}-attribute for \code{Particle}- instead of \code{Parameter}/\code{Coupling}-objects was introduced. Note that a call of \code{setparameters()} (see \cref{sec:tutorial:setparams}) automatically updates the \code{nmass} and \code{nvalue} of all \UFO objects.
    \item A new method \code{UFOBaseClass.dump()}, which returns a string representation to be interpreted by \py, was implemented.
    \item \code{Particle.anti()} now avoids creating duplicate \code{Particle} instances on each call.
    \item \texttt{function\_library.py} was completely rewritten to avoid the call of \code{__exec__} of functions. This yields a significant performance boost for large models of up to \textit{3 orders of magnitude} in run-time.
\end{itemize}
It should be noted that if any of the auxiliary files mentioned above is present in the \UFO model directory, it is going to be ignored upon the model initialisation since only the modified \anyBSM methods are used.


\section{Caching in \anyBSM}
\label{app:cache}
One advantage of the \anyBSM framework compared to many other similar tools is that it provides a sophisticated caching mechanism. This allows us to significantly improve
the run-time of a phenomenological study. There are two levels of caching that can be set via \eg
\begin{minted}[bgcolor=bg]{python}
from anyBSM import anyBSM
SM = anyBSM('SM', caching = <cache-level>)
\end{minted}
with \code{<cache-level>=1} or \code{2}.
Alternatively one can set \code{SM.caching=1,2} after the initialisation. In the command line mode, \code{caching=n} is equivalent to the number of "c" options (\ie, \texttt{anyBSM SM -cc} corresponds to \code{<cache-level>=2} in the example above).

Setting \code{caching}$\geq$\code{1} writes the found particle insertions for all generic topologies for a specific model to disk. The insertions for all Feynman diagrams are determined using a brute-force algorithm. Therefore, this step can take up to several minutes for models with many particles. However, if \code{caching}$\geq$\code{1}, this step is only required once. The option \code{caching=2} (default) additionally caches the obtained analytic results to disk, such that the insertion of Feynman rules (\ie, definitions in the \texttt{vertices.py} and \texttt{couplings.py}) also have to be done only once. The analytic result saved to the cache is always expressed in terms of the abbreviations for the couplings defined in the \UFO ~model files rather than the Feynman rules themselves.

The exact behaviour of the program depends on the chosen evaluation mode. If \code{caching=2} but \code{evaluation_mode='numerical'} (\code{'analytical'}), the code internally first calculates the results using \code{evaluation_mode='abbreviation'} (if no result was found in the cache, otherwise the cached result is read from disk), saves it to the disk-cache and then evaluates it numerically (analytically). However, if \code{caching}$\textless$\code{2} the individual diagrams are evaluated directly using the representations for the couplings/masses that correspond to the active evaluation mode.

One useful example choice may be if the program encounters an infra-red (IR) divergent loop function and returns a corresponding warning message. In some cases, this may not be critical if the loop function is multiplied by a vanishing coupling (\ie, the diagram does actually not exist but was computed because the information that the couplings vanish is not manifest in the \UFO model). Thus, a good cross-check is to run the parameter point again without caching, which should yield the same numerical result but without a warning (since \anyBSM does not calculate a diagram if any of the couplings vanish numerically).

\section{\pyCollier}
\label{app:pycollier}
\pyCollier is a \py interface for the \COLLIER \texttt{Fortran} library~\cite{Denner:2016kdg}. In the current \pyCollier version, many but not all \COLLIER functions are available. The \pyCollier source code is hosted at
\begin{center}
  \url{https://gitlab.com/anybsm/pycollier}.
\end{center}
Running the code requires at least \py version 3.5. The code is most easily used by installing the corresponding \py package by running
\begin{minted}[bgcolor=bg]{bash}
pip install pyCollier
\end{minted}
which will automatically download and install \pyCollier as well as most necessary dependencies. One necessary requirement which is not automatically handled by \texttt{pip} is the presence of a \texttt{Fortran} compiler (required for the compilation of the \COLLIER library) such as \texttt{gfortran} or \texttt{CLANG} which can be installed from the system's package repository.

The \pyCollier module is loaded via
\begin{minted}[bgcolor=bg]{python}
import pyCollier
\end{minted}
Loop integrals can then be evaluated e.g. via
\begin{minted}[bgcolor=bg]{python}
pyCollier.set_renscale(125**2)
pyCollier.a0(125**2)
\end{minted}
where in the first step the renormalisation scale is set to $125\gev$. In the second step, the value of the $A_0$ scalar integral for an internal mass of $125\gev$ is calculated.

A detailed documentation of all available functions can be found at
\begin{center}
\url{https://anybsm.gitlab.io/pycollier/pyCollier.html}.
\end{center}

\clearpage
\printbibliography

\end{document}